\documentclass{elsart}

\usepackage[dvips]{graphicx}
\graphicspath{{figures/}}
\usepackage{subfigure}

\usepackage{amssymb}

\usepackage{times}

\usepackage{xspace}
\usepackage{cite}

\usepackage{units}

\usepackage{url}

\begin{document}

\begin{frontmatter}

{\onecolumn

\title{The magnetized steel and scintillator calorimeters of the MINOS 
experiment}

\collab{The MINOS Collaboration}

\newcommand{\Cambridge}{Cavendish Laboratory, Univ. of Cambridge, Madingley Road, Cambridge CB3 0HE, UK}
\newcommand{\FNAL}{Fermi National Accelerator Laboratory, Batavia, IL 60510, USA}
\newcommand{\RAL}{Rutherford Appleton Laboratory, Chilton, Didcot, Oxfordshire, OX11 0QX, UK}
\newcommand{\UCL}{Dept. of Physics and Astronomy, University College London, London WC1E 6BT, UK}
\newcommand{\Caltech}{Lauritsen Lab, California Institute of Technology, Pasadena, CA 91125, USA}
\newcommand{\ANL}{Argonne National Laboratory, Argonne, IL 60439, USA}
\newcommand{\Athens}{Department of Physics, University of Athens, GR-15771 Athens, Greece}
\newcommand{\NTUAthens}{Dept. of Physics, National Tech. Univ. of Athens, GR-15780 Athens, Greece}
\newcommand{\Benedictine}{Physics Dept., Benedictine University, Lisle, IL 60532, USA}
\newcommand{\BIMC}{Dept. of Rad. Oncology, Beth Israel Med. Center, New York, NY 10003, USA}
\newcommand{\BNL}{Brookhaven National Laboratory, Upton, NY 11973, USA}
\newcommand{\Beijing}{Inst. of High Energy Physics, Chinese Academy of Sciences, Beijing 100039, China}
\newcommand{\CdF}{APC -- Universit\'{e} Paris 7 Denis Diderot, 
F-75205 Paris Cedex 13, France}
\newcommand{\Columbia}{Physics Department, Columbia University, New York, NY 10027, USA}
\newcommand{\GEHealth}{GE Healthcare, Florence SC 29501, USA}
\newcommand{\DOE}{Div. of High Energy Physics, U.S. Dept. of Energy, Germantown, MD 20874, USA}
\newcommand{\Harvard}{Department of Physics, Harvard University, Cambridge, MA 02138, USA}
\newcommand{\HolyCross}{Holy Cross College, Notre Dame, IN 46556, USA}
\newcommand{\IIT}{Physics Division, Illinois Institute of Technology, Chicago, IL 60616, USA}
\newcommand{\Indiana}{Indiana University, Bloomington, IN 47405, USA}
\newcommand{\ITEP}{High Energy Exp. Physics Dept., ITEP, 
117218 Moscow, Russia}
\newcommand{\JMU}{Physics Dept., James Madison University, Harrisonburg, VA 22807, USA}
\newcommand{\JINR}{Joint Inst. for Nucl. Research, Dubna, Moscow Region, RU-141980, Russia}
\newcommand{\LASL}{Nucl. Nonprolif. Div., Threat Reduc. Dir., Los Alamos National Laboratory, Los Alamos, NM 87545, USA}
\newcommand{\LBL}{Physics Div., Lawrence Berkeley National Laboratory, Berkeley, CA 94720, USA}
\newcommand{\Lebedev}{Nuclear Physics Dept., Lebedev Physical Inst., 117924 Moscow, Russia}
\newcommand{\LLL}{Lawrence Livermore National Laboratory, Livermore, CA 94550, USA}
\newcommand{\MIT}{Lincoln Laboratory, Massachusetts Institute of Technology, Lexington, MA 02420, USA}
\newcommand{\Minnesota}{University of Minnesota, Minneapolis, MN 55455, USA}
\newcommand{\Crookston}{Math, Science and Technology Dept., Univ. of Minnesota -- Crookston, Crookston, MN 56716, USA}
\newcommand{\Duluth}{Dept. of Physics, Univ. of Minnesota -- Duluth, Duluth, MN 55812, USA}
\newcommand{\Oxford}{Sub-department of Particle Physics, Univ. of Oxford, Oxford OX1 3RH, UK}
\newcommand{\PSU}{Dept. of Physics, Pennsylvania State Univ., University Park, PA 16802, USA}
\newcommand{\Pittsburgh}{Dept. of Physics and Astronomy, Univ. of Pittsburgh, Pittsburgh, PA 15260, USA}
\newcommand{\IHEP}{Inst. for High Energy Physics, Protvino, Moscow Region RU-140284, Russia}
\newcommand{\RoyalH}{Physics Dept., Royal Holloway, Univ. of London, Egham, Surrey, TW20 0EX, UK}
\newcommand{\Carolina}{Dept. of Physics and Astronomy, Univ. of South Carolina, Columbia, SC 29208, USA}
\newcommand{\SLAC}{Stanford Linear Accelerator Center, Stanford, CA 94309, USA}
\newcommand{\Stanford}{Department of Physics, Stanford University, Stanford, CA 94305, USA}
\newcommand{\Sussex}{Dept. of Physics and Astronomy, University of Sussex, Falmer, Brighton BN1 9QH, UK}
\newcommand{\TexasAM}{Physics Dept., Texas A\&M Univ., College Station, TX 77843, USA}
\newcommand{\Texas}{Dept. of Physics, Univ. of Texas, 1 University Station, Austin, TX 78712, USA}
\newcommand{\TechX}{Tech-X Corp, Boulder, CO 80303, USA}
\newcommand{\Tufts}{Physics Dept., Tufts University, Medford, MA 02155, USA}
\newcommand{\UNICAMP}{Univ. Estadual de Campinas, IF-UNICAMP, CP 6165, 13083-970, Campinas, SP, Brazil}
\newcommand{\USP}{Inst. de F\'{i}sica, Univ. de S\~{a}o Paulo,  CP 66318, 05315-970, S\~{a}o Paulo, SP, Brazil}
\newcommand{\Washington}{Physics Dept., Western Washington Univ., Bellingham, WA 98225, USA}
\newcommand{\WandM}{Dept. of Physics, College of William \& Mary, Williamsburg, VA 23187, USA}
\newcommand{\Warsaw}{Faculty of Physics, Warsaw University, Hoza 69, PL-00-681 Warsaw, Poland}
\newcommand{\Wisconsin}{Physics Dept., Univ. of Wisconsin, Madison, WI 53706, USA}
\newcommand{\deceased}{Deceased.}


\author[Caltech]{D.G.~Michael\thanksref{deceased}},
\author[FNAL,UCL,Sussex]{P.~Adamson},
\author[Wisconsin]{T.~Alexopoulos\thanksref{alexopoulos}},
\author[Oxford]{W.W.M.~Allison},
\author[RAL]{G.J.~Alner},
\author[FNAL]{K.~Anderson},
\author[RAL,Athens]{C.~Andreopoulos},
\author[FNAL]{M.~Andrews},
\author[FNAL]{R.~Andrews},
\author[Stanford]{C.~Arroyo},
% ...

% ...
\address[Athens]{\Athens}
% ...
\address[Caltech]{\Caltech}
% ...
\address[FNAL]{\FNAL}
% ...
\address[UCL]{\UCL}
% ...
\address[Oxford]{\Oxford}
% ...
\address[RAL]{\RAL}
% ...
\address[Stanford]{\Stanford}
\address[Sussex]{\Sussex}
% ...
\address[Wisconsin]{\Wisconsin}

\thanks[cospokesmen]{Co-Spokesperson}
\thanks[corresponding]{Corresponding author. {\it Email address:}
  ahabig@umn.edu (A.~Habig)}
\thanks[deceased]{Deceased}
\thanks[alexopoulos]{Now at Dept. of Physics, National Tech. Univ. of Athens, GR-15780 Athens, Greece} 
\thanks[beall]{Now at Cleveland Clinic, Cleveland, OH 44195, USA}
\thanks[bergfeld]{Now at GE Healthcare, Florence SC 29501, USA}
\thanks[cabrera]{Now at APC -- Universit\'{e} Paris 7 Denis Diderot, 10, rue Alice Domon et L\'{e}onie Duquet, F-75205 Paris Cedex 13, France}
\thanks[brajesh]{Now at Dept. of Physics \& Astrophysics, Univ. of
  Delhi, Delhi 110007, India}
\thanks[demuth]{Now at Math, Science and Technology Dept., Univ. of Minnesota -- Crookston, Crookston, MN 56716, USA}
\thanks[desanto]{Now at Physics Dept., Royal Holloway, Univ. of London, Egham, Surrey, TW20 0EX, UK}
\thanks[frohne]{Now at Holy Cross College, Notre Dame, IN 46556, USA}
\thanks[ho]{Now at Dept. of Rad. Oncology, Beth Israel Med. Center, New York, NY 10003, USA}
\thanks[kim]{Now at Centre for Particle Physics, Univ. of Alberta,
  Edmonton, Alberta T6G 2G7 Canada}
\thanks[kostin]{Now at Fermi National Accelerator Laboratory, Batavia, IL 60510, USA}
\thanks[roylee]{Now at Lincoln Laboratory, Massachusetts Institute of Technology, Lexington, MA 02420, USA}
\thanks[wylee]{Now at Physics Div., Lawrence Berkeley National Laboratory, Berkeley, CA 94720, USA}
\thanks[mcgowan]{Now at Physics Department, St.~John Fisher College,
  Rochester, NY 14618, USA}
\thanks[reyna]{Now at Radiation and Nuclear Detection Systems, Sandia
  National Laboratories, Livermore, CA 94551, USA}
\thanks[schoessow]{Now at Euclid Techlabs LLC, Solon, OH 44139, USA}
\thanks[thron]{Now at Nucl. Nonprolif. Div., Threat Reduc. Dir., Los Alamos National Laboratory, Los Alamos, NM 87545, USA}


\begin{abstract}

  The Main Injector Neutrino Oscillation Search
  (MINOS) experiment uses an accelerator-produced neutrino beam to
  perform precision measurements of the neutrino oscillation parameters
  in the ``atmospheric neutrino'' sector associated with muon neutrino
  disappearance.  
  This
  long-baseline experiment measures neutrino interactions in
  Fermilab's NuMI neutrino beam with a near detector at Fermilab and
  again \unit[735]{km} downstream with a far detector in the Soudan
  Underground Laboratory in northern Minnesota.  The two detectors are
  magnetized steel-scintillator tracking calorimeters. They are designed
  to be as similar as possible in order to ensure that differences in
  detector response have minimal impact on the comparisons of event rates, energy
  spectra and topologies that are essential to MINOS measurements of
  oscillation parameters.  The design, construction, calibration and 
  performance of the far and near detectors are described in this paper.

\end{abstract}


} 

\end{frontmatter}



\section{Introduction}
\label{sec:intro}

The Main Injector Neutrino Oscillation Search (MINOS) experiment is
designed to perform precise measurements of neutrino oscillation
parameters from $\nu_\mu$ disappearance using an accelerator-produced 
muon neutrino beam.
MINOS uses two detectors, called ``near'' and ``far,'' to measure the
characteristics of an intense Fermilab neutrino beam over a baseline
distance of \unit[735]{km}. The two detectors are designed to be as 
similar as possible so that many details of their responses will 
cancel in comparisons of neutrino event characteristics between
the near and far ends of the baseline. The purpose of this paper is to
describe the design, construction, calibration and performance of the
detector systems used in the MINOS experiment. 
Brief overviews of the neutrino beam and the detectors are given below,
concluding with an outline of the detector system 
presentations which constitute the core of this paper.  

MINOS utilizes \unit[120]{GeV} protons from the Fermilab Main Injector
to create the high-intensity NuMI (Neutrinos at the Main Injector)
neutrino beam~\cite{Anderson:1998zz}.  The beamline is precisely aimed
in the direction of the Soudan Underground Laboratory in northern
Minnesota. The NuMI beam provides a high flux of neutrinos at the
end of the decay volume in the energy range $1 < E_\nu <
\unit[30]{GeV}$, but the flux at Soudan is reduced by a factor of about
$10^6$ due to the intrinsic divergence of the beam. 
The relative rates of neutrino charged-current interactions in the MINOS 
near detector at Fermilab 
are approximately 92.9\%~$\nu_\mu$, 5.8\%~$\overline{\nu}_{\mu}$, 1.2\%~$\nu_e$
and 0.1\%~$\overline{\nu}_{e}$ for the low-energy beam configuration. 
With the parameters for $\nu_\mu$ to $\nu_\tau$ oscillations measured by
Super-Kamiokande~\cite{Ashie:2005ik} and other experiments ($\Delta
m^2_{32} \simeq 2.5 \times \unit[10^{-3}]{eV^2}$ and $\sin^2
2\theta_{23} \simeq 1.0$), 
the Fermilab-to-Soudan distance implies that
the neutrino interactions of most interest will be in the
$1<E_\nu<\unit[5]{GeV}$ range.  Details of this experiment's measurements are
published elsewhere~\cite{Michael:2006rx,Adamson:2007gu}. 

The MINOS experiment uses two detectors to record the interactions of
neutrinos in the NuMI beam.
A third detector, called the calibration detector, was
exposed to CERN test beams in order to determine detector response.
The near detector at Fermilab is used to characterize the
neutrino beam and its interactions and is located about \unit[1]{km}
from the primary proton beam target, the source of the neutrino parent
particles. The far detector performs similar measurements
\unit[735]{km} downstream.  The essence of the
experiment is to compare the rates, energies and topologies of events at
the far detector with those at the near detector, and from those
comparisons determine the relevant oscillation parameters. The energy
spectra and rates are determined separately for $\nu_\mu$ and ~$\nu_e$
charged-current (CC) events and for neutral current (NC) events.

All three MINOS detectors are steel-scintillator sampling calorimeters with
tracking, energy and topology measurement capabilities. This is achieved
by alternate planes of plastic scintillator strips and \unit[2.54]{cm}
thick steel plates. 
The near and far detectors have magnetized steel planes.
The calibration detector was not magnetized as the particle momenta were
selected a priori.
The \unit[1]{cm} thick by \unit[4.1]{cm} wide extruded polystyrene
scintillator strips are read out with wavelength-shifting fibers and
multi-anode photomultiplier tubes. All detectors provide 
the same transverse and
longitudinal sampling 
for fiducial beam-induced events.  

\begin{figure*}[htpb]
  \begin{center}
    \includegraphics[width=\textwidth,keepaspectratio=true]{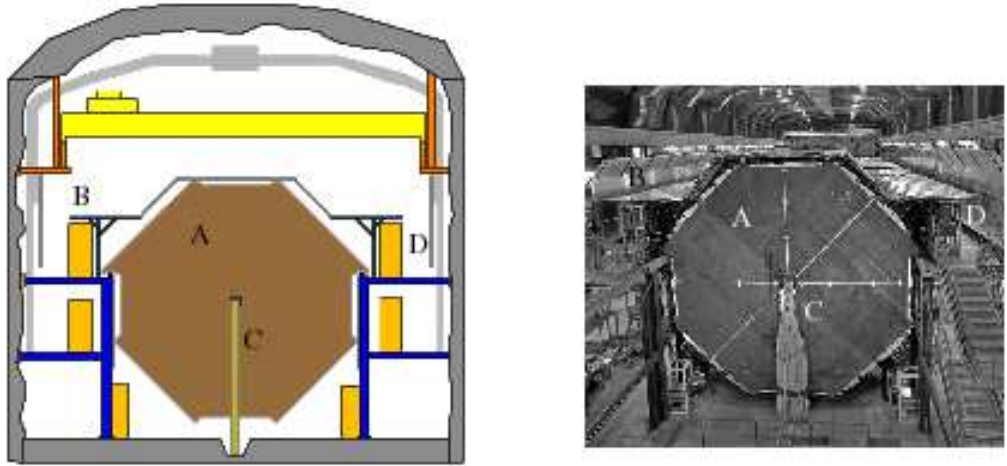}
    \caption{End views of the second far detector supermodule, looking
      toward Fermilab. The drawing (left) identifies detector
      elements shown in the photograph (right): `A' is the furthest
      downstream steel plane, `B' is the 
      cosmic ray 
      veto shield, `C' is the end
      of the magnet coil and `D' is an electronics rack on one of
      the elevated walkways alongside the detector. The
      horizontal structure above the detector is the overhead crane
      bridge. }
    \label{fig:MINOSfar}
  \end{center}
\end{figure*}

The far detector,
shown in Fig.~\ref{fig:MINOSfar}, 
is located in Soudan, MN (\unit[47.8$^\circ$]{N}
latitude, and \unit[92.2$^\circ$]{W} longitude), \unit[735.3]{km} from
the NuMI beam production target at Fermilab, 
in an inactive iron mine currently
operated as a State Park by the Department of Natural
Resources of the State of Minnesota.  Much of the infrastructure used in
the mining days is still in service and is 
used to support the
operation of the experiment.  The detector is housed in a specially
excavated cavern, \unit[705]{m} underground
(\unit[2070]{meters-water-equivalent}), \unit[210]{m} below sea level.
The far detector consists of 486 
octagonal steel planes, with edge to edge dimension of \unit[8]{m},
interleaved with planes of plastic scintillator strips.  This
\unit[5,400]{metric ton} detector is constructed as two 
``supermodules'' axially separated by a \unit[1.15]{m} gap.  Each
supermodule has its own independently controlled magnet coil.  The first
(southernmost) supermodule contains 249 planes and is \unit[14.78]{m} in length while the second supermodule
is comprised of 237 planes and has a length of \unit[14.10]{m}.  The
most upstream planes in each supermodule (planes 0 and 249) are uninstrumented.
The north end view of the second supermodule is shown in
Fig.~\ref{fig:MINOSfar}.

\begin{figure*}[htpb]
  \begin{center}
    \includegraphics[width=\textwidth,keepaspectratio=true]{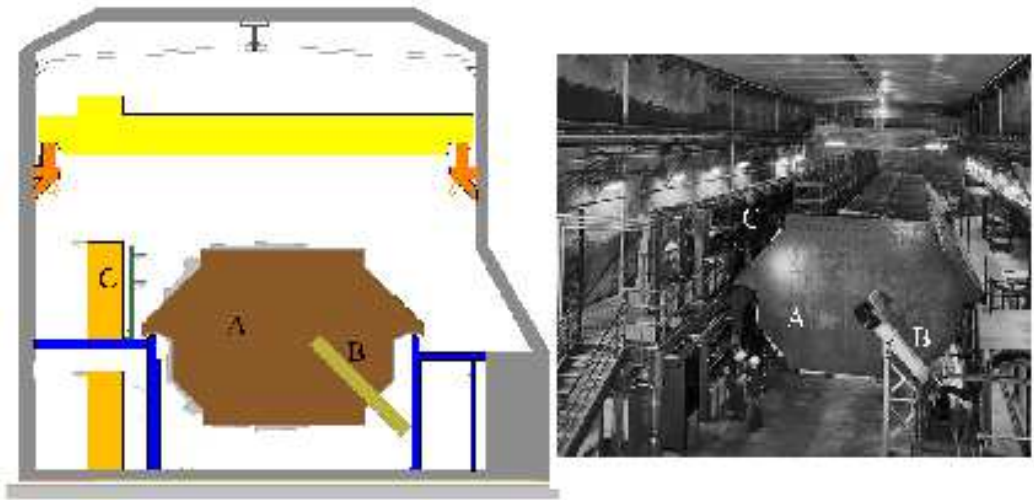}
    \caption{End view of the near detector, looking toward Soudan. The
      drawing (left) identifies detector elements shown in the
      photograph (right): `A' is the furthest upstream steel plane,
      `B' is the magnet coil, and `C' is an electronics rack on the
      elevated walkway.
      Above the detector is the overhead crane bridge.  The
      NuMI beam intersects the near detector near the ``A'' label.}
    \label{fig:MINOSnear}
  \end{center}
\end{figure*}

The \unit[282]{plane}, \unit[980]{metric ton} MINOS near detector, shown
in Fig.~\ref{fig:MINOSnear}, is located at the end of the
NuMI beam facility at Fermilab in a \unit[100]{m} deep underground
cavern, under a \unit[225]{mwe} overburden.  The design of the near detector takes advantage of the high
neutrino flux at this location to define a relatively small target
fiducial volume for selection of events for the near/far comparison. The
upstream part of the detector, the calorimeter section,
contains the target fiducial volume in which every plane is
instrumented. The downstream part, the spectrometer
section, is used to measure the momenta of energetic muons and has only
every fifth plane instrumented with scintillator.

The much smaller calibration detector
was used to measure the detailed
responses of the MINOS detectors in a charged-particle test beam. This
\unit[12]{ton} detector consisted of \unit[60]{planes} of unmagnetized
steel and scintillator, each 1$\times$1 m$^2$. 
It was exposed to
protons, pions, electrons and muons in test beams at the CERN
PS~\cite{Adamson:2005cd} to measure the energy and topological responses
expected in the the near and far detectors. In order to include
response variations due to the different electronics used in the
detectors, the calibration detector acquired data with both near and far
detector electronics.  The energy responses of the three MINOS detectors
were normalized to each other by calibrating with cosmic-ray muons.


The choice of solid scintillator as the MINOS detectors' active system
was the result of a three year research and development process which
also evaluated the possibility of liquid scintillator, Iarocci tubes (or
their variants) and RPCs~\cite{Ables:1995wq,minosrandd}.  Solid
scintillator was chosen for a number of reasons (not all unique to this
technology): good energy resolution, excellent hermiticity, good
transverse segmentation, flexibility in readout, fast timing, simple and
robust construction, potential for distributed production, long-term
stability, ease of calibration, low maintenance, and
reliability.  The MINOS Detectors Technical Design
Report~\cite{:1998zzb} summarizes the decision making process as follows:
\begin{quote}
  The development program has included extensive laboratory tests of
  different active detector technologies, test beam work, Monte Carlo
  simulations of reactions of interest to MINOS, and evaluation of the
  costs of different options. We believe that this baseline design
  represents the best experimental approach, in light of the current
  knowledge of neutrino oscillation physics, and also offers a high
  probability of being able to react effectively to potential future
  physics developments.
\end{quote}
Safety and practicality of construction were also important criteria.
After the solid scintillator decision was made, a second phase of design
optimization occurred, where parameters such as steel thickness, width
of scintillator strips, and degree of readout multiplexing were set
based on Monte Carlo studies~\cite{:1998zzb}.  These were trade-offs
between cost and performance.  The channel that is most sensitive to
these choices is
$\nu_\mu\rightarrow\nu_e$ appearance. Narrower strips and thinner steel
plates would improve $e$ identification ability, however that gain was
counterbalanced by the loss in statistics (for the same construction
cost). Another consideration for strip width was muon energy resolution;
but that was dominated by Coulomb scattering (or range measurement) and
not very sensitive to the strip width. The steel thickness is also relevant
for shower energy resolution but \unit[2.5]{cm} was adequate. Note
that the ability to react to developments in the field was tested when
the high $\Delta m^2$ value hinted at by the original Kamiokande
measurements~\cite{Fukuda:1994mc} was superseded by the current low
$\Delta m^2$~\cite{Fukuda:1998mi} after the civil construction had begun
and these design decisions had already been made.


The MINOS detectors required a significant scale-up in size from previous
fine-grained scintillator sampling calorimeters, hence creative 
reductions in costs per unit of the scintillator and electronics
systems resulted.  The final design includes advances in detector 
technology which will be of interest to future detector
applications requiring large areas of plastic scintillator.

The MINOS near and far detectors have now been been operating for
several years, both with cosmic-ray events and with
the accelerator neutrino beam. The far detector started commissioning
data collection in September~2002, and has been fully operational for
cosmic-ray and atmospheric-neutrino data since July~2003. The
near detector has been operating since January~2005.  The NuMI beam
started providing neutrinos to the MINOS experiment in March 2005. 

This paper summarizes the considerations that have driven
the detector designs, provides details of individual subsystems, describes
the construction and installation issues, and presents performance data from
operational experience and from bench measurements of subsystems. It also
 provides a framework for more detailed publications,
either already in print or in preparation, which discuss specific
detector subsystems.

The remainder of this paper is organized as follows:
Section~\ref{sec:steel} describes the steel detector planes, the
magnetic coils, and the resulting detector magnetization. Section~\ref{sec:scint} contains a detailed
description of the scintillator system that is the heart of the MINOS
detectors. It includes the design and fabrication of the scintillator
strips and the characteristics of the wavelength-shifting fibers and
photomultiplier tubes that read them out. It describes the assembly of
extruded plastic scintillator strips into modules and the performance of
those modules. Section~\ref{sec:elec} covers the electronics and data
acquisition systems for the near and far detectors; these require
different front-end designs because of the very different counting rates
at the two detector locations. Section~\ref{sec:calib} gives detailed
descriptions of the calibrations of the two detectors and their
electronic readout systems. 
Section~\ref{sec:install}
describes the facilities of the underground laboratories in which the
near and far detectors are located. It also covers the installation of
the detectors and the survey techniques used to determine the direction
of the far detector from Fermilab as required for the precise
aiming of the neutrino beam. Section~\ref{sec:ops} documents the overall
performance of the detector systems in the MINOS experiment and also
includes a brief description of the computer software used to measure
performance and analyze MINOS data.  
Section~\ref{sec:conclude} concludes with a brief summary of detector 
performance as observed in data-taking currently underway.


\section{Steel Planes, Magnet Coils and Magnetic Fields}
\label{sec:steel}

The MINOS near and far detectors are sampling calorimeters that utilize toroidally magnetized, \unit[2.54]{cm} thick steel
planes~\cite{Nelson:2001gm} as the passive absorber material.  The differences in beam sizes and neutrino interaction
rates at the near and far detector sites led to substantially different magnetic designs for the two detectors and allowed 
the near detector to
be much smaller and less costly than the far detector. This section summarizes the specifications, designs, and performance 
of the steel and coils.

\subsection{Magnet design} 
\label{sec:steel-req}

The MINOS magnets are designed to provide a measurement of muon momentum based on curvature with 
resolution of $\sigma_P/P\sim$12\% for muons with energies greater than \unit[2]{GeV}, and to facilitate the containment of negatively charged muons.  
 The average fields in the near and far detectors were required to have similar strengths to minimize systematic uncertainties 
arising from near/far detector differences.  The field strength averaged over the
fiducial volume in 
the near detector \unit[1.28]{T}, 
compared to \unit[1.42]{T}  in the far detector.  
One of the design goals of the 
magnet system is that the average magnetic field in each toroid be known to better than 3\%. Monte Carlo studies indicate that 
uncertainties in the magnetic field strength at this level result in detector acceptance uncertainties of significantly less 
than 1\% at all muon energies of interest, and an average uncertainty in the energy of exiting tracks of less than 2\%.   The magnetic
calibration specifications require that stochastic variations in field
strength between different steel planes not significantly degrade 
the overall momentum resolution. Monte Carlo studies of the effect of plane-to-plane field variations on momentum resolution 
provide a specification of stochastic residual variations (after global calibration) of less than 15\%. 

\subsection{Steel planes}

\label{sec:steel-planes}

\subsubsection{Far detector configuration}
\label{sec:steel-planes-fd}

\begin{figure*}[htpb]
  \centering
  \includegraphics[width=0.9\textwidth]{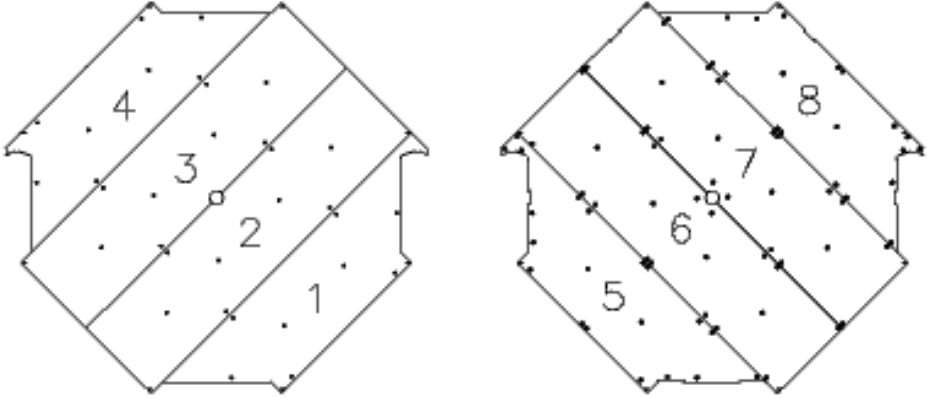}
  \caption{Arrangement of steel plates in the two layers of a far
    detector plane, showing the plate numbering scheme. A drawing of
    the bottom (downstream) layer is on the left and the top (upstream) layer
    is on the right, both seen looking toward Fermilab. ``Top'' and
    ``Bottom'' refer to how they were stacked when being assembled.  
    The dots indicate
    holes for the plug welds or handling fixtures. Numbering is along the U (left) and
    V (right) axes. The installed planes are supported by the
    ``ears'' on plates 1, 4, 5 and 8.}
  \label{fig:steelparts}
\end{figure*}

The MINOS far detector has 486 steel planes, each one constructed of eight component plates.  All detector components
were moved underground through the existing mineshaft, which limited
dimensions to \unit[9$\times$2$\times$1]{m$^3$} and weights to
\unit[5.5]{metric tons}
or less. Each \unit[8]{m} wide octagonal steel plane was constructed underground by plug-welding together eight \unit[2]{m} wide,
\unit[1.27]{cm} thick plates.  After attaching scintillator-strip modules to one side, the planes were mounted vertically with a
\unit[5.95]{cm} center-to-center spacing.  The basic far detector steel plane construction is shown in Fig.~\ref{fig:steelparts}. 

\subsubsection{Steel plane construction}
\label{sec:steel-planes-construction}

Each plane was assembled on a steel lift frame (called a ``strongback''), which was used to lift the
completed plane onto the support structure.  Each plate is identified by a part number specifying where it fits in the
octagon, a ``heat number'' specifying the batch of steel from which it is made, and a serial number unique to that plate.  

The construction of each of the 486~detector planes began with the placement and alignment of four steel sheets on a strongback 
to form the bottom of two layers.  The top layer was then placed and
aligned in the orthogonal direction.  The plates were placed to minimize gaps between sheets.  The eight sheets 
of a complete plane were then welded together via seventy-two~\unit[2.5]{cm}
diameter plug-weld holes in the top set of sheets 
(Fig.~\ref{fig:steelparts}).  
Surviving gaps of greater than \unit[2]{mm}  
were measured and recorded.  Most of these gaps were in the range of \unit[2--4]{mm} and at 
most \unit[9]{mm}.  Fewer than half of the seams had recordable gaps, 
typically located at the outer edge of the seam and about \unit[30]{cm}
to \unit[50]{cm} 
long. 
Following the assembly of the steel planes, the scintillator detectors were mounted on the plane and the full assembly 
lifted into place. 

The assembled planes are supported on two rails, one on each side of the detector.  Each plane is bolted to the
previously-installed plane with 8~axial bolts around the periphery and 8~additional bolts around the central coil hole.  The steel planes are
magnetically isolated from the steel support structure by \unit[1]{cm} thick stainless steel strips between the plates and the support rails.
Plumbness and plane to plane alignment were obtained by checking each plane as it was installed using a laser
survey device (Ref.~\cite{Bocean:2004dj} and Sec.~\ref{sec:install-align}), occasionally adding shims as needed when new planes were bolted to a
supermodule to maintain the specification of \unit[6.4]{mm} plumbness.

\subsubsection{Steel characteristics}
\label{sec:steel-planes-mass}

The steel plates were made from low-carbon (AISI 1006 designation)
hot-rolled steel.  They were required to have flatness to better than
\unit[1.5]{cm} -- half the ASTM~A-6
specification~\cite{astma6}.  The carbon content was specified to be
(0.04$\pm$0.01)\%.  Samples from each of the 45 foundry runs (called ``heats'') were
tested to ensure that their radioactivity was less than \unit[0.15]{$\gamma$/kg/sec} for $\gamma$-rays above \unit[0.5]{MeV}.
From block samples of the various heats, the average steel density is found to be \unit[7.85$\pm$0.03]{g/cm$^3$}. 

As steel was delivered over the course of construction, each plate was individually weighed using a scale with a least count of \unit[0.9]{kg},
and this value was compared to a nominal weight for that part number.   The scale calibration was checked and verified to be stable during construction.  An uncertainty of \unit[1]{kg} in the plate masses implies a plane-mass uncertainty of
$\unit[\sqrt{8}]{kg}\simeq\unit[3]{kg}$.  Deviations from the nominal weight were found to be correlated with variations in the 
thickness of the steel. 
The first 190 (upstream) planes had an average mass of \unit[10,831]{kg}
and the remaining 296 (downstream) planes had an average mass of
\unit[10,718]{kg}.  The rms mass variation within each group of planes
is 0.35\%, which grows to 0.62\% if the detector is considered as a whole.

Requirements on the accuracy of the target mass and on muon range measurements imposed the specification that the fiducial masses of the
near and far detectors be known to 1\%.  The average thickness of the near detector planes was measured to be \unit[2.563$\pm$0.002]{cm},
compared to \unit[2.558$\pm$0.005]{cm} for the far detector. 

\subsubsection{Near detector steel}
\label{sec:steel-planes-nd}

The near detector was assembled from 282~steel planes, fabricated as single plates of \unit[2.54]{cm} thickness from a subset of the same
foundry heats used for the far detector steel.  The near detector target (fiducial) region was chosen to be 
\unit[2]{m} in diameter to give a high rate of fully contained neutrino interaction events in the central region of the beam. The magnet coil
hole in the steel plates was located outside this area.

Plate thickness variations in the near detector planes were found to be $\sim$0.3\% by surveying with an ultrasound probe.
No systematic difference in steel density was found between the two
detectors.  
As was required for the far detector steel, the flatness specification for the near detector plates was set at half
of the ASTM~A-6 flatness standard, or \unit[1.5]{cm}.  

\subsection{Magnet coils}
\label{sec:steel-coil}

The near and far detector steel geometries place somewhat different
requirements on their respective magnet coil designs. The coil designs were optimized separately, taking into account differing detector geometry as well as differences in the laboratory
infrastructures available at Fermilab and Soudan.

\subsubsection{Far detector coil}
\label{sec:steel-coil-fd}

\begin{figure}[htpb]
  \centering
  \includegraphics[width=0.7\columnwidth]{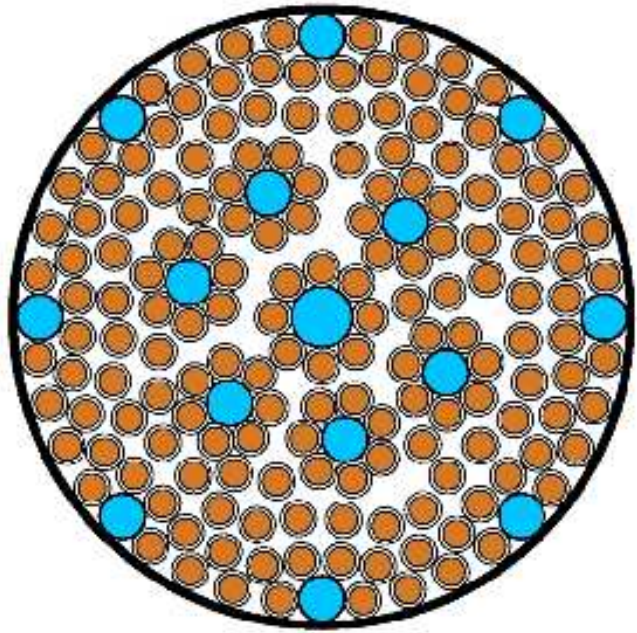}
  \caption{Cross section schematic of one of the far detector supermodule
  coils. The larger diameter circles represent the copper cooling tubes and the
  smaller circles are the 190 turns of 1/0 gauge stranded copper wire. The
  outlines of these conductors are to-scale representations of the insulator
  thickness. The outer circumference of the assembly is a copper-sheet jacket
  cooled by eight cooling tubes.}
  \label{fig:fdcoil}
\end{figure}

Each supermodule is independently magnetized by its own
coil~\cite{Kilmer:1999fd}, as shown by item ``C'' in Fig.~\ref{fig:MINOSfar}.
Each coil consists of a central bore leg running through holes at the center of
each plane, a single return leg located in a trench beneath the detector, and end legs that connect the bore to the return legs. Figure~\ref{fig:fdcoil} shows 
a schematic cross section of the coil in the central bore leg, inside a supermodule. The conductor consists of 190~turns of 1/0 gauge stranded copper wire with
Teflon insulation (National Electrical Code designation TGGT).  The bore leg is housed inside a \unit[25]{cm} diameter, water-cooled copper
jacket. The return leg is also water cooled and the end legs are air-cooled. An \unit[80]{A} power supply gives a \unit[15.2]{kA}-turn
total current that provides an average toroidal magnetic field of
\unit[1.27]{T}.  Each coil dissipates \unit[20]{kW}. 

In order to minimize temperature induced aging of nearby scintillator,
the outer jacket characteristics were designed to ensure a worst case
maximum temperature of \unit[150$^\circ$]{C}.
Each coil's cooling-water system carries \unit[72]{l/min} and was designed to remove up to \unit[25]{kW} of heat per supermodule. A secondary
heat-exchange system removes the heat from the underground laboratory.  Fixtures along the air-cooled end legs of the coil
provide a \unit[15]{cm} separation between the coil and the steel planes to allow air circulation and to reduce distortion of the field in
supermodule end planes by the current in the end-legs of the coil.

\subsubsection{Near detector coil}
\label{sec:steel-coil-nd}
\begin{figure}[htpb]
  \centering
  \includegraphics[width=0.7\columnwidth]{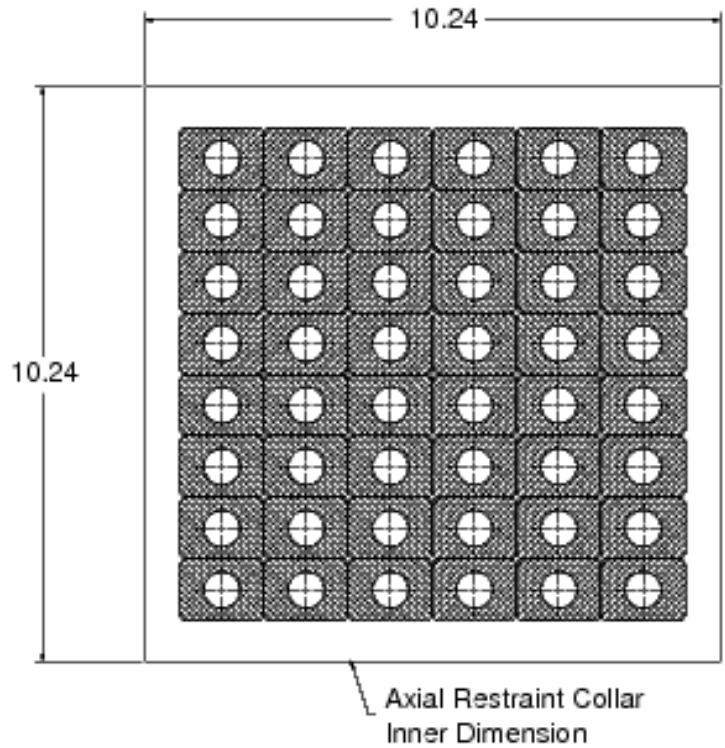}
  \caption{Cross section schematic of the near detector 
  coil. The dimensions
  shown are in inches.}
  \label{fig:ndcoil-xsec}
\end{figure}

\begin{figure}[htpb]
  \centering
  \subfigure{
    \includegraphics[width=0.45\columnwidth]{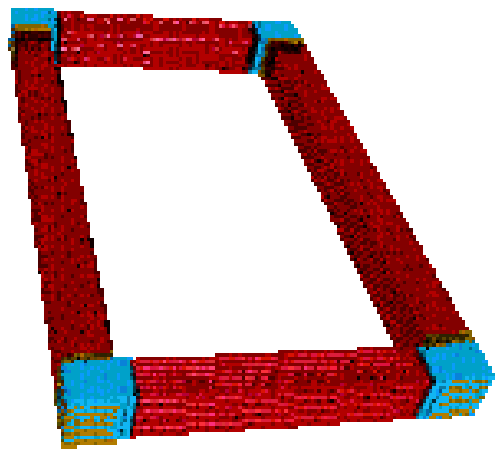}
  }
  \subfigure{
    \includegraphics[width=0.45\columnwidth]{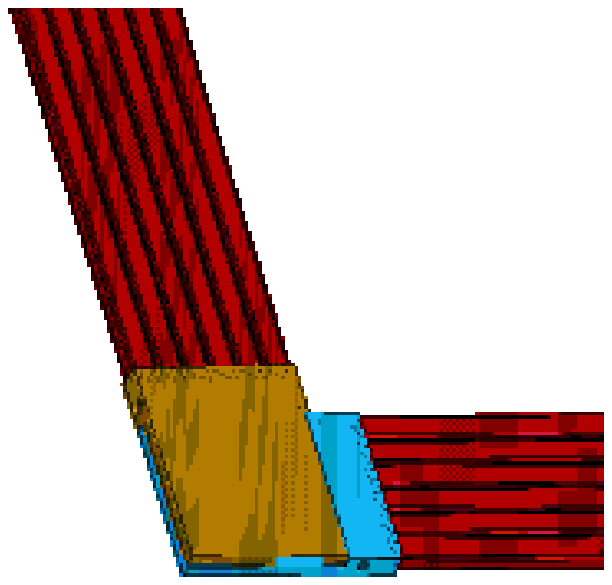}
  }
  \caption{Sketches of the four legs of the near detector coil assembly (left)
  and of one lap joint between two coil planks (right).}
  \label{fig:ndcoil-assy}
\end{figure}

The near detector coil hole is offset \unit[55.8]{cm} from the center of the plane and the detector is placed so that beam is centered halfway
between the hole and the left vertical edge of the plane, as shown in Fig.~\ref{fig:MINOSnear}. 
Because of the squashed-octagon geometry, a \unit[40]{kA-turn} current is required to
achieve sufficient fields. Figures~\ref{fig:ndcoil-xsec} and \ref{fig:ndcoil-assy} show the cross section and the geometry of
the near detector coil, respectively. The coil~\cite{Kilmer:1999nd} consists of eight turns, each with
\unit[18.76]{m-long} bore and return legs and two \unit[2.89]{m}-long end legs that connect the bore and return. The return leg is routed along the
lower east 45$^\circ$ face of the steel plane.  The high current carried by the coil requires substantial cooling, provided by a
closed loop low-conductivity water system that transfers the heat out of
the underground enclosure. There are no photodetectors on the coil-return side of the near
detector by design, so the fringe fields from the return do not affect detector operation. 

The coil conductor is made from cold conformed aluminum and has a \unit[2.79$\times$3.81]{cm$^2$} rectangular cross section with a
\unit[1.65]{cm} diameter central water channel. The 48 conductors are arranged in a six by eight rectangular pattern, with groups of six conductors
formed into ``planks.''  The current runs in parallel through the conductors within a plank. The electrical connections were made with
full-penetration aluminum welds at each end. This offers the potential to disassemble the coil for repair or replacement in case of
failure. The coil is a single eight-turn \unit[5]{kA} electrical circuit
which dissipates a power of \unit[47]{kW}.  Cooling water of less than
\unit[$80^\circ$]{C} flows through the coil at
\unit[380]{l/min}, limiting conductor temperature.

\subsection{Detector plane magnetization}
\label{sec:steel-magnetization}

\subsubsection{Magnetic field determination}
\label{sec:steel-coil-perform-calib}

\begin{figure*}[htpb]
  \centering
  \subfigure{
    \includegraphics[width=0.45\textwidth]{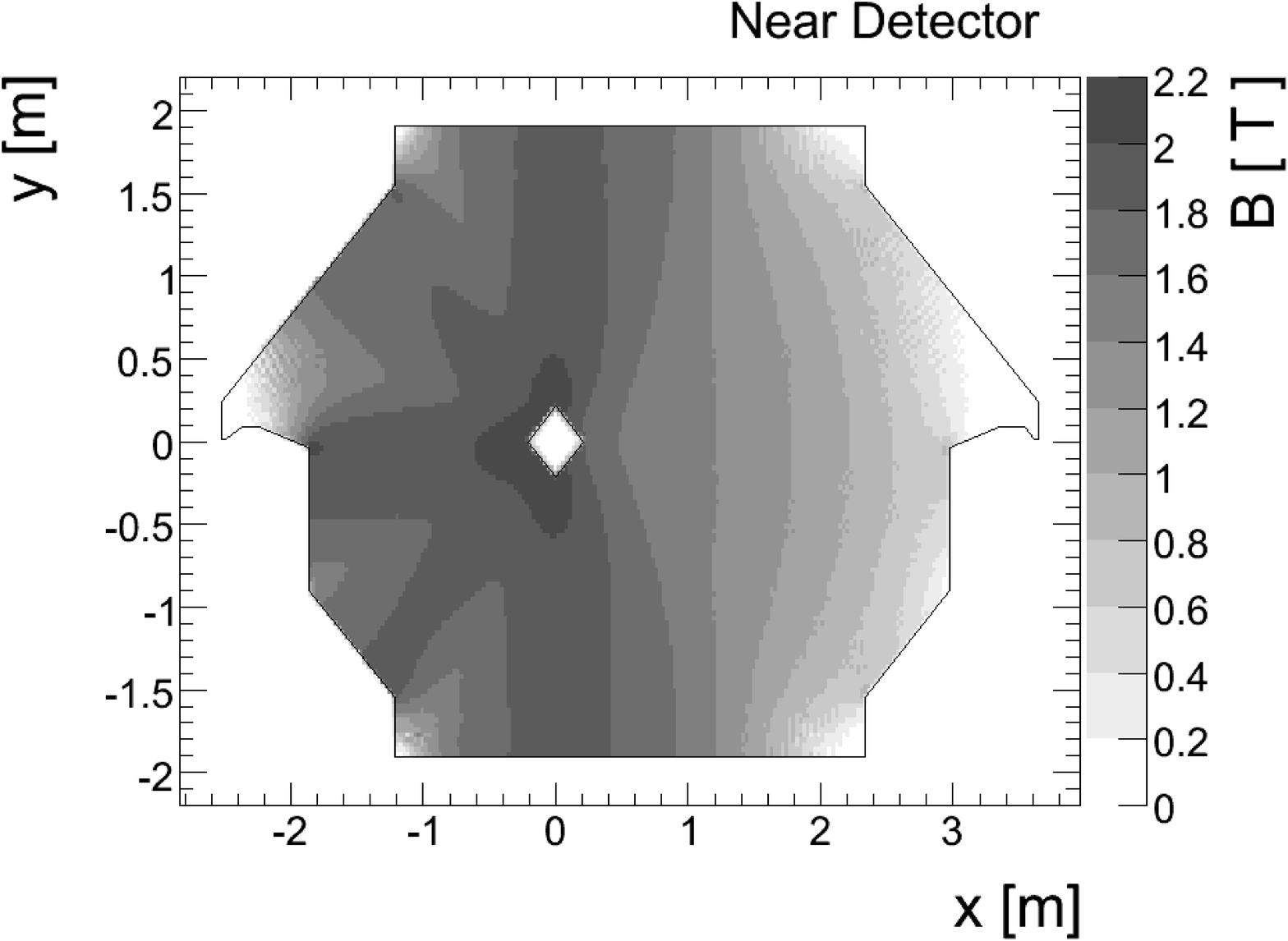}
  }
  \subfigure{
    \includegraphics[width=0.45\textwidth]{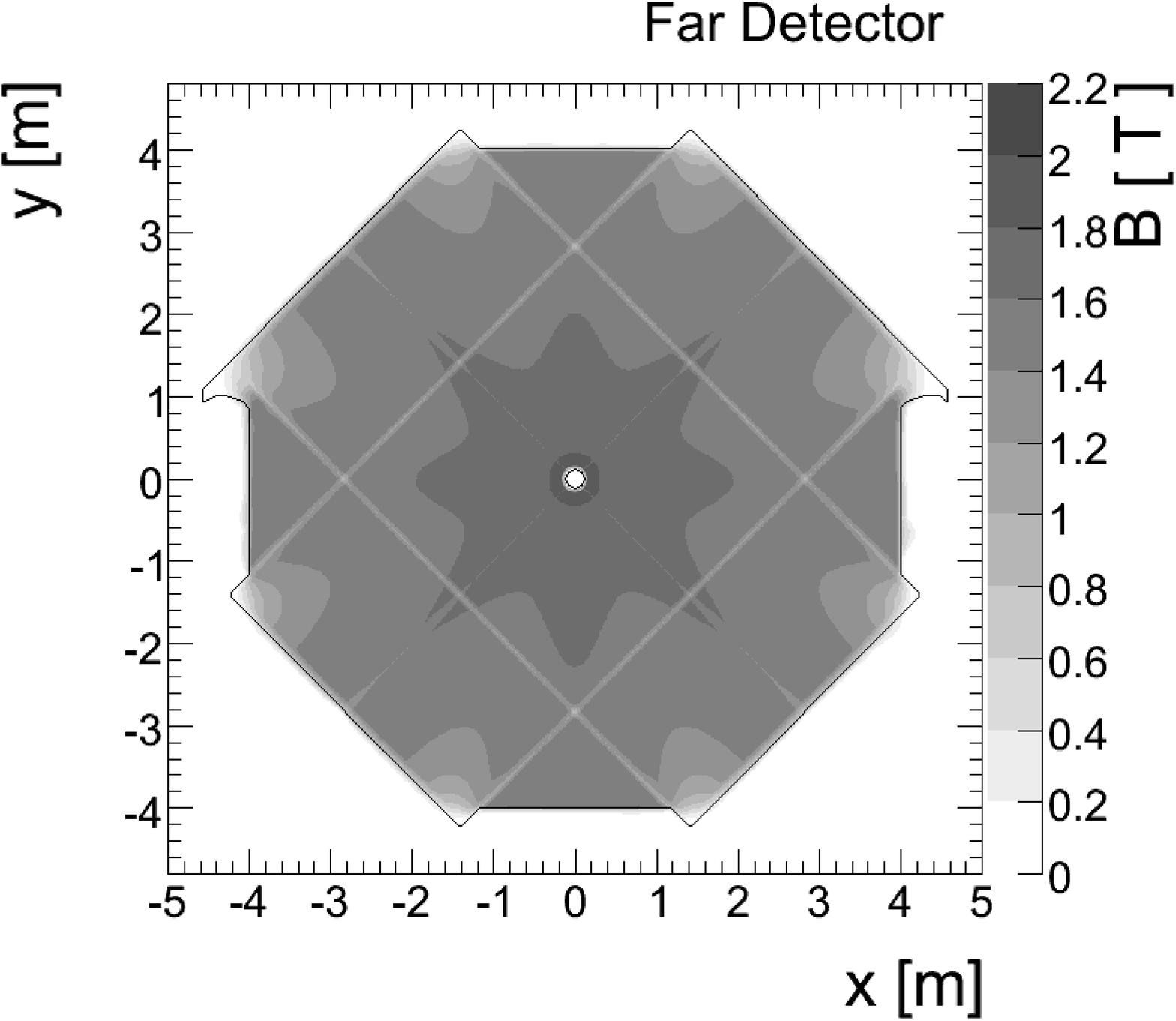}
  }
  \caption{Magnetic field maps for a typical near (left) and far (right)
    detector plane.  The greyscale indicates the magnetic field strength
    $B$ as calculated by finite element analyses using 3-D models.  Note
    that the near detector plane is shown looking upstream to the
    neutrino beam in this figure, whereas in Fig.~\ref{fig:MINOSnear} the view is to
    the downstream direction.  The
    effect on the field of joints between the steel pieces used to make
    the far detector planes is seen as the straight light grey lines
    in the lower figure.}
  \label{fig:steelfig}
\end{figure*}

The finite element analyses (FEA) of both the near and far detectors'
magnetic fields 
were performed with the ANSYS~\cite{ansys} general purpose finite
element program, using a 3-D scalar magnetic potential approach. The
accuracy of the field values depends on the mesh density
(discretization) of the model, the input magnetization (``B-H'') curve, and the
normalization to coil currents set using power-supply
current shunts.  Figure~\ref{fig:steelfig} shows the results of FEA
calculations of the near and far detector magnetic field maps for
detector planes near the
detector centers.  

There are a number of potential sources of
plane-to-plane magnetic field variations, including mechanical and
chemical nonuniformity and field distortion at the ends of the toroids.
The steel for the two detectors was produced in 45 different foundry
heats with slightly different chemical compositions (and hence magnetic
properties).  Test toruses were fabricated from the steel in each heat
and used to measure B-H curves by magnetic induction. The variations in
these B-H curves between heats were found to be small, allowing all plane
field maps to be based upon a single representative B vs H relationship.
FEA calculations confirmed that expected mechanical variations between
planes, such as variations in the gap between steel sheets in the far
detector, yield less than 15\% field differences.  Finally, the presence of
coil end legs introduces field distortion in the end planes of each
toroid.  These end effects were shown to have a negligible effect on
momentum measurements at the level required for early MINOS results, and
for these studies a single field map appropriate to the center of the
near or far detector was used for all planes. For subsequent analysis and
simulation, the
effect of the proximity of the coil return on the field of planes near a
detector or supermodule end is accounted for via interpolation.  The FEA
generated fields for the end plane, the third plane, and an interior plane
are interpolated to the outer ten planes in each detector.  The accuracy
of this interpolation technique was confirmed by comparing  
its values to actual field maps of the last 12 planes. 
The residual RMS field errors from the interpolation
procedure have been shown to be less than \unit[5]{Gauss} for all
intermediate planes.  

To normalize the field maps,
each of the near and far detector planes is equipped
with 50-turn magnetic induction coils that measure the average magnetic flux along a line from the coil to the detector periphery at discrete angles.  
When instrumented with precision analog integration readouts, the induction coils provide a measurement of the average magnetic flux through the steel 
with uncertainties of less than 2\%. 

A second method of determining the magnetic field map comes from
analyzing the trajectories of stopping muons.
For this class of events two redundant measurements of muon momentum are available, one based on range ($P_{range}$), and the other on the 
measured curvature of the track ($P_{fit}$). The systematic error in the range-based momentum arises primarily from uncertainty in the detector mass, 
approximations to the true detector geometry made in the reconstruction software, and uncertainties in the underlying energy loss model used. 
The combination of these effects results in a 2\% systematic error in the
track momentum from range.
The ratio $P_{range}/P_{fit}$ therefore provides a means of assessing the 
consistency of the range-based and curvature-based energy scales for
contained tracks to the same level of precision.   

The comparison of the $P_{range}/P_{fit}$ ratio between the Monte Carlo
(MC) simulation,
where the magnetic field is known perfectly, and the data gives
the estimate of the magnetic field uncertainty in measuring muon
momentum from curvature in the actual detector. The double
ratio $(P_{range}/P_{fit})_{data}/(P_{range}/P_{fit})_{MC}$
directly compares two methods of determining muon momentum for the
data and the simulation and does not depend on reconstruction effects. The
well defined muon tracks produced in the neutrino interactions of beam
neutrinos in the near detector, when analyzed for the measurement of the
magnetic field uncertainty, produce the value
$(P_{range}/P_{fit})_{data}/(P_{range}/P_{fit})_{MC} \approx 1.01$~\cite{Ospanov:2008zza}.
\noindent This result is consistent with both the final measurement of the magnetic
field strength and the uncertainty on the range measurement of
the stopping muon tracks in the data and the simulation.


\section{Scintillator system}
\label{sec:scint}

The MINOS scintillator system consists of approximately 100,000 extruded 
polystyrene scintillator strips, each \unit[4.1]{cm} wide, \unit[1.0]{cm} 
thick and up to \unit[8]{m} long; the total surface area of this system 
is \unit[28,000]{m$^2$}. 
Fiber readout of extruded scintillator was chosen as opposed to direct
readout of cast scintillator because of a nearly 20::1 cost advantage.  Most
of the cost savings comes from the use of wavelength shifting (WLS) fibers
to channel the light to the ends of the strips.  WLS fibers minimize
self-absorption by absorbing light peaked at \unit[420]{nm} and
re-emitting it at \unit[470]{nm}.  One WLS fiber runs down the center of
the wide face of each strip and collects the light from the entire
strip, leading to a reduction in photocathode
area (compared to direct scintillator readout) by a factor of over 300.
Optical summing of the WLS fiber light readout in the far detector led to
further cost saving as the result of reductions in the number of PMTs 
and associated electronics channels.

\subsection{General description of the scintillator system}
\label{sec:scint-des}



We describe here the specifications that led to
the design of the plastic scintillator system.  Because of its large
size, the far detector drove the design features of the system and we
describe it first, then describe how the near detector differs.

\renewcommand\theenumi {\roman{enumi}}

%

\begin {enumerate}
\item
{\bf Geometry:} Each steel octagon (\unit[8]{m} across)
is covered by a plane of scintillator. 
Each plane has one ``view'' of 
strips, with the next plane having the orthogonal
view. The two views are at $\pm 45 ^\circ$ relative to the vertical in order 
to avoid having strip readout connections at
the bottom of the detector. The \unit[4.1]{cm} strip width was the result of 
an optimization that included the response of the
detector to simulated neutrino interactions and cost considerations.

\item
{\bf Modularity:} The planes were built from modules, each consisting of 
groups of scintillator strips
placed side-by-side and sandwiched between aluminum covers. The strips were 
glued to one
another and to the covers to make rigid, mechanically strong, light-tight 
modules. WLS fibers were routed through
manifolds at both ends of the modules to bulk optical connectors. Continuous 
scintillator planes were formed by placing
eight of these modules next to one another on a steel detector plane.  

\item
{\bf Routing of scintillator light:} Light from the end of each WLS fiber 
is carried by a clear fiber to a
Hamamatsu R5900-00-M16 PMT, which has sixteen $\unit[4\times4]{mm^2}$ pixels.  
Eight fibers from non-neighboring 
scintillator strips are mapped onto one pixel, as described in 
Sec.~\ref{sec:scint-pmts}.

\item
{\bf Light output:} In general, strips produce different amounts of light 
when excited by a normally incident, minimum ionizing particle (MIP). 
Only events producing a total of at least 4.7~photoelectrons summed over 
both ends were used in the later analyses. 
In addition, the average light output for a MIP 
crossing at the far end of a strip as seen from the other end should be 
greater than 1.0~photoelectron.

\item
{\bf Uniformity:} After a correction for fiber attenuation, the light output 
is uniform over all the scintillator strips
to within $\pm30$\%. WLS fiber length is the most important cause of 
strip-to-strip differences in light output. 

\item
{\bf Calibration:} The absolute response to hadronic energy deposition was 
calibrated to 6\%. In addition, the relative
response of the near detector to the far detector and between different 
locations within the far detector were calibrated
to 3\%.

\item
{\bf Short-term stability:} It was required that the average light output 
should vary by less than 1\% per month of
operation.  This duration is determined by the time required for a complete 
cosmic-ray muon calibration of the far
detector.  PMT gains which vary more rapidly than this can be corrected by 
equalizing the response to hourly LED light injections, and cosmic ray
muons are used to correct for drifts on a daily basis.

\item
{\bf Long-term stability:} Decrease in light output due to the aging of 
various components is the main effect over the
long term. The design goal was that there should be no more than a 30\% 
decrease in light output over a period of ten
years; this will not significantly degrade the detector's physics 
performance.

\item
{\bf Linearity:} The response of the system depends linearly on hadronic 
energy deposition to within 5\% up to \unit[30]{GeV}.

\item
{\bf Time measurement:}  Time measurements are primarily useful for studies of atmospheric
neutrinos in the far detector because accurate time-of-flight measurements can 
distinguish upward-going neutrino events from
backgrounds induced by downward-going cosmic-ray muons.  The detector has a time 
resolution of better than \unit[5]{ns} for five observed
photoelectrons (assuming that this resolution 
scales as $1/\sqrt{N_{\rm pe}}$).


\end {enumerate}

\begin{figure}[htpb]
  \begin{center}
\includegraphics[width=\columnwidth]{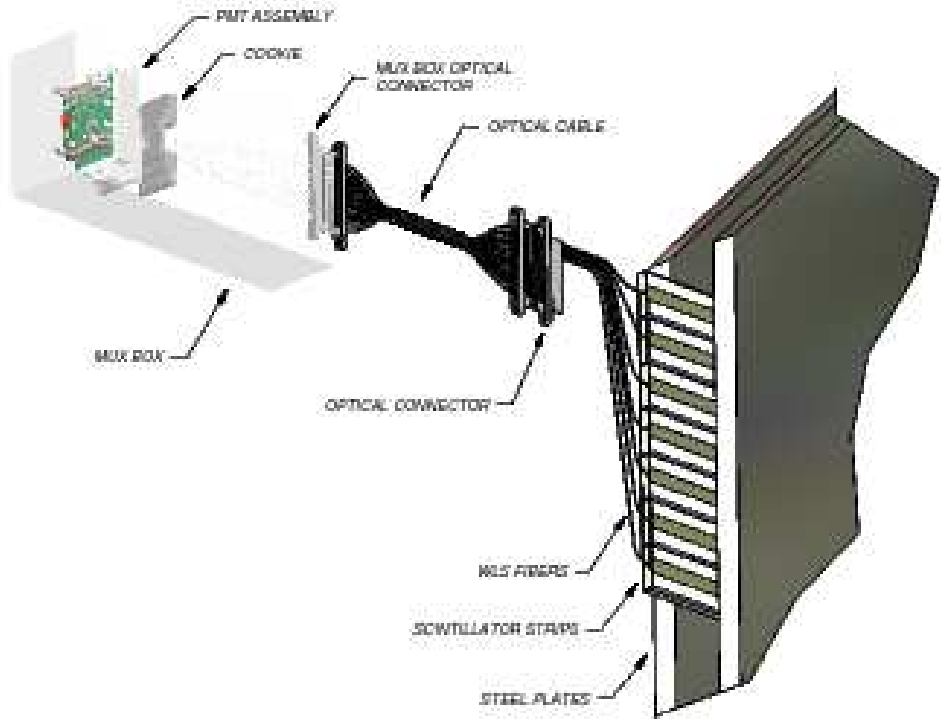}
\caption{Schematic drawing of the scintillator system readout for a module.  
  An edge of a detector plane is on the right side of the sketch, showing 
  several strips extending out of a scintillator module and beyond the edge 
  of the plane for clarity. 
  The light produced in a strip (Fig.~\ref{fig:scincross})
  travels out of the module in a WLS fiber, and is then carried by a clear
  optical fiber (assembled into a cable) to a multiplexing (MUX) box where
  it is routed to a pixel of the photomultiplier tube (PMT) assembly.}
  \label{fig:scinschematic}
  \end{center}
\end{figure}

Figure~\ref{fig:scinschematic} illustrates the light detection and collection
for part of one scintillator module.  Clear fiber cables
connect to the module and transmit light  from the edges of the detector
to centralized locations where the PMTs and readout electronics are mounted
(Sec.~\ref{sec:scint-pmts-fd}). A light injection system illuminates the
the WLS fibers near their ends
with LED-generated UV light to perform the system's primary calibration
(see Fig.~\ref{fig:limconcept} in Sec.~\ref{sec:calib-li}).

The primary task of the far detector is the measurement of
the properties of neutrinos initiated from the Fermilab beam. A secondary 
task is the detection and characterization of atmospheric neutrinos.
This second measurement, however, must contend with large backgrounds from high 
energy gammas and neutral hadrons produced by cosmic-ray muon interactions 
in the rock surrounding the detector. These backgrounds
have been strongly reduced by deploying an active veto
shield made of MINOS scintillator modules.  This anticoincidence layer 
detects hadron shower remnants emerging from the rock above and beside the 
detector. The shield's design and performance are described in
Section ~\ref{sec:vetoshield}.


The near detector is designed to have similar physics
response to neutrino events as the far
detector. However, some differences are unavoidable because the neutrino
event rate per unit mass is a factor of
$10^6$ greater than that in the far detector. The key differences between
the scintillator systems of the near detector and far detector are:

\begin {enumerate}

\item The near detector scintillator modules are much shorter, ranging 
from \unit[2.5]{m} to \unit[6]{m} in length.

\item The long WLS fibers of the far detector (and their corresponding
\unit[$\sim$5]{m} attenuation lengths) required readout of both ends.  
In contrast only 
one end of each near detector WLS fiber is read out.  With a mirrored far
end, the near detector WLS fiber gives approximately the same
light yield as the dual-ended readout in the far detector.  Single-ended 
readout necessitates attaching each Hamamatsu R5900-00-M64 photomultiplier 
pixel to only one scintillator strip.  This PMT has 
sixty-four \unit[2$\times$2]{mm$^2$} pixels but is otherwise very similar in 
construction and response to the R5900-00-M16 PMTs 
used in the far detector.

\item Due to its much higher event rate, the near detector requires 
faster, dead-time free readout electronics.

\end {enumerate}

All other features of the near and far detector 
scintillator systems are identical,
including the strips, the WLS and clear fibers, the light injection systems, 
and construction techniques of the modules.
The resulting physics capabilities are discussed in
Sec.~\ref{sec:calib-energy}. 

\subsection{Scintillator strips}
\label{sec:scint-strips}

Three technologies are crucial to the scintillator system, namely:
i)~low-cost extruded polystyrene scintillator; ii)~high-quality
WLS and clear fibers; and iii)~multi-pixel PMTs.

\begin{figure}[htpb]
  \begin{center}
    \includegraphics[width=\columnwidth]{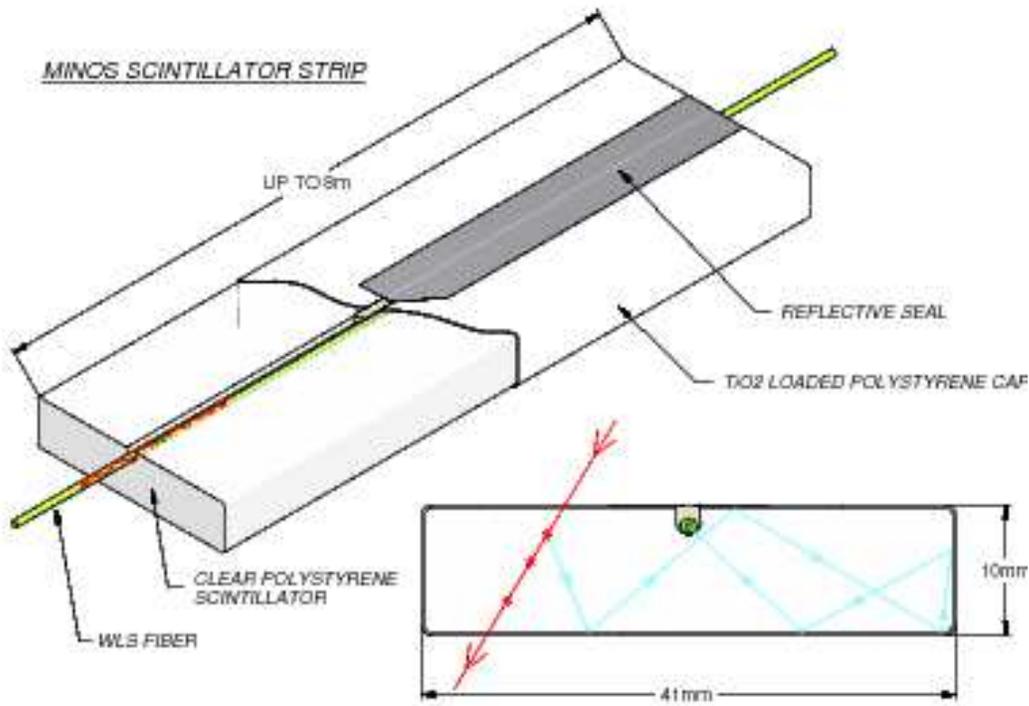}
    \caption{Cutaway drawing of a single scintillator strip. Light produced by
      an ionizing particle is multiply reflected inside the strip by the
      \unit[0.25]{mm}-thick outer reflective coating (shown in the cross-section view).
      Light absorbed by a WLS fiber is re-emitted 
      isotropically. Those resulting waveshifted photons whose directions fall
      within the total internal reflection cones are transported along the 
      fiber to the edges of the detector, subsequently being routed to the 
      photodetectors (Fig.~\ref{fig:scinschematic}).}
    \label{fig:scincross}
  \end{center}
\end{figure}

Figure~\ref{fig:scincross} shows a MINOS custom developed scintillator
strip~\cite{Pla-Dalmau:2001en} with its WLS fiber located in a
\unit[2.3]{mm}-deep by \unit[2.0]{mm}-wide groove in the center of the
``top'' face.  The fiber must be completely contained inside the groove
to ensure efficient light collection
(Sec.~\ref{sec:scint-performance-mappers}).  A specularly reflective
strip of aluminized Mylar tape is placed over the groove after the WLS
fiber has been glued in place.  The scintillator surface is covered by a
thin (\unit[0.25]{mm}) co-extruded titanium-dioxide (TiO$_2$)-loaded
polystyrene layer that serves as a diffuse reflector.  The scintillator
and TiO$_2$ coating are co-extruded in a single process, a standard
technique in the plastics industry.  The TiO$_2$ concentration was
chosen to be as high as possible without posing extruding problems.  In
R\&D tests the highest concentration of TiO$_2$ that did not affect the
quality of the extruded product was 12.5\% by weight, which coincided
with the concentration needed to maximize reflection of scintillator
light.  A 15\% concentration was achieved for scintillator production,
performed by a different extruding manufacturer.  The thickness of the
TiO$_2$ layer was as thin as could be reliably co-extruded
and thick enough that ultraviolet light (comparable to scintillator
light) could not shine through.  Bench tests of light reflection and
propagation were well-matched by models~\cite{Border:2001vk}, with
reflection angles following Lambert's law.  Absolute reflectivity
measurements, known to 1\%, were then tuned in the simulation below that level
to match observations.

The procedure used to fabricate the scintillator strips was as follows:

\begin {enumerate}

\item Polystyrene pellets (Dow STYRON 663W) were placed in a nitrogen gas 
environment to prevent reduction in light yield of the finished product,
which would otherwise result from exposure to atmospheric oxygen during the 
melting process.   

\item Scintillator fluors PPO (2,5-diphenyloxazole, 1.0\% by weight) and 
POPOP (1,4-bis(5-phenyloxazol-2-yl) benzene, 0.03\% by weight) were mixed with 
polystyrene pellets in a nitrogen gas environment.

\item The mixture was loaded into the primary extruding machine, where 
it was melted and pushed into the main port of the forming die. 

\item At the same time a mixture of polystyrene pellets and TiO$_2$ 
(concentration of 15\% TiO$_2$ by weight) was loaded into a secondary extruding
machine where it was melted and pushed into an auxiliary port of the forming 
die to produce the reflective coating. This material was distributed uniformly 
around the outer surface area of the strip except for the groove. 

\item A continuous strip, including its reflective coating, exited the 
die into a sizing and cooling line where its final
shape was defined. The strips were then cut to length by a traveling saw.

\end {enumerate}

The diffuse reflector coating is a unique feature developed for 
MINOS~\cite{Yun:2001pt}. The co-extruded coating is in
intimate contact with the inner clear scintillator; the two are completely 
fused together, forming a single solid piece.  Besides providing the required 
reflective properties, 
the coating protects the inner reflective surfaces from mechanical damage, 
allowing the strips to be shipped with minimal attention to packaging and 
handling. 
Two other secondary features of the reflective coating are the protection of
the sensitive scintillator from environmental chemical attack and the 
prevention of deleterious optical coupling by adjoining mechanical epoxies.
Tests showed the reflectivity of the TiO$_2$ loaded coating to be as
good as or superior 
to that of other, labor-intensive candidate reflective materials, such as
highly polished surfaces, Bicron TiO$_2$ paint, or finished surfaces
wrapped in Tyvek or Mylar.

Quality assurance feedback is a key component to successful scintillator 
production. For example, the light output increased by about 20\% compared 
to pre-prod\-uction test runs when uniform quality-control
processes and production extrusion conditions were established. This 
illustrates that large-scale production runs can help to
assure consistent performance for this type of scintillator.

\subsection{Wavelength shifting fibers}
\label{sec:scint-fibers}

The WLS fiber is \unit[1.20$^{+0.02}_{-0.01}$]{mm} diameter, 
double-clad polystyrene fiber with \unit[175]{ppm} of Y11 (K27)
fluor produced by Kuraray, Inc. in Japan. The cladding consists of an inner 
layer of acrylic 
and an outer layer of polyfluor. The fiber used in MINOS is ``non-S'' type, 
with a nominal S-factor of 25 compared to 75
for some ``S-type'' fibers (in which the polystyrene chains are oriented along 
the fiber direction). The polystyrene core
of non-S-type fiber is optically isotropic and more transparent than the core 
of S-type fiber, resulting in a 10\% greater
attenuation length. However, the transverse polystyrene chain orientation 
results in an increased vulnerability to cracking from bending or rough 
handling.

The fiber was chosen following a series of measurements of light output versus 
various fiber properties. Fiber from an alternate manufacturer
was tested but it did not satisfy our requirements. The final composition of 
Kuraray fiber was selected to give the highest light output from the far 
end of the longest 
scintillator strips (a length of about \unit[9]{m} of
WLS fiber coupled to \unit[3]{m} of clear fiber).  Hence, the long attenuation 
length properties of the fiber were of particular
importance. 
The fiber diameter was chosen to maximize the coverage of the 
\unit[4$\times$4]{mm$^2$} PMT pixel by eight fibers.

The fiber was flexible enough to allow delivery on spools of \unit[1]{km} 
each, making automated use of the fiber particularly
easy. Testing of the fiber was done using blue LEDs in an apparatus that 
illuminated different points along a fiber
wrapped around a cylinder. Production quality assurance tests were done 
relative to a set of ``reference fibers'' which had been previously
shown to satisfy our requirements. Two fiber samples from each spool, one 
from the beginning and the other from the end,
were taken for testing. Kuraray made similar measurements prior to shipping. 
In addition, Kuraray provided data on the
fiber diameter every \unit[10]{cm} along its length, automatically recorded 
during production. A spool of fiber was considered to
be acceptable as long as both test fibers had light output of at least 85\% 
of the reference fiber at all locations
along it and the spool had only a small ($<$10) number of spots with diameter 
variations outside the nominal specifications. MINOS
rejected only \unit[3]{km} out of \unit[730]{km} of WLS fiber, 0.4\% of that 
delivered.

After installation in the detector, an unexpected level of single
photoelectron spontaneous light emission was observed in the WLS fiber.
Bench tests~\cite{Avvakumov:2005ww} confirmed the source of this light
and also showed that clear fibers yield negligibly smaller noise rate.
This WLS-induced light had an initial rate of several Hz/m (contributing
about 1/3 of the single-photoelectron noise rate in the detector) and
has decayed exponentially with a time constant of several months.

\subsection{Scintillator modules}
\label{sec:scint-modules}

Integrating scintillator strips into modules provides 
several advantages.  The details of scintillator
module design and construction are contained in the following two sections.

\subsubsection{Scintillator module design}
\label{sec:scint-modules-design}

The packaging of scintillator strips into modules provides the 
following functionalities:

\begin {enumerate}

\item
a mechanically strong structure that holds strips together 
and which is sufficiently robust for shipping 
and mounting to the steel plates,

\item
a light-tight enclosure,

\item
a package that mitigates the risk of the polystyrene contributing fuel for a 
fire in the vicinity of the detector,

\item
a means of connecting of the WLS fibers to clear fibers for transmission of 
the light signal to the PMTs,

\item
a unit that allowed much assembly work 
to be done away from the detector sites 
and still fit into the vertical elevator
shaft at the Soudan mine (Sec.~\ref{sec:install-fd}).  
(The elevator shaft constraints ultimately determined 
both the length and width of scintillator modules.)

\end{enumerate}

\begin{figure}[htpb]
  \begin{center}
\subfigure[Exploded view.]{
   \includegraphics[width=0.7\columnwidth,keepaspectratio=true]{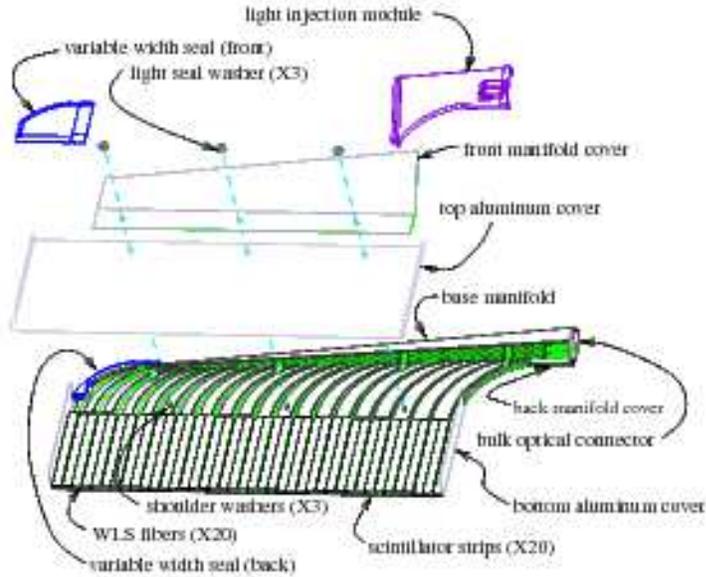}
}
\subfigure[Assembled view.]{
 \includegraphics[width=0.7\columnwidth,keepaspectratio=true]{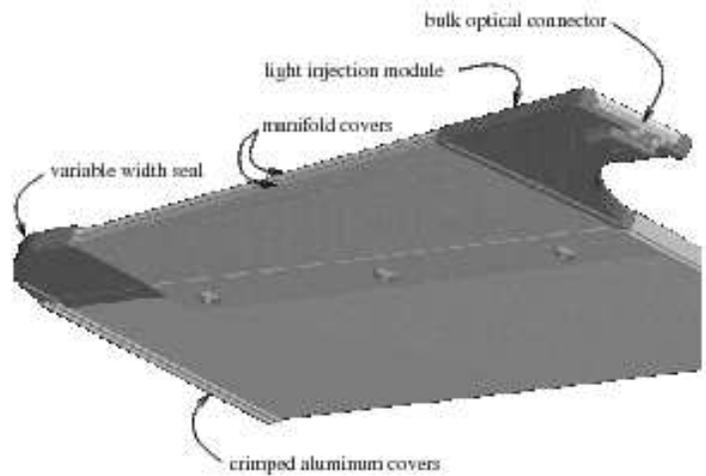}
 }
\caption{A typical MINOS module manifold assembly 
          wherein WLS fibers from the
         scintillator strips are routed to bulk optical connectors (a) and
         enclosed by protective light tight aluminum covers (b). }

\label{fig:moduleend}
\end{center}
\end{figure}

The last \unit[20]{cm} of the active portion
of a typical MINOS module, including a manifold assembly, 
is illustrated in Fig.~\ref{fig:moduleend}.
The function of each module component is explained below.

The foundations for a module are the aluminum covers, consisting of two 
flat sheets with formed perpendicular flanges on the left and right sides.
The scintillator strips are laminated to the bottom cover by
epoxy adhesive,
as described in Section~\ref{sec:scint-modules-construction}.
Following lamination, the flanges of the top aluminum cover
are nested within the flanges of the bottom cover and the two
are then crimped together.

The lamination of the scintillator strips to the aluminum covers 
provides the basic structural strength of the modules.
The crimped flanges complete a light-tight seal
around the scintillator strips.
Furthermore, the covers implement a fire seal
around the strips.

The base manifolds serve to route the WLS fibers
from the scintillator strips to the bulk optical connectors.
Each fiber is channeled through an individual groove.  This
design feature eliminates registration errors
between the scintillator strips and the optical connector
and also protects the fiber ends.
The grooves are designed to guarantee that the WLS fiber 
bend radius exceeds \unit[12]{cm} 
despite tolerance stack-up on the extrusions.

The front and back manifold covers
shield the WLS fibers in the manifold from ambient light,
and fire-seals most of the manifold as well.
The light injection module serves two purposes:
It enables the implementation of a light injecting fiber calibration system
described in Section~\ref{sec:calib-li} and
provides a light seal over the somewhat complex
geometry of the manifold in the region of  the bulk optical connector.
The light-tight washers (Fig.~\ref{fig:moduleend}) provide alignment references during
module construction and installation
(Sec.~\ref{sec:install-fd-plane-surv}).

The scintillator strips in each module are close-packed
to minimize inactive zones between strips.
As a consequence the widths of the modules 
vary slightly, but by less than 0.5\%.
The aluminum covers are individually formed
to exactly match the cumulative width of the extrusions which they enclose.
The variable width seal serves to light seal the gap
between the fixed-width manifold parts and the variable-width aluminum covers.

\begin{figure*}[htpb]
\centering
  \subfigure{
    \includegraphics[width=0.45\textwidth]{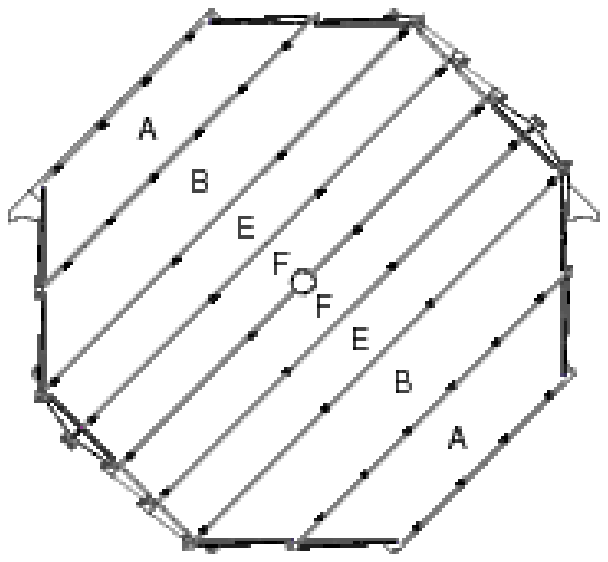}
  }
  \subfigure{
    \includegraphics[width=0.45\textwidth]{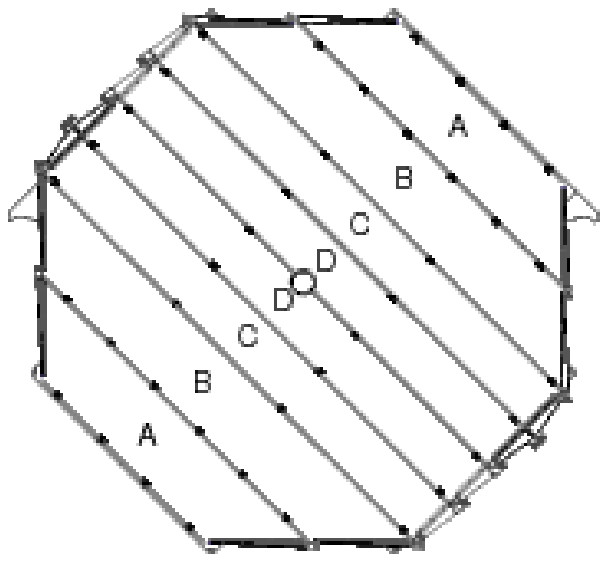}
  }
\caption{Layout of $U$ (top) and $V$ (bottom) modules on far detector planes.
       $U$- and $V$-type planes are interleaved. ``A'' and ``B'' module types 
         have 28 scintillator strips and the other types have 20 strips.
         The first (upstream) scintillator plane of each supermodule is of 
         the $V$-type.}
\label{fig:farmodlayout}
\end{figure*}

Some scintillator modules must provide clearance for the detectors' magnet 
coils (for example, see module types $D$ and $F$ in 
Fig.~\ref{fig:farmodlayout}).  A semi-circular hole of radius \unit[197]{mm} is cut into the aluminum 
covers of the affected modules, and short lengths of scintillator strips
passing through the hole are also cut away.
Due to the rectangular nature of the strips, the region of missing
scintillator is larger than the coil hole itself, as long as
\unit[598]{mm} for the most central strips, decreasing in stepwise
fashion with each neighboring strip.  
However, the WLS fibers passing through the affected strips are not cut.
Rather, a ``bypass'' channel routes them around the hole.
This bypass protects the fibers from physical damage
and ambient light.

Twenty-two variations of MINOS modules were built
which all share the basic architecture
described here.
The layout of modules utilized 
on different planes of the far and near detectors
is described below.

\begin{figure}[htpb]
  \begin{center}
    \begin{minipage}[t]{0.9\columnwidth}
      \includegraphics[width=\textwidth]{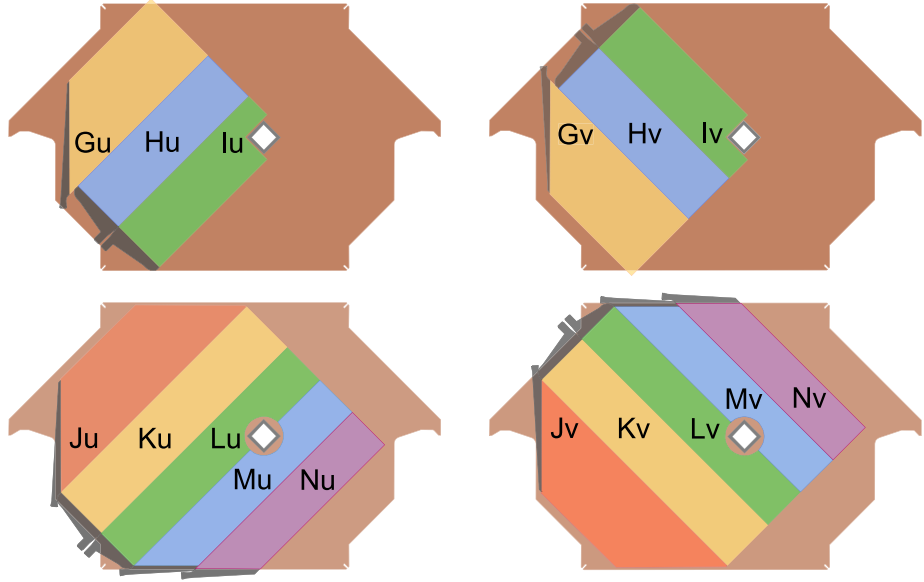} 
      \caption[Near detector scintillator module
      layout]{\label{fig:NearInstFig1} The four different configurations of 
planes used in the near detector, showing the different layouts of the 
scintillator modules.  The upper two figures show partially instrumented 
planes (``calorimeter region'') while the lower two figures show the fully 
instrumented ones (``tracking region'').
Strips oriented in the $U$ direction are on the left,
and $V$, on the right. These orientations alternate in the detector 
to provide stereo readout. The G-N notations
denote the different shapes of the scintillator modules.
The $U$ and $V$ planes require slight variations
on each shape, leading to a total of 16 module types.
The beam is centered midway between the coil hole and the left side of the 
plane, hence the scintillator need only cover that area in the target region. }
    \end{minipage}
  \end{center}
\end{figure}

\subsubsection{Design of the far detector planes}
\label{sec:far-mods-active-planes}

The far detector utilizes 484 active planes.
The layout of eight modules on a far detector plane 
is illustrated in Fig.~\ref{fig:farmodlayout}.
Module widths were designed so that 
no module crossed a vertex of an octagon,
thereby simplifying the module-end cuts to either both
perpendicular or both 45$^\circ$.

The far detector required six types of modules to configure 
the $U$ and $V$ scintillator planes, as shown in Fig.~\ref{fig:farmodlayout}.
The center four modules are different for the $U$ and $V$ plane types,
whereas the outer modules are all the same.
The center four modules contain twenty \unit[8]{m} long strips.
The outer modules contain 28 strips
varying in length from \unit[3.4]{m} to \unit[8]{m}.
The ends of all scintillator strips
are cut perpendicular to the length of the strips.
The strip-ends of the outer modules, 
whose aluminum covers have 45$^\circ$ ends,
follow the edge of the steel plate like steps of a staircase.
Each plane has 192 strips covering
about 99\% of the steel octagon surface.

\subsubsection{Design of the near detector active planes}
\label{sec:near-mods-active-planes}

An important distinction exists between
far detector and near detector modules.
WLS fibers in far detector modules
terminate in optical connectors at both ends.
WLS fibers in near detector modules
terminate in an optical connector
on the near end
and are mirrored at the far end.
Therefore, the far ends of near detector modules
are closed with simple plates.

Only 153 of the 282 planes comprising 
the near detector are active. Active
planes are instrumented with 
four distinct scintillation module patterns:
full $U$-view (FU), full $V$-view (FV), 
partial $U$-view (PU), and partial $V$-view (PV)
(Fig.~\ref{fig:NearInstFig1}).
This required 16 types of modules 
that contain either 14, 20 or 28 scintillator strips.
The area of partial coverage is set to ensure
the complete measurement of neutrino events
occurring in the near detector fiducial volume.
The beam centerline is located at the horizontal midpoint
between the left edge of the steel plate and the coil hole,
as shown in Fig.~\ref{fig:NearInstFig1}.
The full-view coverage extends around the coil hole
in order to track long range muons 
downstream of neutrino interactions.

The upstream 120 planes of the near detector comprise the calorimeter section
and are all instrumented in order to yield a high resolution view
of the neutrino interactions.  This section was assembled using a repetitive 
10-plane pattern: FU-PV-PU-PV-PU-FV-PU-PV-PU-PV.  For data analysis,
the calorimeter section is divided into three longitudinal sections:
planes 1-20 are the ``veto'' section,
used to exclude tracks that originate upstream of the detector;
planes 21-60 represent the ``target'' region,
as all neutrino-induced showers which occur here
are contained within the length of the detector;
planes 61-120 complete the calorimeter section
and are used to contain and measure 
the hadronic showers of neutrino events in the target region.

The spectrometer section of the near detector, planes 121-281,
uses the same 10-plane pattern 
but with partial-view scintillator modules removed.
That is, a full-view plane is included in every fifth plane only. This 
downstream section is used solely to track muons from neutrino interactions.

\subsubsection{Scintillator module assembly facilities}
\label{sec:scint-modules-assembly-facilities}

  The construction of scintillator modules was the single largest production 
  job in the MINOS experiment. The modules were fabricated at three assembly 
  facilities located at collaborating institutions and operated by
  staff members of those institutions.
  Twenty-strip wide rectangular modules, types C, D, E, and F in 
  Fig.~\ref{fig:farmodlayout}, were constructed at the California Institute 
  of Technology while the 28~strip wide trapezoidal modules (types A and B 
  in Fig.~\ref{fig:farmodlayout}) 
  were built at the University of Minnesota, Twin Cities.
  Each factory produced about four modules per day
  and each was staffed by a crew of nine technicians working 40 hours
  per week. Each factory produced a total of about 2,000 modules over
  a two-year period. At Caltech, the in-house staff was augmented by a
  number of temporary employees without previous experience building
  particle physics detectors. The Minnesota facility was mainly staffed
  by part-time undergraduate workers. 
       
  The near detector factory was located at Argonne National Laboratory
  and was staffed by three Argonne technicians. It produced about 600
  modules at the rate of about eight modules per week over a two-year
  period.     

  Each module assembly facility was operated under a local 
  manager with oversight by
  a local MINOS collaboration physicist. The local institution was
  responsible for worker health and safety but the MINOS
  construction project provided advice and oversight with the help of
  Fermilab Environment, Safety and Health (ES\&H) professionals. 
  The most significant health issue was the development
  of worker sensitivity to epoxy vapors during the production start-up
  phase, apparently initiated by skin contact with
  liquid epoxy, particularly during the gluing of fibers in strips. It was
  easily mitigated by improved ventilation and the use of gloves and other
  protective equipment to prevent skin contact with liquid.

  The quality of modules produced by all three facilities was excellent.
  Only a very small number of constructed modules had to be discarded,
  most due to broken fibers. 
  After finished modules were delivered to the detector sites, 
  a small fraction were found to have light leaks 
  despite careful light tightness verification prior to
  shipping. Rapid feedback to the assembly facilities 
  eliminated the causes of this problem, which were primarily transport
  related stresses at the long seam between the light injection modules
  and the aluminum module skins.  These light leaks were remedied by 
  applying a combination of tape, epoxy, and RTV at the junctions.



\subsubsection{Scintillator module construction steps}
\label{sec:scint-modules-construction}


Module components were purchased
commercially or fabricated at special purpose facilities operated by
the collaboration. 
The assembly process required a period of four days to complete 
each scintillator module.  Most of this time was needed to cure 
structural and optical epoxies at different stages of the 
assembly process.   Each module was built on a dedicated 
assembly support panel which was moved from one assembly 
station to another by means of roller tables.  Assembly 
facilities were supplied with the following materials to build
scintillator modules.  

\begin {enumerate}

\item {\bf Aluminum covers.} 
Rolls of aluminum sheets (\unit[0.5]{mm} thick) 
were purchased cut to the proper nominal width for assembling the various types
of modules. Aluminum sheets were unrolled and cut to length. Because the 
actual width of every module varied from the nominal due to scintillator strip 
width tolerances, the long edges of aluminum covers 
were trimmed with a slitting tool and then bent $90^\circ$ by hand-operated 
forming tools to accommodate the crimp-sealing procedure. 

\item {\bf Scintillator strips.}  
  The extruder supplied \unit[8.18]{m} long strips for rectangular module types C, D, E and
  F (Fig.~\ref{fig:farmodlayout}).  They were trimmed to
  \unit[8.00]{m} at the module factories.  Precut \unit[11.48]{m} long strips
  were supplied for the trapezoidal module types A and B.  The strips were cut to length
  in a fixture designed to hold all 28 strips used in one module
  simultaneously.  One end of the fixture had a stop positioned at an
  angle of 26.57$^\circ$ ($\tan^{-1}(0.5)$) relative to the length direction
  of a module.  Positioning the ends of the stock extrusions against
  this stop enabled cutting all 28 strips to their correct length with a
  single perpendicular cut.  The \unit[11.48]{m} length was chosen so
  that the unused portions of the strips cut for a ``B'' module provided
  the raw stock for an ``A'' module, thereby minimizing scrap.
  Scintillator strips for the near detector were supplied in several
  lengths due to the larger number of near detector module types.

\item {\bf WLS fiber spools}, described in Sec.~\ref{sec:scint-fibers}. 

\item {\bf Manifold components}, described in 
Sec.~\ref{sec:scint-modules-design}.  

\item {\bf Bulk optical connectors}, 
described in Sec.~\ref{sec:scint-connect}.  

\item {\bf Adhesives.  Structural:} 3M~2216 translucent two-part epoxy  
was used to laminate the aluminum skins to scintillator strips.  
{\bf Optical:} Epon~815C resin with Epicure~3234 teta hardener (in a six to
one ratio by weight, respectively) was used to bond WLS fibers into 
scintillator strip grooves in all modules and to bond the 
reflective-tape mirrors to the far ends of WLS fibers in near detector 
modules.  {\bf Manifolds:} 3M~DP~810 epoxy was used as the adhesive for the 
assembly of manifold components, while black GE RTV~103 was used to create 
light seals at all joints.

\item {\bf Tape.}  
3M~850 aluminum-coated Mylar reflective tape, \unit[1.27]{cm} wide, 
was used to cover fibers in scintillator grooves and as mirrors
to terminate the far ends of WLS fibers in near detector modules. Black 
vinyl electrical tape (3M~33) was used to double-seal certain areas of each 
module against light leaks. The vinyl tape was covered with acrylic 
adhesive backed aluminum tape (McMaster-Carr 7631A32, \unit[0.003]{in} thick) 
to prevent creeping.  

\end {enumerate}

\begin{figure}[htpb]
\begin{center}
\includegraphics[width=0.6\columnwidth,keepaspectratio=true]{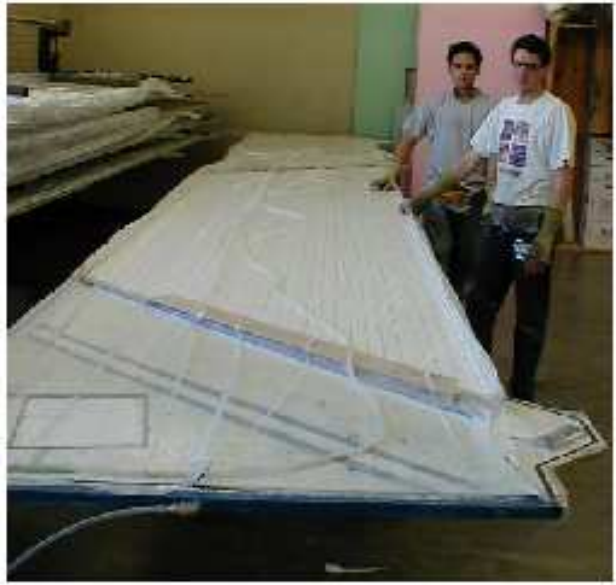}
\caption{Scintillator strips and manifolds being vacuum bonded (laminated) 
           to the bottom aluminum cover.}
\label{fig:laminating}
\end{center}
\end{figure}

Special equipment, listed below, was used in each of the three assembly
facilities.  The equipment, designed and constructed at various 
MINOS institutions, allowed the semi-automatic
fabrication and assembly of module components in an efficient and
repeatable manner.

\begin{figure*}[htpb]
\begin{center}
\includegraphics[width=\textwidth,keepaspectratio=true]{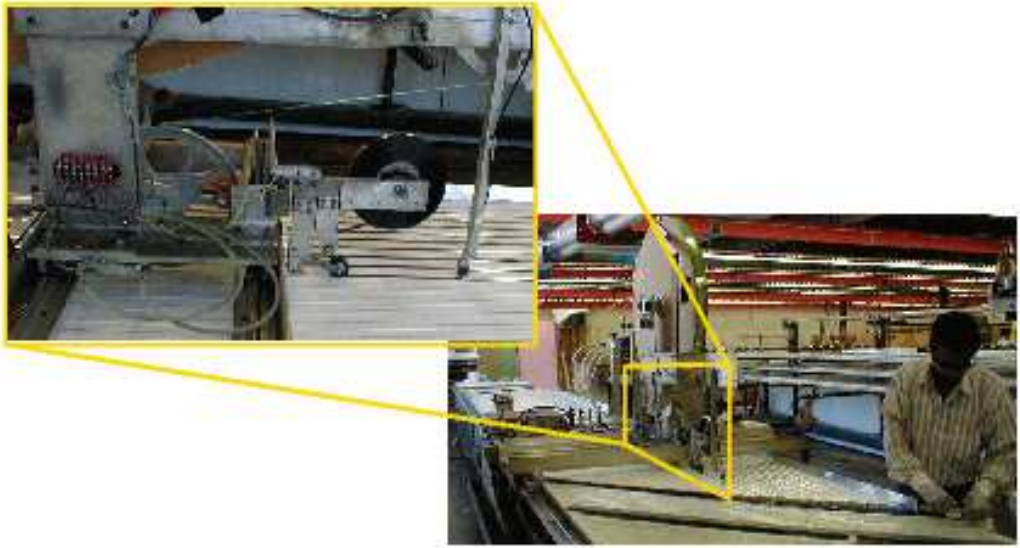}
\caption{The WLS fiber gluing machine.
         The operator threads a fiber through the manifold and into the 
	 optical connector while the head moves down the length of the module.
         The inset shows the moving head which dispenses 
         optical epoxy in the U-shaped groove,
         inserts the WLS fiber and covers the groove with reflective tape.
       }
\label{fig:glueMachine}
\end{center}
\end{figure*}

\begin{figure*}[htpb]
\begin{center}
\includegraphics[width=\textwidth,keepaspectratio=true]{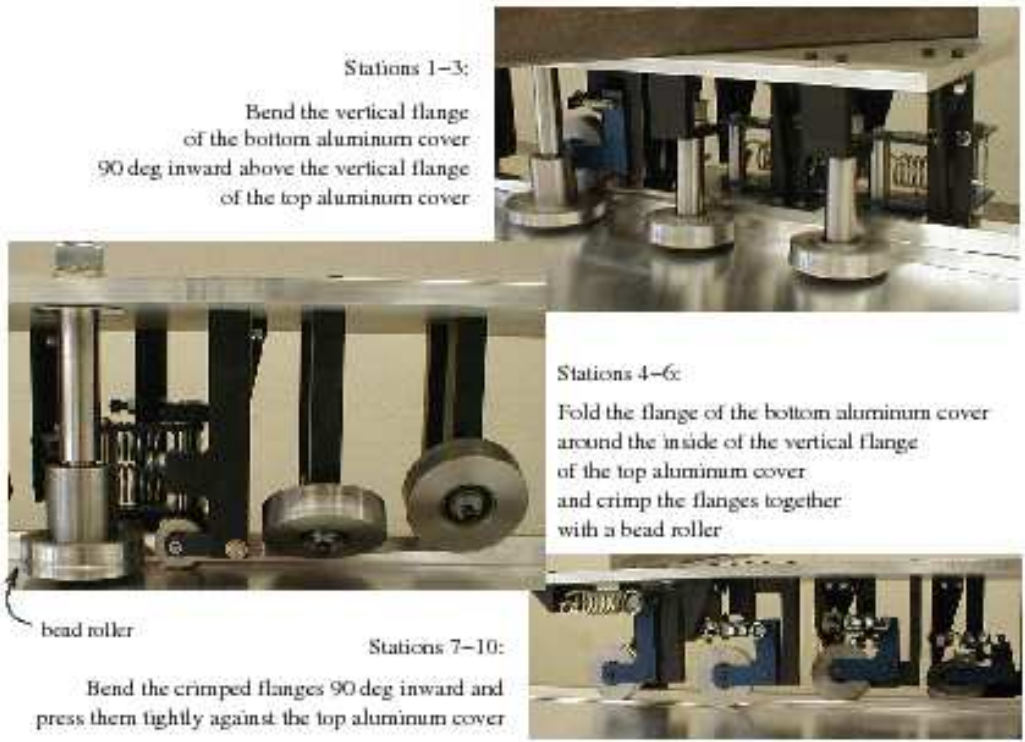}
\caption{The 10-station head of the crimping machine.  The insets show
  the function of each set of rollers to form a light-tight seam between
  the top and bottom aluminum covers of each module.  All sharp
  edges are concealed following completion of the crimp.
}

\label{fig:crimpHead}
\end{center}
\end{figure*}

\begin{figure}[htpb]
\begin{center}
\includegraphics[width=\columnwidth,keepaspectratio=true]{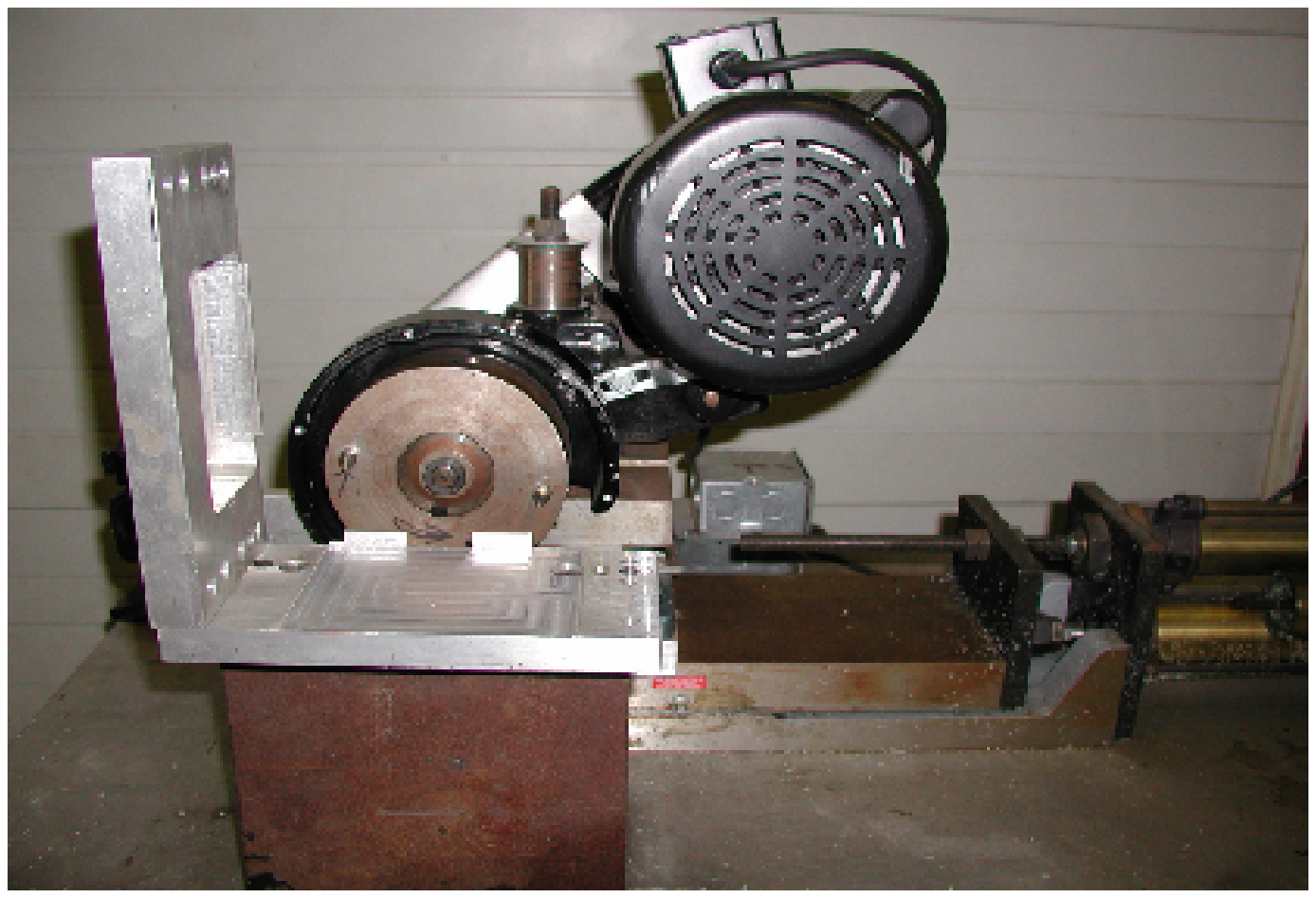}
\caption{The fly-cutting machine,
          shown with its safety shield removed for clarity.  
	   The flywheel with diamond bits can be seen behind an L-shaped 
           clamping fixture used to hold the 
	   optical connector at a precise location.  Smooth transverse motion 
	   across the optical face is provided by the pneumatic drive
           located at the lower right in this photograph.
      }
\label{fig:flycutter}
\end{center}
\end{figure}

\begin {enumerate}

\item {\bf Assembly support trays.} 
Each module was assembled on a dedicated
support tray that measured \unit[1.2]{m} wide and \unit[9.1]{m} long.  
The trays were 
moved to various stations throughout the four-day assembly process.  
Support trays 
were made by joining two \unit[1.2]{m} $\times$ \unit[4.6]{m}, 
commercially available,
honeycomb sandwich panels, 
approximately \unit[2.54]{cm} thick. Support trays were equipped with 
fixturing holes for 
aligning the modules and with vacuum ports for the lamination process. 
During lamination, 
a polyethylene sheet was placed over the assembly tray and sealed along
the tray's perimeter with vacuum sealant tape as illustrated 
in Fig.~\ref{fig:laminating}.


\item  {\bf Fiber gluing machine.}
  A semi-automated gluing machine was developed to insert, 
glue and cover a WLS fiber into each scintillator strip. 
In one continuous operation, the machine 
injected optical epoxy to fill the \unit[2.3$\times$2.0]{mm} groove, inserted the 
\unit[1.2]{mm} diameter WLS fiber, 
pushed it to the bottom of the groove and applied reflective 
tape to cover the fiber and groove.  The fiber gluing machine 
(Fig.~\ref{fig:glueMachine})
consisted of a head that traversed the length of a stationary table, 
upon which a partially assembled
module was placed. The head carried a glue-mixing dispenser, a spool of WLS
fiber and a roll of the reflective tape.  

\item {\bf Storage rack.}
  Assembly support trays were placed into storage racks during the 
(overnight) epoxy curing time.  One storage rack could store five support 
panels, which were loaded by sliding the panels from an assembly station 
sideways onto a storage shelf.  The shelves were raised and lowered by 
electric motors.  The racks were mounted on casters that allowed them to be 
moved around the assembly facility.  

\item {\bf Crimping machine.}
  A manually-operated machine was used to bind the long edges of the 
lower and upper aluminum covers together.  The bent edge of the upper cover 
was nestled just inside 
the bent edge of the lower cover.  The crimping machine's roller system 
gradually folded both 
edges over each other to form a seal as the head, containing 10 crimping 
stations, traversed the length of the table.
The function of each crimping station is 
shown in Fig.~\ref{fig:crimpHead}.  

\item {\bf Fly cutter.}
 A custom-built cutting/polishing machine was used to shave the surface of 
 the bulk optical connector and its embedded WLS fibers to produce an 
optically smooth surface 
for good light transmission.  The machine consisted of an x-y table equipped 
with a rotating flywheel. The flywheel was driven by a motor with high-quality 
bearings.
The cutting was done by a pair of diamond bits mounted to diametrically 
opposite sides of the flywheel.
Motion along the face of the connector was controlled by a pneumatic drive.  
The depth of cut was precisely set by manual operator control prior to each
pass.  A clear shield covering the flywheel protected the operator from
getting too close to the sharp, spinning bits when the machine was operating.
 Figure~\ref{fig:flycutter} shows the fly cutter without its clear protective 
shield. 

\item {\bf Module mapper.}
  An automated x-y scanning table was used to measure the response of each 
module to a \unit[5]{mCi} $^{137}$Cs source.  The Cs source was encased in a 
lead pig that was attached to a traveling 
x-y scanning carriage with a range of \unit[1.3]{m} in width and \unit[8]{m} 
in length.  Optical fiber cables were connected
to the bulk optical connector(s) on every module and the transverse and 
longitudinal response of every 
scintillator strip was determined.  This procedure is described in detail in 
Sec.~\ref{sec:scint-performance-mappers}

\end {enumerate}

\begin{figure*}[htpb]
\begin{center}
\includegraphics[width=\textwidth,keepaspectratio=true]{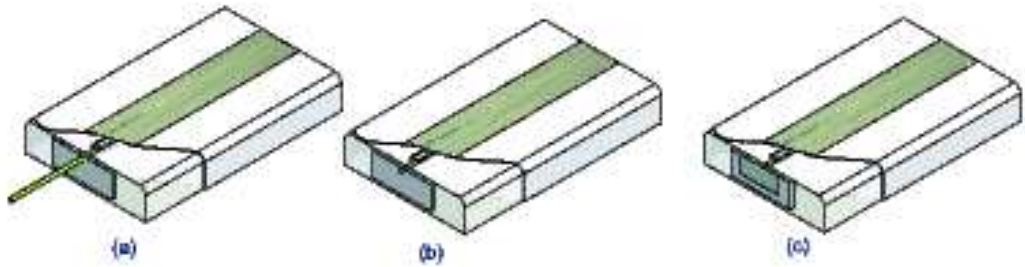}
\caption{Mirroring near detector fibers.  The WLS fiber is pushed through 
a hole in the aluminum tab and the tab is glued to the end of the scintillator 
strip (a).  A hot knife is used to cut the fiber flush to the plane of the tab 
(b).  Optical epoxy is applied to the fiber and tab, which are then covered by 
reflective tape (c).}
\label{fig:ndstrip-mirror}
\end{center}
\end{figure*}

Assembly of scintillator modules was carried out as follows:

\begin {itemize}
\item[Day 1:] Cut scintillator strips to proper length; cut and form the 
bottom aluminum cover; apply structural epoxy adhesive  
to the inside of aluminum cover.  Install scintillator strips and fiber 
manifold(s) with optical connector(s) inside the epoxied cover.  
(Figure~\ref{fig:moduleend} 
shows the end of one particular type of module illustrating the details of the 
scintillator-manifold interface.)  Apply sealing 
compound to the periphery of the assembly support tray and vacuum seal 
with a polyethylene sheet to provide uniform pressure for lamination.  
(This is shown in Fig.~\ref{fig:laminating}.)  Move the assembly tray to a 
storage rack, maintain the vacuum four hours and allow to cure overnight.

\item[Day 2:] Move the assembly support tray from the storage rack to the 
fiber-gluing machine.  Remove the polyethylene lamination sheet to prepare the 
assembly for gluing WLS fibers to the scintillator.  Use the fiber gluing 
machine to apply optical epoxy into the groove, insert fiber to the bottom 
of the groove, and cover the fiber with reflective aluminized Mylar tape.
Extend the WLS fiber ends sufficiently beyond the scintillator strips to route 
fibers to the optical connector through the guides in the manifold tray.  

For near detector 
modules, apply the aluminum tab to the far end of the scintillator strip and pull 
WLS fiber through the hole in the tab.  Cut the fiber flush against the tab with 
a heated knife edge.  Apply optical epoxy over the cut-fiber end and tab.  
Place reflective 
tape across the tab and fiber and secure the tape until epoxy cures.
This process produced 1.71$\pm$0.02 times the light output of simply
polishing the cut end and covering it with a dab of optical grease and
black paper.  
Figure~\ref{fig:ndstrip-mirror} illustrates the mirroring process.

Remove the support tray from the fiber gluing machine and place it in a storage 
rack overnight to allow time for the epoxy to cure.  
At the end of each day, pot the fiber optic connectors with optical epoxy 
(Epon 815C with 2\%
carbon black additive to prevent light transmission to the fibers).
 
\item[Day 3:] Cut and form the top aluminum cover; apply structural epoxy to 
the inside 
of the top aluminum cover.  Remove the assembly support tray from the storage rack and 
place the top aluminum cover over the assembly.  Apply sealing compound to the 
periphery of the support tray for lamination, and vacuum-seal the assembly 
with a polyethylene sheet.  Move the assembly support tray to a storage rack 
to cure overnight.

\item[Day 4:] Move the module from the assembly support tray onto the 
crimping table.  (At this point the assembly tray is placed in a rack to 
begin assembly of another module.)
Crimp the sides of the bottom and top aluminum covers together to form a 
light-tight enclosure.
Flycut the fiber-embedded optical connector to a flat, polished finish.  
Assemble the manifold and variable-width end seals with 3M~DP~810 
epoxy and black RTV103 sealant.  Cover any RTV with black vinyl tape and 
cover that with aluminum
tape. Move the module to the mapper table.  Attach optical fiber cables 
to the module's bulk optical connector(s) and check for light leaks with a 
bright hand-held 
lamp.  Repair any leaks and verify that the module is light tight. Use the 
mapper to 
measure the response of the module to a series of transverse scans at 
predetermined longitudinal 
positions with a $^{137}$Cs source to verify the module performs within 
specifications. 
Remove the module from the mapper and pack it in a shipping crate.
 
\end {itemize}

\subsection{Photodetectors and enclosures}
\label{sec:scint-pmts}

The MINOS detectors are read out by Hamamatsu 64-anode (M64) PMTs for the 
near detector and 16-anode (M16) PMTs for the
far detector. The PMTs are housed in light-tight, steel enclosures containing 
clear fiber bundles which are interlaced from
cable connectors to PMT pixels (Fig.~\ref{fig:scinschematic}). 
In the near detector each M64 resides in an individual
enclosure. In the far detector each enclosure (called a 
``MUX box'') houses three M16 PMTs. This box also implements the optical
summing (multiplexing) of eight fibers onto each PMT pixel. The fibers are 
held in place on the PMT face with a precision
of \unit[25]{$\mu$m} relative to alignment marks etched by the manufacturer 
on the first dynode. 

\begin{figure}[htpb]
  \begin{center}
    \includegraphics[width=\columnwidth]{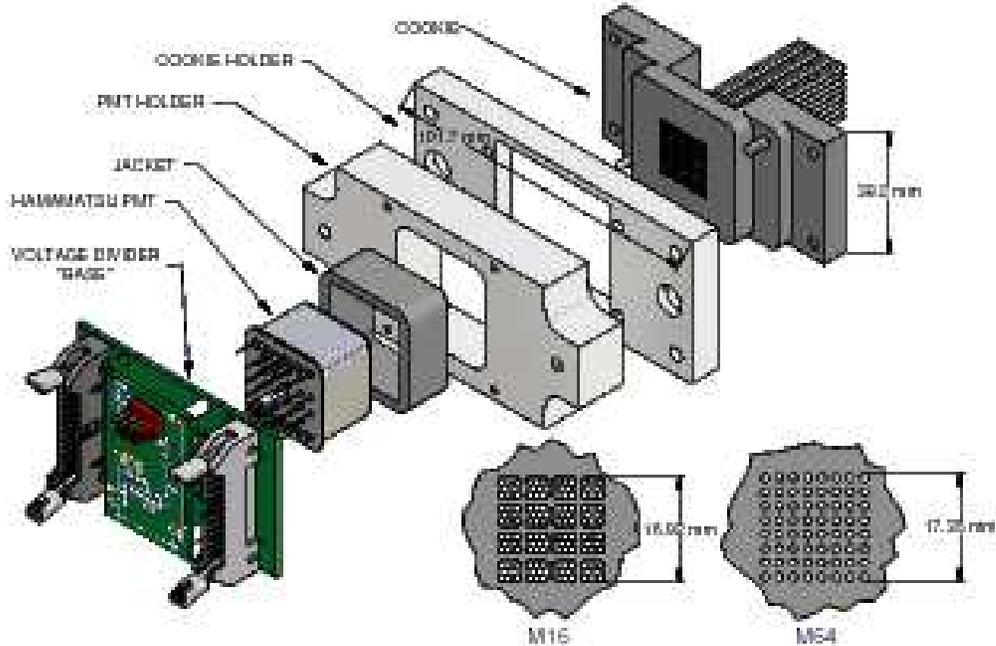}
    \caption{\label{fig:mux} The M16 PMT mounting assembly, one of three
      inside each far detector MUX box.  The near detector M64 mounting 
      assembly is identical except that only one fiber is placed on each 
      pixel.  The fiber ``cookie'' layouts
      used in the two cases are shown face-on in the lower right of the
      figure.}
  \end{center}
\end{figure}

Inside a PMT enclosure, each clear fiber from the connector  is terminated 
in a ``cookie,'' a plastic part machined with a
diamond-bit fly-cutter to attach fibers to the PMT.  Figure~\ref{fig:mux} 
shows the mechanical assembly that mates the PMT to the cookie. 
In addition to keeping the PMTs dark, the MUX boxes shield the PMTs 
from stray magnetic fields and the base
electronics from ambient electronic noise.


Each PMT has a divider network mounted on a printed-circuit board. This base
is designed for a negative potential photocathode and contains
the voltage divider recommended by Hamamatsu Photonics. Although the
anodes are electrically isolated, only one dynode chain serves all anodes.
The base provides a signal from the last dynode, a 
positive-polarity sum of the response from all the tube's pixels, which is used as a 
trigger for the PMT readout
(Sec.~\ref{sec:elec-fe-far}). In the near detector, this dynode signal 
is ignored during the beam spill when all anodes are
continuously read out (Sec.~\ref{sec:elec-fe-near}).

Each PMT, mounted in its base, was bench-tested before assembly to 
ensure that it met MINOS requirements for quantum 
efficiency, gain, crosstalk, linearity, and dark 
noise~\cite{Tagg:2004bu,Lang:2005xu}. The tests showed that for both
M16 and M64 PMTs, no 
crosstalk contribution came from sources external to that particular PMT. 
The tests demonstrated that the main sources of crosstalk in these PMTs 
are light sharing between pixels caused by
refraction in PMT windows, imperfect photoelectron focusing, and 
electromagnetic coupling of dynodes.

\begin{table}
\begin{tabular}{l|c|c} \hline
Feature     & M16                        & M64                   \\
\hline
\parbox[t]{0.2\columnwidth}{\raggedright Pixel Gain Variation} &
\parbox[t]{0.4\columnwidth}{Max/min pixel gain $<$3 \\ rms
  between pixels 23\% }   & \parbox[t]{0.4\columnwidth}
  {Max/min pixel gain $<$3\\rms between pixels 25\%}\\\hline
\parbox[t]{0.2\columnwidth}{\raggedright Quantum Efficiency}     & $>$12\% at \unit[520]{nm}   & $\sim$12\% at \unit[520]{nm}\\\hline
\parbox[t]{0.2\columnwidth}{\raggedright Dark Noise (\unit[1]{p.e.})} & $\unit[\sim25]{Hz}$ per pixel & $\unit[\sim4]{Hz}$ per pixel\\\hline
\parbox[t]{0.2\columnwidth}{\raggedright Linearity (typical at nominal gain)}  & $<$ 5\% below \unit[100]{p.e.}  & $<$ 5\% below \unit[50]{p.e.}     \\\hline
\parbox[t]{0.2\columnwidth}{\raggedright PE Crosstalk: (pixels)}  &   &   \\
\parbox[t]{0.2\columnwidth}{\raggedright nearest-neighbor} & 1.92\% & 2.58\% \\
\parbox[t]{0.2\columnwidth}{\raggedright diagonal-neighbor} & 0.38\% & 0.47\% \\
\parbox[t]{0.2\columnwidth}{\raggedright non-neighbor} & 0.90\% &0.69\% \\
\parbox[t]{0.2\columnwidth}{\raggedright total (all pixels)} & 3.20\% & 3.74\% \\
\hline
\end{tabular}
\caption{Main characteristics of the M16 PMTs used in the far detector and 
M64 PMTs used in the near detector. 
Further details can be found in Refs.~\cite{Tagg:2004bu,Lang:2005xu}.}{}
\label{t:m64}
\end{table}

\subsubsection{Near detector photodetection details}
\label{sec:scint-pmts-nd}

Strips in the upstream 120 scintillator planes of the near detector (the 
``calorimeter section'') are read out
individually. In the downstream muon spectrometer section, sets of four anode 
pads are connected in parallel to reduce the
number of front-end electronics channels.  The four summed strips are about 
\unit[1]{m} apart 
so that muon tracking is unambiguous.  The near detector uses a total of 194 
PMTs.

Pre-installation tests of the M64s established the operating high
voltage for each tube based upon the requirement of a gain of 0.8$\times$10$^6$ averaged over all pixels.  A typical PMT required
\unit[$\sim$800]{V} to reach this gain.

\subsubsection{Far detector photodetection details}
\label{sec:scint-pmts-fd}

In the far detector, there are a total of 484 MUX boxes housing the 1452 
PMTs that read out signals from both ends of each
scintillator strip. An additional 64 PMTs are used in the cosmic-ray veto
shield covering the far detector (Sec.~\ref{sec:vetoshield}).

Calibrations established the operating high voltage for each PMT. The
requirement was a gain of 
$1 \times 10^6$ for the highest gain pixel on each M16.  A typical value
of this setting was \unit[$\sim800$]{V}.

During the detector design phase of MINOS an investigation showed that
the uniformity of response over the area of each pixel was sufficient 
to accommodate eight fibers in a close-packed arrangement and that fiber 
position shifts as large as $\unit[200]{\mu m}$ 
from nominal could be tolerated.
A pixel's eight fibers come from scintillator plane strips separated by 
approximately \unit[1]{m}.  The fiber assignment on the opposite end
of the strips was determined so as to provide an unambiguous 
3-dimensional (``demultiplexing'') reconstruction of charged particle 
tracks and showers.

The multiplexing pattern~\cite{Rebel:2004mm} is designed to minimize 
the effect of crosstalk in the M16s.   
The demultiplexing algorithm~\cite{Rebel:2004mm} is based on two 
assumptions: first, the approximate position of a hit along a strip is 
a function of the ratio of the signal amplitudes at its two ends, 
and second, on any plane in the detector, the
width of an event spans \unit[1]{m} or less.  The algorithm uses information 
from neighboring planes to choose the overall
best solution, including the effect of crosstalk.

\subsection{Connectors and cables}
\label{sec:scint-connect}

Cables of clear fibers transport photons from the WLS fibers 
(Sec.~\ref{sec:scint-fibers}) to the MUX boxes
(Sec.~\ref{sec:scint-pmts-fd}).    
The clear fiber from Kuraray has attenuation lengths in the 
\unit[11--15]{m} range for green LED light.
Except for the \unit[175]{ppm} fluor concentration in the WLS fibers,
the clear and WLS fibers are geometrically and compositionally
identical. As a result, at the fiber-to-fiber interface
the directions of signal photons are preserved within the total 
internal reflection angle
and none should escape.  However losses do arise from mismatching 
of refractive indices in the coupling medium at the interface
and from concentric misalignment of fiber ends. 
By using an optical couplant (Dow Corning Q2-3067), the transmission 
loss at each interface was improved from 10-15\% to 5-10\%. 

\begin{figure}[htpb]
  \begin{center}
    \includegraphics[width=0.9\columnwidth]{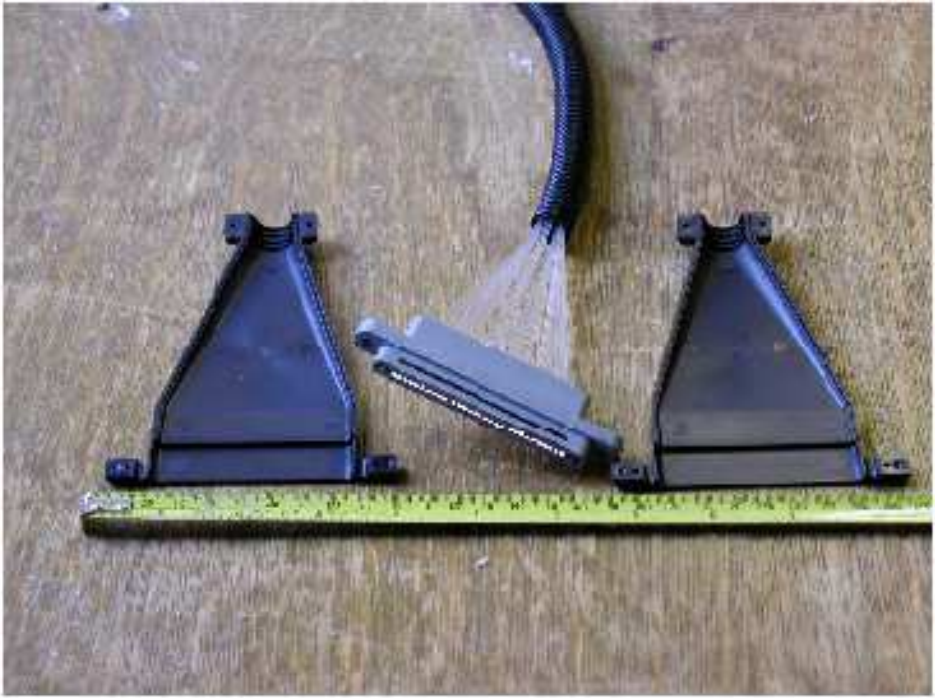}
    \caption{\label{fig:connector} 
      Components of a clear fiber cable connector assembly.}
  \end{center}
\end{figure}

A finished cable is made of four components, namely clear fiber, 
connectors, opaque cable conduit, and opaque shroud
(injection molded clamshell) which allows conduit to be joined to connector.  
The various pieces are shown in Fig.~\ref{fig:connector}.
The injection-molded Noryl connectors at the ends of cables mate with 
nearly identical connectors on the module manifolds and
MUX boxes.  The optical surface of a connector is a linear array of 34 holes, 
starting and ending with pairs of holes
for connecting screws and aligning dowel pins, with 30 fiber holes in between.
The center-to-center alignment of fibers holes was measured on a few hundred
connector pairs and found to be uniform to better than $\unit[30]{\mu m}$. 
The flexible corrugated black nylon conduit used for light-tightness was a 
commercial product (Kabelschlepp PRF16/BL). 
Rare pinholes discovered in the checkout procedure were covered with black 
vinyl tape.  The shroud consists of identical
mating custom injection molded clamshell pairs screwed together with the seams
sealed with RTV103 (black) silicone rubber.  Custom made black neoprene
rubber bands covered the interface region between mating connectors to
complete the light tightness.

\subsection{Cosmic ray veto shield}
\label{sec:vetoshield}
In order to reduce the
background in the measurement of atmospheric neutrinos~\cite{Adamson:2006an} 
arising from cosmic rays entering the top and
sides of the far detector, a veto shield covering these areas was installed.
The surface area of the far detector's octagon sides is 40\% air gaps, 40\% 
steel plate edges and 20\% scintillator edges.
Although 80\% of the surface area is uninstrumented, most cosmic ray tracks 
entering the detector through
this area have large enough angles to the detector planes that they do not 
penetrate very far before
passing through scintillator.  
The veto shield was designed to detect all cosmic ray tracks entering the 
detector, including those that penetrate
deep inside before passing through a scintillator plane. It was assembled 
from 168 of the same C- and
E-type scintillator modules used in the main detector 
(Sec.~\ref{sec:scint-modules-design}) and is read out using the same
front-end electronics and data acquisition system described in
Section~\ref{sec:elec}.  
Unlike signals from the body of the far detector, however, 
the veto shield signals are not included in the 
trigger logic (Sec.~\ref{sec:elec-fe-far}). 

\begin{figure}[htpb]
\begin{center}
\includegraphics[width=\columnwidth]{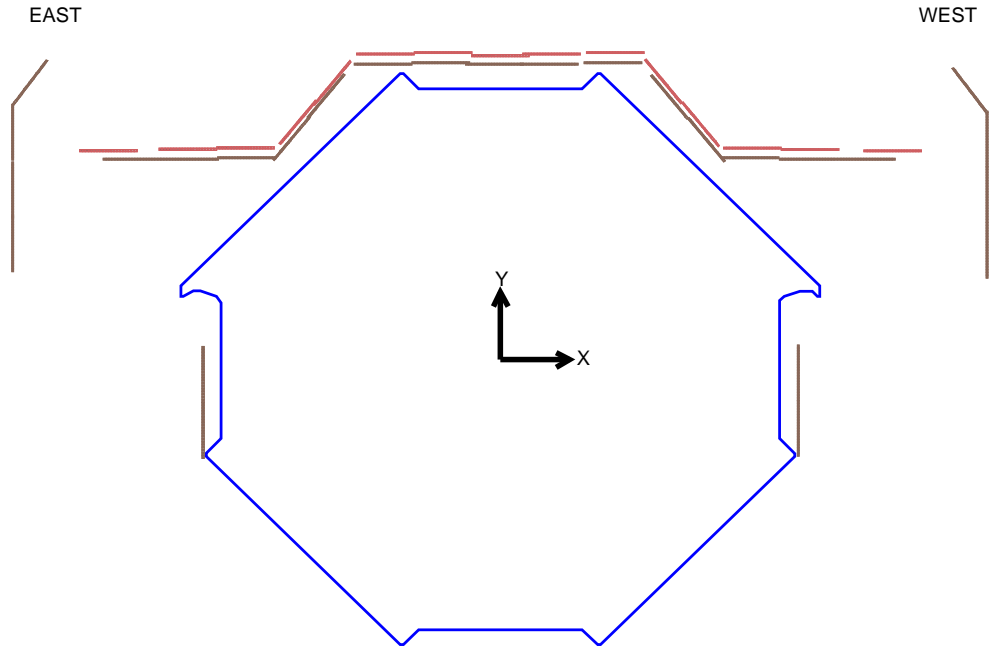}
\caption[A schematic of the MINOS far fetector veto shield]{Schematic of 
 the MINOS far detector veto shield in the $XY$ plane (not to scale).
 Individual scintillator modules that comprise the shield are shown, 
 including the double layer above the detector. }
\label{fig:VetoShield}
\end{center}
\end{figure}

The veto shield scintillator strips are aligned parallel to the long ($Z$) 
axis of the detector and read out at both
ends. Each supermodule is covered by two shield sections, each 8\,m long. 
Since a supermodule is approximately \unit[15]{m} long,
the two sections overlap by about a meter at the center of the supermodule. 
Figure~\ref{fig:VetoShield} shows a vertical
cross section through the shield (in the $XY$ plane).  The horizontal and
diagonal shield sections that cover 
the top of the detector have two layers of
scintillator while the vertical shield modules along the east and west
sides of the far detector (known as ``walls'') have just one.  A Monte Carlo 
simulation~\cite{Howcroft:2004un} shows that more than 99.9\% of
the cosmic ray muons that deposit energy in eight or more far detector 
planes produce signals in the veto shield, excepting those which enter
through the north and south faces where no veto shield is present.

The output signals from eight adjacent strips are summed together and read 
out by a single electronics channel on each
end. The summing pattern is the same on both ends so it is not possible to 
demultiplex the hits in the shield in the same
way as is done for the body of the far detector.  
Therefore the transverse hit location resolution of the shield is
approximately \unit[12]{cm}, as calculated by dividing the width of eight
\unit[4.1]{cm} wide strips by $\sqrt(12)$.

To reduce false cosmic ray muon tagging that could arise from single 
photo\-electron noise, the dynode threshold is set
between one and two photoelectrons. The singles rate in the shield is 
predominately due to background $\gamma$
radiation from the cavern walls and noise from the wavelength shifting 
fibers~\cite{Avvakumov:2005ww}.

A cosmic ray muon entering the detector is identified by a shield signal in 
time coincidence with an event observed in the
main detector.  Cosmic ray muons passing through the detector were used to 
measure the timing resolution of a single hit in
the shield to be \unit[(4.2$\pm$0.3)]{ns}~\cite{Howcroft:2004un}.
The inefficiency for tagging cosmic ray muons is dominated by three factors: 
high singles rates in the shield causing
readout dead-time, cosmic ray muons passing through small gaps in the 
shield, and muon signals falling below the \unit[1--2]{photoelectron}
threshold. The efficiency of the veto shield for cosmic ray 
muons that stop in the detector
is measured to be 95-97\%, depending on the the algorithm 
used~\cite{Howcroft:2004un, Blake:2005nr}. 

\subsection{Scintillator performance}
\label{sec:scint-performance}

The performance of the scintillator system components was optimized during 
the R\&D period and monitored during construction and operation. 
Section~\ref{sec:scint-performance-mappers}
describes the measurement of strip response in every module prior to 
installation. The effects of aging in scintillator system components
(Sec.~\ref{sec:scint-performance-aging}) was also measured during the MINOS 
R\&D program. Ongoing performance is monitored continuously during routine 
experiment operation using cosmic ray muon tracks
(Sec.~\ref{sec:scint-performance-variation} 
and~\ref{sec:scint-performance-timeres}).

\subsubsection{Module maps}
\label{sec:scint-performance-mappers}

A module mapper was used at each module production facility and at Soudan 
for quality control and calibration.  The
mapper used a well-defined $\gamma$ beam from a \unit[5]{mCi} $^{137}$Cs 
source ($\gamma$ energy \unit[662]{keV}) to illuminate a
\unit[4$\times$4]{cm$^2$} square on the scintillator module. A computer
controlled x-y drive moved the source across the face of the module.
The PMT signal from each strip was integrated for \unit[10]{ms} once 
every \unit[40]{ms} and recorded. This duty cycle was due to the integration
of the RABBIT PMA card~\cite{Drake:1987ff,Drake:1985iw}
in current monitoring mode and the limit of reliable readout of the data 
through the Rabbit-CAMAC-GPIB system. Since the
strip thickness is a small fraction of the $\gamma$ absorption length, 
the scintillator is fully illuminated.
These measurements provided a detailed map of the response of each 
scintillator module to ionizing radiation.
Measurements were made every \unit[8]{cm}
along the length of each strip with a precision and reproducibility  
of $\sim$1\%. Besides providing these pulse height performance calibrations, 
the locations of the edges of the scintillator strips were determined to a 
precision of \unit[1]{mm} by the response curve as the
source was scanned transversely across each strip. 

\begin{figure}[htpb]
  \begin{center}
    \includegraphics[width=\columnwidth,keepaspectratio=true]{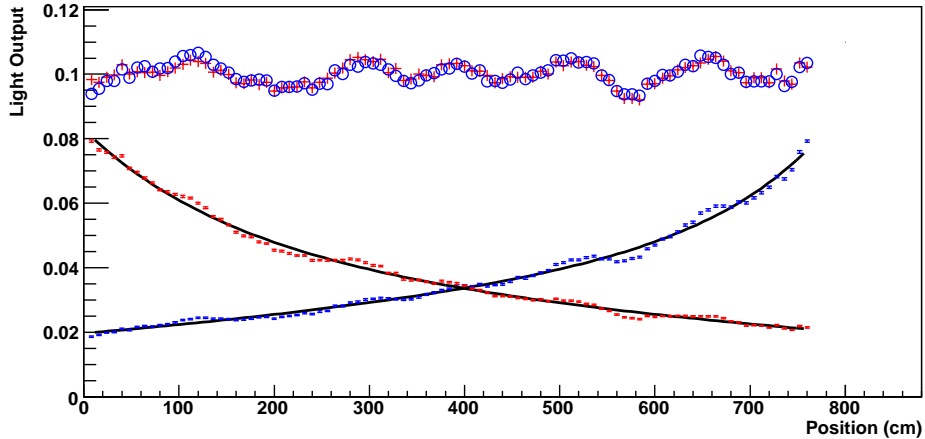}
    \caption{ Module mapping data from a typical far detector module.
      The lower graphs show responses from each end of a single
      scintillator strip and their least squares fits to the sum of two
      exponentials. The responses from the two ends are consistent.  The
      vertical scale was used to compare the absolute light yield of
      different modules, and is approximately 1/100 of the number of
      photoelectrons at the PMT for a cosmic ray muon passing normally
      through the module plane. The top (wavy) series of datapoints
      shows the ratios of these data to their respective fits,
      normalized to 0.1 for display purposes.}
  \label{fig:modulemap}
  \end{center}
\end{figure}

\begin{figure}[htpb]
  \begin{center}
    \includegraphics[width=\columnwidth,keepaspectratio=true]{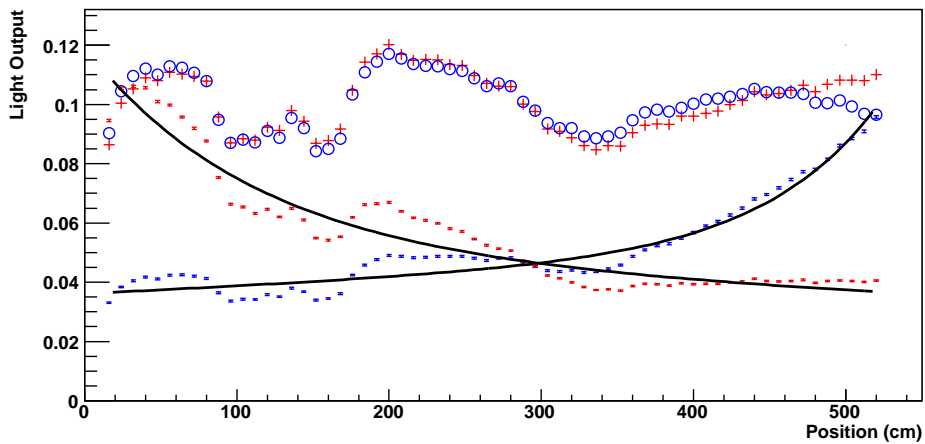}
    \caption{ Mapping data from a module with a badly glued fiber, to be
      compared to Fig.~\ref{fig:modulemap}.  Bubbles in
              the nozzle inhibited glue flow in the region of about 
              \unit[100--200]{cm} and  \unit[300--350]{cm}.
   }
  \label{fig:modulebadglue}
  \end{center}
\end{figure}

\begin{figure}[htpb]
  \begin{center}
    \includegraphics[width=\columnwidth,keepaspectratio=true]{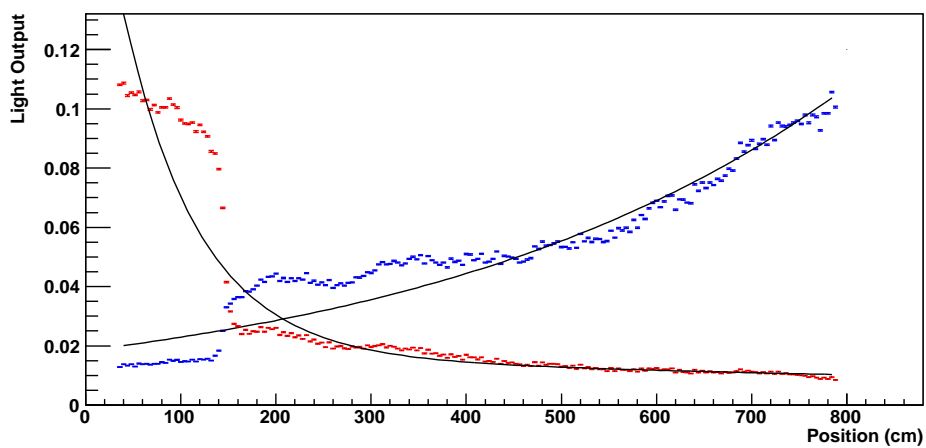}
    \caption{ Mapping data from a module containing a rare damaged 
     fiber. The associated strip had a fiber that was 
     damaged at about \unit[150]{cm} from one end as seen by abrupt steps 
     in the otherwise moderately sloping attenuation curves.  
}
  \label{fig:modulemapbadfiber}
  \end{center}
\end{figure}

Figure~\ref{fig:modulemap} shows the mapper's response for a typical strip 
in a production module. The small variations in
the light output from a smooth exponential behavior are real and result 
from variations in scintillator, fiber depth, and
gluing. Figure~\ref{fig:modulebadglue} shows the map for a strip with 
imperfect fiber gluing (the fiber was not completely
surrounded by the optical epoxy), resulting in local drops in light output 
due to poor optical coupling between
scintillator and fiber. Figure~\ref{fig:modulemapbadfiber} shows a rare 
example of a strip in which the WLS fiber has a
defect within the module, causing poor transmission at one location and a 
discontinuity in both response curves.

A subset of roughly 1000 modules was mapped at each of the production factories 
(Caltech and Minnesota) and again at Soudan.
Typically the maps taken at the two factories using two different 
mappers vary by less than 2\% at all locations. Larger
variations could be traced to shipping damage of the epoxy around 
fibers caused by improper operation of the
fiber gluing machine. No properly built module was found to suffer a
significant change in light output measured at Soudan compared to 
that measured at the production factory, with the
exception of one shipment that was exposed to \unit[-30$^\circ$]{C} 
winter weather, which caused some fibers to separate from their
strips. Even these few damaged modules were still within nominal 
acceptance tolerances for use in the detector.


\subsubsection{Accelerated aging tests}
\label{sec:scint-performance-aging}

In order to verify the long-term stability of scintillator light output, 
several accelerated aging studies were performed:
\begin {enumerate} 
\item Tests on small samples of module material were conducted 
  in four different aging environments: 
\begin{enumerate}
  \item typical room conditions
  ($\sim$\unit[22$^\circ$]{C}, relative humidity $\sim$50\%);
  \item \unit[50$^\circ$]{C} at 50\% relative humidity;
  \item varying temperature from
  \unit[-30$^\circ$]{C} to \unit[+50$^\circ$]{C} in a 1-hour cycle;
  \item  and 
  \unit[50$^\circ$]{C} with 100\% relative humidity.
\end{enumerate}
\item Mechanical stress tests were performed on scintillator strips 
  with fibers glued in place.
\item Full-length scintillator modules and clear fiber cables 
  were exposed to four different temperatures for six months: 
  20$^\circ$, 28$^\circ$,
  38$^\circ$, and \unit[50$^\circ$]{C}. 
\end {enumerate}

The most significant results of these tests were:
\begin {enumerate}
\item The scintillator showed a non-reversible reduction in light 
  output with age, associated with yellowing. The change in
  light output is -1.2\%/year at \unit[20$^\circ$]{C}.
\item Although optical epoxies typically become visibly yellow with age, 
  the impact on light collection was found to be negligible because there is
  only a thin epoxy layer between fiber and scintillator.
\item The scintillator showed a reversible loss in light output of 
\unit[-0.3\%/$^\circ$]{C} as temperature was raised (but see item (v) below).
\item A non-reversible decrease in the attenuation length of the WLS 
  fiber as a function of time was measured.
  The change in light output at the end of an 8~m fiber was found to be 
  -1.2\%/year at \unit[20$^\circ$]{C}.
\item Increased temperature resulted in faster aging of the scintillator
  and WLS fiber. The rate of aging is parametrized exponentially with
  temperature as  $\exp(\Delta T/\tau)$ with $\tau=\unit[10^\circ]{C}$.
\item Temperature cycling (over $>$10$^4$ cycles) induced no measurable
  aging effect beyond that caused by the part
  of the cycle that was above room temperature.
\item High humidity combined with high temperature accelerates aging.
\item ``Normal'' mechanical stresses of production, assembly, 
  shipping, and installation  had no effect on light output. Extreme
  stresses can break fibers or strips and
  exposed fibers are susceptible to cracking (which can significantly
  reduce the light transmission), but handling with normal industrial 
  practice did not damage fibers in any measurable way.

\end {enumerate}

\renewcommand\theenumi {\arabic{enumi}}

The overall conclusion is that after 10 years the light output of the MINOS
scintillator system will be 65-75\% of its initial value. 
One note of possible importance is that the aging tests were performed on
scintillator that had been extruded about six months prior to the aging 
measurements.
Subsequently, other measurements have suggested that the scintillator 
undergoes an initial drop in light output
 on the order of 10-15\% in the first few months before stabilizing to 
the behavior reported above. 

\subsubsection {Variation in light output from module to module}
\label{sec:scint-performance-variation} 

\begin{figure}[htpb]
  \begin{center}
    \includegraphics[width=0.9\columnwidth]{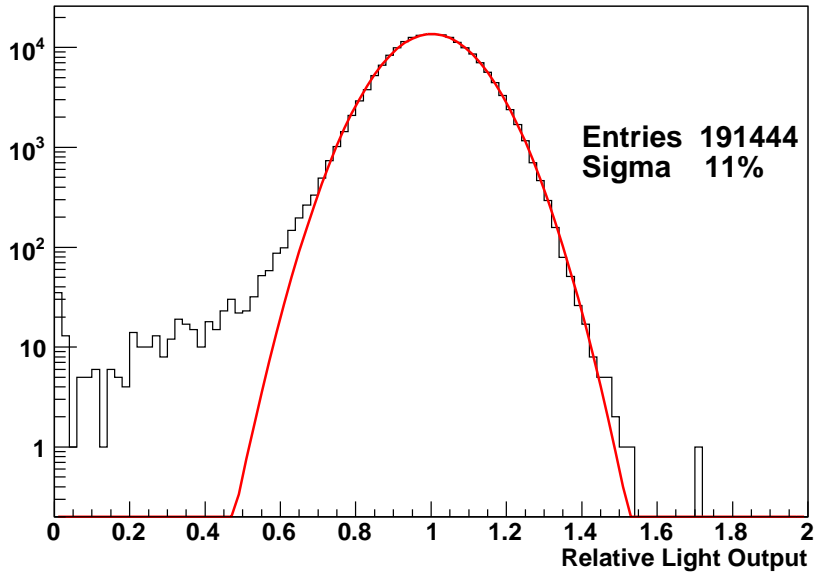}
    \caption{Distribution of light output from all far detector strips 
     for a \unit[662]{keV} $\gamma$ source at the strip center as measured
     by the module mappers. }
    \label{fig:modmaplightvariation}
  \end{center}
\end{figure}

Differences in module light output result from a convolution of variations 
in several components and in the quality
of the construction. Factors include the light output from the scintillator 
strips, the quality of the WLS fiber, 
the quality of gluing the WLS fiber into the modules, damage to the 
fibers, and the quality of finishing the 
fibers at the connector ends. Figure~\ref{fig:modmaplightvariation} shows 
the variation in light output at the center
of all far detector scintillator modules as measured by the module mappers. 
The Gaussian fit has a width of
11\% and only 0.16\% (317 out of 191,444) have light output less than 50\% 
of the average. Typically the fiber in
these abnormal strips had suffered damage during construction.

\begin {figure}[htpb]
  \begin{center}
    \includegraphics[width=0.9\columnwidth]{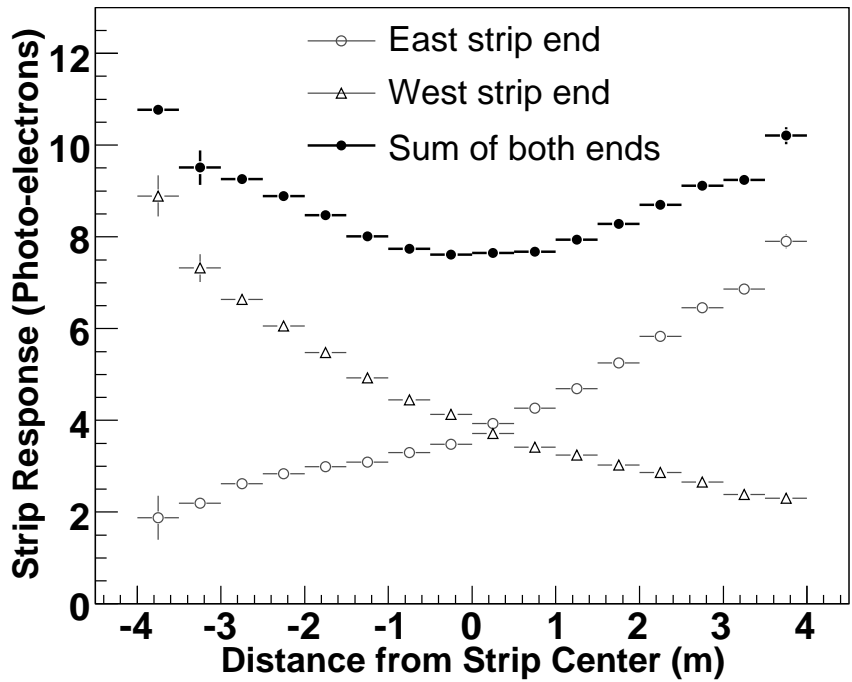}
    \caption {\label{fig:fardetsignals}  Average light output from 
     in-situ far detector strips as a function of
     distance-from-center for normally incident MIPs. 
     The data are from stopping cosmic ray muons, for which
     containment criteria cause lower statistical precision at the ends
     of the strips.  
   }
  \end{center}
\end {figure}

After correcting for known systematic effects, average photoelectron yields 
for minimum ionizing particles (MIPs) of installed modules also vary because
of differences in transmission at the two connector interfaces,
variation in transmission at the PMT interface, and pixel to pixel
quantum efficiency variations in the PMT. In addition, different gains
of PMT pixels produce further variation in average MIP signal size. The
average light yields at strip center for normally incident MIPs in the
near and far detectors are about seven and six
photoelectrons per plane respectively, summed over both ends.  
Stopping cosmic ray muons are used as the MIP so their location on
the Bethe-Bloch curve in relation to the minimum-ionizing point can
be accounted for (Sec.~\ref{sec:calib-muons-stopping}).  
However, it should be
noted that the variation across the face of the detector, both with
strip number and along the strip length, is significant.
Figure~\ref{fig:fardetsignals} shows the average light output for both
ends of the readout for muons crossing at different positions along the
strips in the far detector. The data shown are an average from a subset
of strips that are all \unit[8]{m} long and in the $U$-view planes. The
light output varies across the face of the detector because of differing
strip and clear fiber lengths.  While the signal from one end of a strip
varies by a factor of 4-5 along the length of the strip, the sum of both
strip ends varies by only $\sim$25\%. In the near detector, the
relatively short strip length and far end mirrors provide some
compensation for the single-ended readout.

\subsubsection{Time resolution}
\label{sec:scint-performance-timeres}

\begin{figure}[htpb]
  \begin{center}
    \includegraphics[width=0.9\columnwidth]{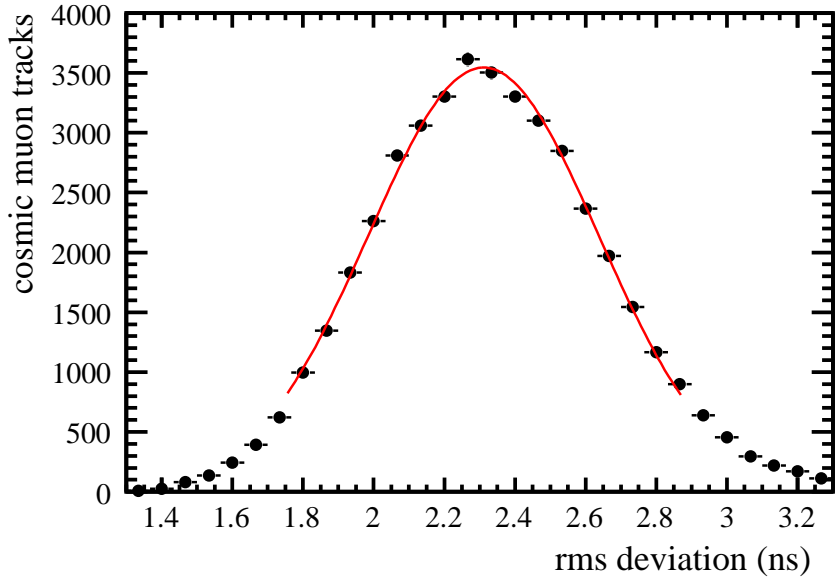}
    \caption{Distribution of rms deviations from linear timing fits 
      to cosmic ray muon tracks. The data (solid
      circles) are well-described by a Gaussian distribution
      (curve) with \unit[2.3]{ns} standard deviation.}
    \label{fig:TimeRes}
  \end{center}
\end{figure}

The time resolution of the scintillator system was measured using 
cosmic ray muons at the far detector. The time offsets
between readout channels were first calibrated as described in 
Sec.~\ref{sec:calib-muons-timing}. The times and
positions of the hits on each muon track were then compared and a 
linear timing fit performed under the assumption
that the muons travel at the speed of light. The rms deviation of 
the measured times from the fitted times was calculated for
each muon track. Figure~\ref{fig:TimeRes} shows the distribution of 
these rms deviations and a Gaussian fit giving
a mean resolution of \unit[2.3]{ns}.

\section{Electronics and Data Acquisition}
\label{sec:elec}

While considerable care was taken to make the near and far detectors as
similar as possible, the same electronics could not be used for the two
detectors.  The far detector is \unit[700]{m} underground and the rate
of cosmic muons and neutrino interactions is around \unit[0.5]{Hz}.
The data volume is therefore entirely dominated by
detector noise of \unit[$\sim10$]{kHz} per plane
side~\cite{Avvakumov:2005ww}. In contrast, there are several neutrino
interactions per \unit[8--10]{$\mu$s} long spill in the near detector (3.5 
reconstructed events per 10$^{13}$ protons on target~\cite{Adamson:2007gu}) .
Consequently different requirements led to the design of different electronics
systems. To minimize the systematic uncertainties associated with these
differences, the response of the calibration detector was measured
separately with both near and far detector front-end electronics
(Sec.~\ref{sec:calib-near-far-elec}).

The design goals of the electronics and data acquisition systems
(``DAQ'') are summarized below.  The near detector and far detector
front end electronics are described separately followed by a section
describing the DAQ common to both.  Note that the relatively high
electrical power used by the near detector front-end electronics
(Sec.~\ref{sec:elec-fe-near}) requires a chilled water cooling system
for the racks housing this system (Sec.~\ref{sec:ops-lab-nd}).  The
low-power far detector front-end electronics racks
(Sec.~\ref{sec:elec-fe-far}) require only air cooling by rack-mounted
muffin fans, as do the DAQ racks (Sec.~\ref{sec:elec-daq}) at both
detectors.

\subsection{Electronics design goals}
\label{sec:elec-req}

The primary goals of the electronics are to provide adequate
information for the separation of neutral and charged
current neutrino interactions and to enable the measurement of
the energies of accelerator-neutrino and cosmic ray particle
interactions with minimal bias. These goals led to the following
specifications:

\renewcommand\theenumi {\roman{enumi}}
\begin{enumerate}
\item The electronics should digitize the charge and the time of each
  pulse from the photodetectors with a (programmable) threshold of
  \unit[0.3]{photoelectrons}. The lowest gain pixels in a PMT should have 
  a gain of
  3$\times$10$^5$, requiring the electronics to have a readout threshold
  of \unit[16]{fC}.
\item A minimum ionizing particle should produce \unit[2--10]{photoelectrons} 
  at each
  side of the scintillator strips. This signal has to be measured with a
  1--2\% accuracy, which requires the least significant bit to
  correspond to \unit[2]{fC}.
\item Signals can be as large as \unit[25]{pC}
  for a high-gain (10$^6$) PMT pixel. The dynamic range of
  the system therefore has to be around 12--13 bits.
\item The near detector has to separate signals coming from different
  interactions during the \unit[8--10]{$\mu$s} beam spill. A timing
  accuracy of around \unit[20]{ns} is sufficient, given the maximum
  instantaneous event rate of \unit[20]{events} per spill, and is
  coincident with the \unit[53.1]{MHz} RF structure of the beam.  The far
  detector, however, is also used to study atmospheric neutrino
  interactions. A timing resolution on the order of \unit[3--5]{ns} is
  needed to separate upward from downward going muons or neutrinos.
\item There should be negligible dead time during the neutrino
  spill and at least an 80\% live time out of spill to record
  atmospheric neutrino and cosmic ray muon interactions. 
\end{enumerate}
\renewcommand\theenumi {\arabic{enumi}}

\subsection{Near detector front-end electronics}
\label{sec:elec-fe-near}

\begin{figure}[htpb]
  \centerline{
    \includegraphics[width=0.9\columnwidth,bb=220 140 465 405]{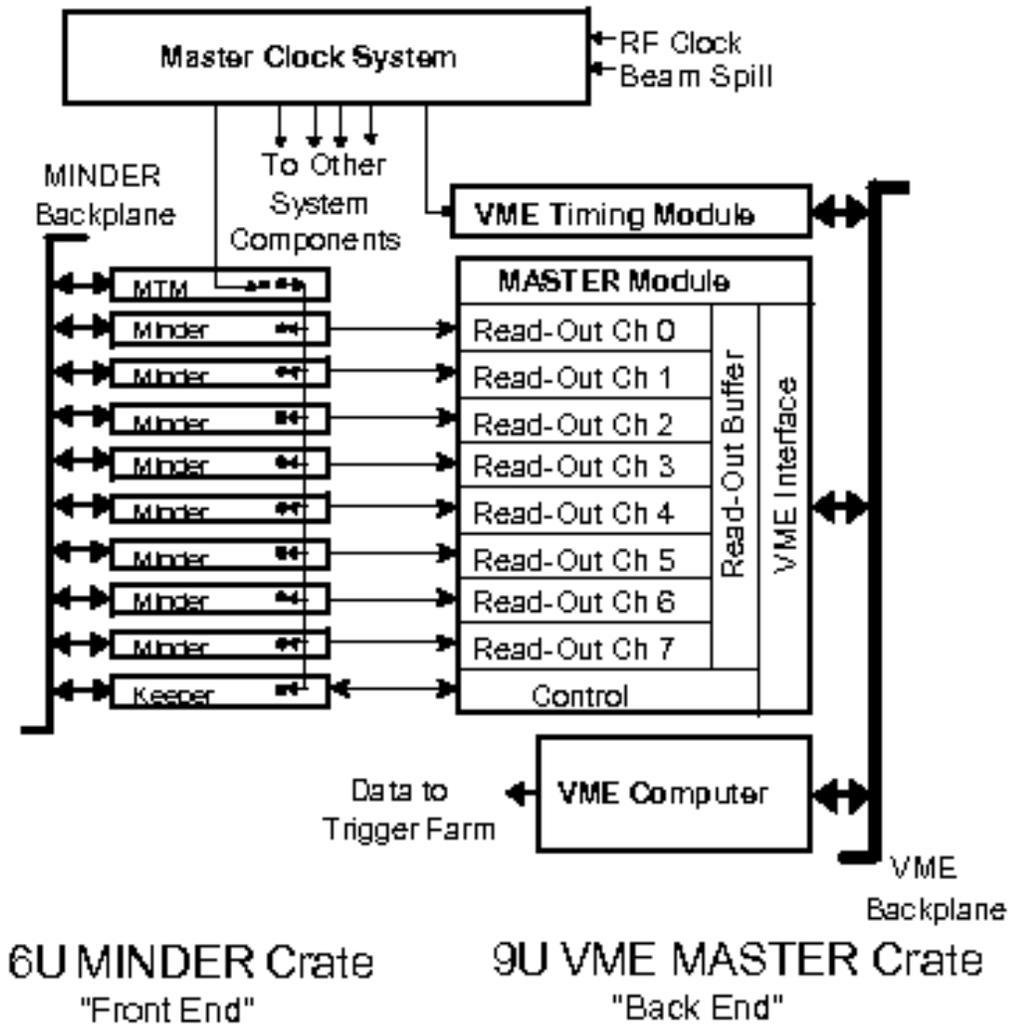}
  }
  \caption{\label{fig:nd-elec-overview}Schematic overview of the MINOS
    near detector readout electronics.  One 64-pixel PMT is read out by
    four ``MINDER'' cards.  At the end of a trigger, a ``MASTER'' card
    in a different VME crate reads out eight MINDERS, applies low-level
    calibrations, and transfers the resulting data across the backplane
    to a VME computer, where it is then passed on to a trigger farm for
    futher processing.  The ``Keeper'' card in the MINDER VME crate
    distributes trigger signals to the MINDERs, and a clock system keeps
    all components in sync.}
\end{figure}

In the near detector front-end electronics~\cite{Cundiff:2006yz}, each
PMT pixel is digitized continuously at the frequency of the beam RF
structure of \unit[53.103]{MHz} (\unit[18.83]{ns}).
This is achieved with an individual front-end channel unit consisting
of a small mezzanine
printed circuit board (PCB) called a ``MENU''.  The principal MENU
components are an ASIC named the {\it Charge Integration Encoder}
(abbreviated ``QIE'')~\cite{yarema:1992dp}, a commercial flash analog-to-digital converter
(``FADC''), and a data buffer.  Sixteen MENUs reside on a VME type-6U PCB
(called a ``MINDER''), with four MINDERs required for each fully used
M64 PMT.  An overview of the near detector electronics system is shown
in Fig.~\ref{fig:nd-elec-overview}.

The QIE input signal current, $I$, is split into eight binary-weighted
``ranges'' with values $I/2, I/4, I/8, \ldots$, and integrated onto a
capacitor for each range.  A bias current is added to ensure that the
capacitor voltage on one and only one range is within the predetermined
input limits of the FADC.  The QIE selects that voltage for output to
the FADC, and also outputs a 3-bit number representing the range value.
Four independent copies of the current splitter, integration, and output
circuits in the QIE permit continuous dead-timeless operation.  The
combined information of an 8-bit FADC value and 3-bit range value
provide an effective dynamic range of sixteen bits.  For the purpose of
calibration, an additional 2-bit code allows the identification of the
current splitter and capacitor circuit (known as the ``CapID'') used
for each RF cycle.

\begin{figure}[htpb]
\centerline{\includegraphics[clip,width=\columnwidth]{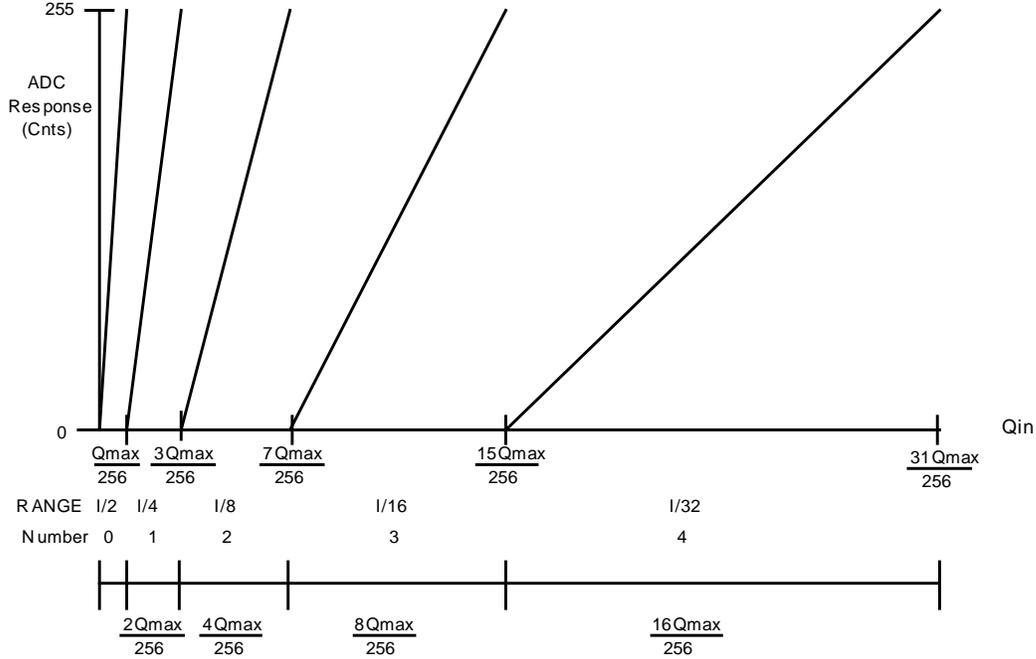}
}
\caption{Response of the QIE-chip based electronics: The x-axis shows
  the amount of charge injected into the chip, while the y-axis shows
  the ADC response for the different ranges. Only the first five out of
  the eight ranges are shown.}
  \label{fig:QIEDiagram}
\end{figure}

Data are stored locally on each MENU during a trigger gate and
subsequently read out.  A beam-spill gate of up to 1000 RF periods
(\unit[18.83]{$\mu$s} maximum, \unit[13]{$\mu$s} typical) is formed
from a signal provided by the Fermilab Accelerator Control Network
(ACNET).  Outside of beam spills, cosmic ray and PMT singles-rate data
are acquired individually for each PMT for 8 RF periods (\unit[150]{ns})
whenever the charge on the dynode exceeds the equivalent of
approximately $\frac{1}{3}$ of a photoelectron with a \unit[10]{ns} shaping
time.  Charge injection calibration data are acquired using a
calibration trigger gate of 256~RF periods.

At the end of a trigger gate, data are transfered from the front-end to
VME type-9U modules called ``MASTERS'', which read out up to eight
MINDERS each.  A dead-time of \unit[$\sim$600]{ns} is imposed on the front-end for
each integration cycle of data read out per MINDER.  Given the roughly
\unit[1]{kHz} of dark noise from the PMTs, this yields a typical
dead-time of 0.5\% in cosmic ray data on a PMT-by-PMT basis.  Beam-spill
gates supersede the storage and readout of dynode triggered data and
therefore suffer no dead-time.

Each data word from the front-end, consisting of range, FADC value, and
CapID, is linearized in the MASTER using a lookup table which
represents the results of a charge injection calibration of each MENU.
The resulting linearity is better than 0.5\% over the entire dynamic
range.  The linearized digitizations exceeding a pedestal suppression
threshold of $\sim$20\% of a photoelectron are attached to
channel-identification and time-stamp data and stored as 64-bit words in
a dual-port readout buffer.  The buffers are read out every
\unit[50]{ms} in a single DMA block transfer performed by the local VME
processor.

To provide uniformity, a centralized near detector clock system is used
to distribute a continuous \unit[53]{MHz} reference, spill signals, and other
control signals to all front-end modules.  Clock signals are also used
to synchronize the readout of data by the VME processors and the DAQ
system.  The near detector clock is synchronized to the Fermilab
accelerator but the phase of timing signals relative to an independent
GPS system is used to allow accurate reconstruction of the absolute
UTC event time.  
\subsection{Far detector front-end electronics}
\label{sec:elec-fe-far}

\begin{figure}[htpb]
  \centerline{
    \includegraphics[width=0.9\columnwidth]{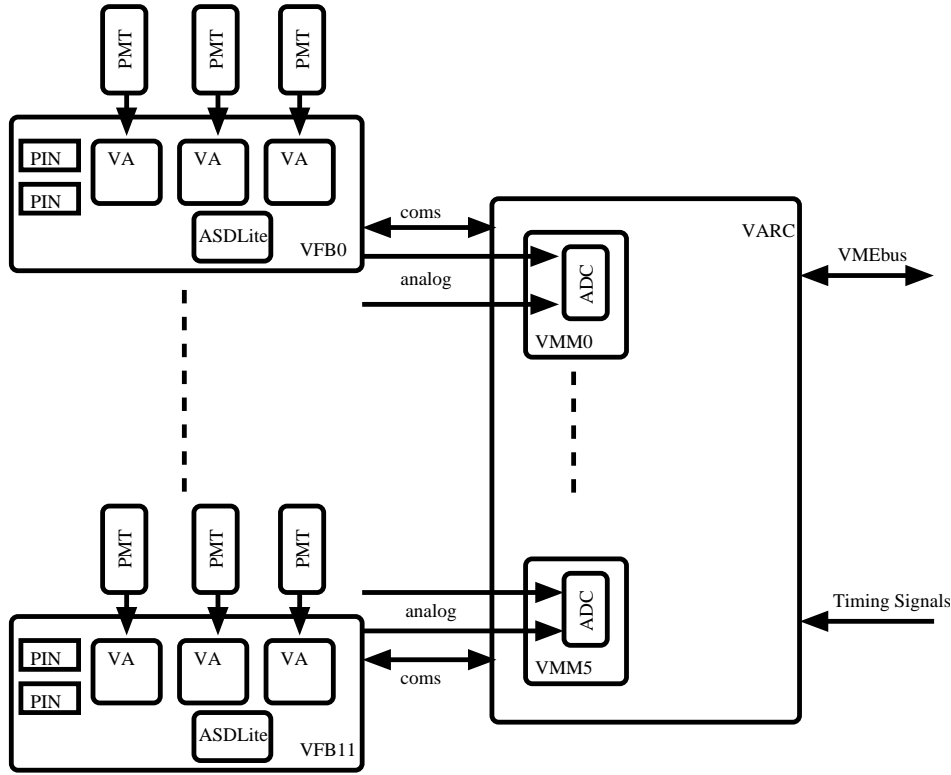}
  }
  \caption{\label{f:fd-elec-overview}Schematic overview of the MINOS far
    detector readout electronics. Three PMTs are connected with short
    flat ribbon cables to the VFB, which also houses
    two PIN diodes to monitor the light level of the light injection
    system. The VA ASIC amplifies and holds the PMT signals, which are
    multiplexed via an analog link onto an ADC on the VARC. The VFB
    is controlled through a digital link (coms) by the VARC.}
\end{figure}

The far detector electronics~\cite{Oliver:2004ek} were 
specifically designed for the low rate underground environment. 
The neutrino beam
generates only a handful of events per day and the cosmic muon rate
\unit[700]{m} underground (\unit[2070]{m} water-equivalent depth) 
is \unit[0.5]{Hz}. The signal rate is
therefore dominated by detector noise and is \unit[3--6]{kHz} per PMT.
Given this low rate, commercial \unit[10]{MHz} digitizers operated at
less than \unit[5]{MHz} are able to serve
many channels at once and still operate with very low dead time.  A
block diagram of the readout structure is shown in
Fig.~\ref{f:fd-elec-overview}.

The readout is based on the front-end ASIC VA32\_HDR11 (short VA chip),
developed in collaboration with the Norwegian company IDEA
ASA~\cite{ideas}. Three VA chips are mounted onto the VA front-end
board (VFB).  There are 32~channels on the chip of which only 17 are
used, each channel consisting of 
a charge sensitive preamplifier, a shaper, a track and
hold stage, and an output switch. The output from a selected channel can
be switched to a differential output buffer which drives a \unit[$\sim$6]{m}
long cable to a remote ADC. There is a second multiplexer that can
route charge to a selected channel for test and calibration.

Three VA chips are mounted on the VFB,
located on the outside of the PMT MUX box 
(Sec.~\ref{sec:scint-pmts-fd}).  The VFB provides support circuitry for
power distribution and biasing of the VA chips. It also houses the
ASDLite~\cite{ASDLite} ASIC, which compares the dynode signals from the
PMTs with a common programmable threshold to provide a discriminated
signal for time-stamping and readout initiation.

A serial slow control interface is provided to adjust VA bias levels,
enable critical voltage regulators, and monitor voltages and
temperature. The VFB is operated in slave mode
and fully controlled by the VA readout controller (VARC) described
below.

The MINOS light injection system (Sec.~\ref{sec:calib-li}) monitors the
stability and linearity of the PMTs. The light injected into the detector is
monitored by PIN (Positive Intrinsic Negative) photodiodes 
mounted on the VFB.  This signal undergoes a
pre-amplification and shaping before being fed into a spare channel on
the VA chip.  The PIN diode signals are read out in coincidence with the 
PMT signals.

The analog signals from the VA chip are multiplexed onto an ADC, which
is located on a VA Mezzanine Module (VMM).  
Two VFBs are connected to
each VMM.  The VARC houses 6~VMMs and controls
the signal digitization, triggering, time-stamping and bias of the VA
chips.  Each VARC can thus service up to 36~PMTs of 16~channels each.

The VARC is implemented as a 9U~VME card.  Three VARCs, a timing card (see
below), and a Motorola VME processor share a single VME crate. There are
a total of 16~VME crates to read out the 22,000 electronics channels of
the detector.  The VARC receives the discriminated dynode signal of each
PMT. It time-stamps these signals with an effective \unit[640]{MHz} TDC,
implemented in a Xilinx~\cite{xilinx} \unit[80]{MHz} field programmable
gate array, and then generates the hold signal for the VA ASIC. The delay
time is programmable and is around \unit[500]{ns} after the trigger
signal is received. The signals held in the VA ASIC are then multiplexed
to a commercial 14-bit~\unit[10]{MHz}~ADC\cite{fdadc}, which is operated
at \unit[5$\times$10$^6$]{samples/s}.  The resulting dead time is 
\unit[5]{$\mu$s}
per dynode trigger. The digitization sequence is started if the VARC
receives at least two discriminated dynode signals from different PMTs
in a \unit[400]{ns} window.  This so-called 2-out-of-36 trigger reduces
the dead time due to dark noise in the PMTs and fiber noise in the
scintillator, without compromising the recording of physics events.

The system has been tuned such that \unit[1]{ADC count} corresponds to
\unit[2]{fC}. The electronic noise in the fully installed system is
typically around \unit[2.5]{ADC counts} (\unit[5]{fC}) and changes less
than one ADC count over \unit[24]{hours}.  The entire detector readout
is synchronized by a \unit[40]{MHz} optical timing distribution signal
slaved to a GPS clock from TrueTime~\cite{gps}.

Once the data are digitized they are transmitted to a local FIFO
and stored there for further processing. The pedestal is subtracted
and data above an individually settable sparsification threshold are
written to an on-board VME memory.  The VARC also controls pedestal and
charge injection calibration runs.  This memory is read out by the DAQ
system described in Sec.~\ref{sec:elec-daq}.

\subsection{Data acquisition}
\label{sec:elec-daq}
\begin{figure*}[htpb]
  \centerline{
    \includegraphics[width=\textwidth]{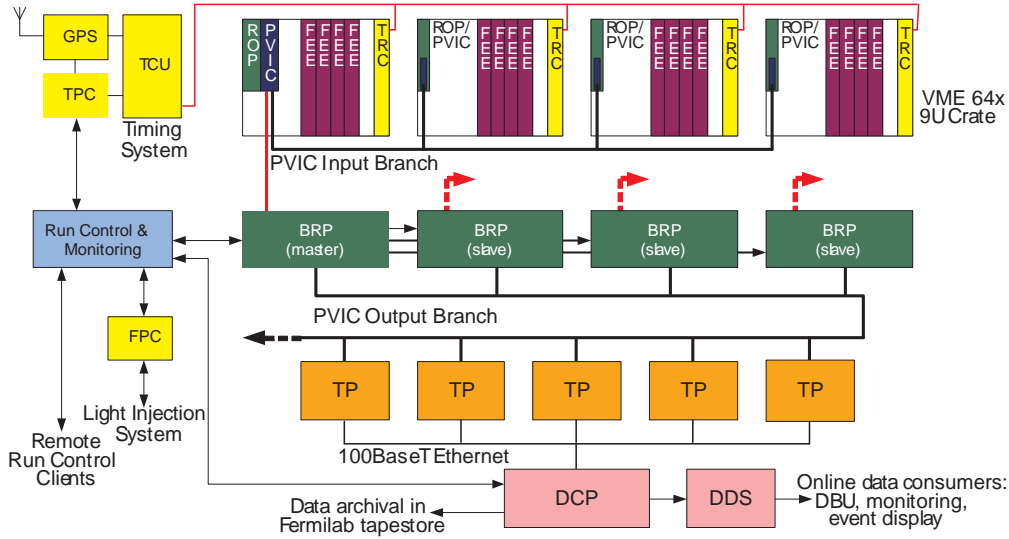}
  }
  \caption{\label{f:daq-architecture-fd}The architecture of the MINOS
    DAQ system.  The far detector front-end electronics and clock system
    are shown as an example.  For clarity, only one of the six PVIC
    input branches is completely illustrated, with three more shown in summary.  }
\end{figure*}

The near and far detector data acquisition systems are functionally
identical, with appropriate front-end software accommodating the
differences of the front-end electronics of the two detectors.  The
main requirements for the DAQ systems are 
({\it i}) to continuously read out the
front-end electronics in an untriggered, dead time free manner and 
({\it ii}) to
transfer the data from all front-end modules to a small farm of PCs
where software algorithms build and select events of interest
and perform monitoring and calibration tasks.  The DAQ system allows
for diagnostic and calibration data taking runs and can be operated 
remotely and
unmanned.  Figure~\ref{f:daq-architecture-fd} shows the layout of the
DAQ system, using the far detector as an example.  The system is
constructed entirely from commercially available components.  A
detailed description of the DAQ system architecture and components can
be found in~\cite{Belias:2004bj}.

The digitized data in the front-end electronics buffers (time blocks)
are read out with single block DMA transfer by a VME computer in each
crate, the readout processor (ROP).  The ROP is a PowerPC running
VxWorks that encapsulates all the readout specific software for the
different front-end electronics systems of the detectors.  There are 16
ROPs serving the far detector and 8 serving the near detector.  The
interrupt-driven readout of each ROP (typically \unit[50]{Hz}) is
synchronized by the timing system through local timing system cards in
each crate.  Each ROP assembles consecutive time blocks in memory into
convenient units, called time frames, which are nominally one second in
length.  The time frames are overlapped by one time block to circumvent
processing problems at the time frame boundaries.  Each time frame
carries a header containing a full self description.  Data blocks from
monitoring tasks or real-time electronics calibrations carried out in
the ROPs are appended to the time frame as necessary.

The transfer of data from the multiple ROPs along the detector uses
PCI Vertical Interconnection (PVIC) buses~\cite{pvic:2004} which
allow transfer rates of up to \unit[40]{MB/s}.  Groups of ROPs are
interconnected with differential PVIC cables to form a chain which is
connected by an optical PVIC cable to an off-detector DAQ PC, the
Branch Readout Processor (BRP).  The near detector is configured with
four such chains (termed {\it input branches}) and the far detector with six.
All PCs in the MINOS DAQ system are Intel Pentium-based and run the
GNU/Linux operating system.

Time frames are buffered by the ROPs until requested by their BRP.  They
are then transferred into the memory of the BRP by
DMA across the PVIC.  One BRP acts as the master, coordinating the
transfer of data by issuing transfer instructions to the BRPs.  Each BRP
is also connected to an output branch, a differential PVIC bus which
connects the BRPs to a small farm of trigger processors (the TPs).  When
a time frame has been successfully transferred to the BRPs, the master
BRP selects a TP and instructs each BRP to send (by DMA) its time frame
to an address in the TP memory via the PVIC output branch.  In this way
a full time frame of data from the entire detector is available in a
single TP for processing.  The TPs, BRPs and ROPs are all able to
buffer and queue multiple time frames.  Typically, in a physics run the
raw data rate into the trigger farm is \unit[8]{MB/s} at the far
detector and \unit[5]{MB/s} at the near detector.  A farm of five
\unit[3.2]{GHz} Pentium4 PCs at each detector comfortably copes with
these rates.

After various data integrity checks the TP performs a number of
processing tasks on the time frame data.  Monitoring statistics, such
as data flow rates, are accumulated. Data from the
interspersed pulsing of the Light Injection (LI) system
(Section~\ref{sec:calib-li}) are
identified as LI events and analyzed.  Summaries of the detector
response to the pulses are generated and output for offline detector
calibration.

In parallel with the LI processing, the TP applies software
triggering algorithms to the time frame to locate events of physics
interest.  The primary triggers are summarized below.  Since each of
these triggers can, in principle, gather multiple events (notably the
spill triggers), the output from the trigger is termed a {\it snarl}.  The
triggers fall into three categories: special triggers for debugging and
calibration, bias-free triggers based on spill signals or spill times to
gather beam events, and triggers based on the clustering of hits in the
detector to gather out-of-spill events.  In this last case, a candidate
snarl is first identified as a temporal cluster of hits bounded by
\unit[150]{ns} gaps of detector inactivity.  The non-spill-based trigger
algorithms are applied to these candidates.

\renewcommand\theenumi {\roman{enumi}}
\begin{enumerate}
\item {\bf Spill trigger:} At the near detector, each digitization that occurs
  within the spill gate is tagged by the front-end electronics
  (Sec.~\ref{sec:elec-fe-near}).  These are identified, extracted from
  the time frame and output as a single spill event with no further
  selection.
\item {\bf Remote spill trigger:} At the far detector a direct spill
  signal is not available so a remote spill trigger is applied.  The
  near detector GPS system is used to generate time-stamps of the spill
  signals. These are transmitted to the far detector over the internet
  where they are stored and served to the TPs on request.  All readout
  within a configurable time window around each spill is extracted and
  written out as a spill-event.  Since the DAQ has considerable
  buffering capability there is ample time to wait for spill information
  to arrive.
\item {\bf Fake remote spill trigger:} Fake spill times are generated
  randomly between spills to provide random sampling of detector
  activity.
\item {\bf Plane trigger:} $M$ detector planes in any set of $N$ contiguous
  planes must contain at least 1 hit.  Nominally $M$=4, $N$=5.
\item {\bf Energy trigger:} $M$ contiguous planes of the detector have a
      summed raw pulse height greater than E and a total of at least $N$
      hits in those planes.  Nominally $M$=4, E=1500 ADC counts,
      $N$=6.  This trigger is not normally used at the near detector.
\item {\bf Activity trigger:} There must be activity in any $N$ planes of
  the detector.  Nominally $N$=20.
\item {\bf Special triggers:} A variety of special runs are available to
  perform detector and electronics calibration or debugging.
\end{enumerate}
\renewcommand\theenumi {\arabic{enumi}}

The integrated trigger rates at the near and far detectors are typically
\unit[4]{Hz} and \unit[30]{Hz}, respectively, and are dominated by cosmic
rays and noise.  The output rate to disk from all sources, including LI
and monitoring summaries, is approximately \unit[20]{kB/s} and
\unit[10]{kB/s} for the near and far detectors, respectively.

All data output by the TP are transmitted via TCP/IP to the Data
Collection Process which collects and merges the output streams
from all the TPs.  The output stream is ordered and formatted as a
ROOT tree~\cite{Brun:1997pa} before being written to disk.  The
active output file is shared with a Data Distribution System (DDS)
which serves the data to various online consumers such as online
monitoring and event displays.  A data archival task transfers
completed data files to the Fermilab mass storage facility by
Kerberized FTP over the internet.

Overall control of the DAQ is provided by the Run Control system
through a well defined state model implemented by all DAQ components.
The system employs a client-server model with a single server
controlling the DAQ.  Multiple graphical user interface (GUI) 
clients are used to connect to the
server over TCP/IP allowing remote operation.  A system of control
exchange avoids contention between clients.  The GUI client is based
on the ROOT framework.  This displays the current state and
performance of the DAQ as well as allowing the operator to modify the
system configuration and control data taking.  Automated run sequences
are used to facilitate unmanned operation.

\subsection{Inter-detector timing}
\label{sec:ops-comm}

Each of the near and far detectors incorporates a GPS
receiver~\cite{gps} used as a clock for the front-end electronics and
DAQ computers. The GPS receivers are located
underground and connected by optical fiber to surface antennas.  The
worst-case resolution of each receiver is approximately \unit[200]{ns} 
during normal operation.

The near detector uses the GPS unit to record the time at which Main Injector
protons arrive at the target.  The far detector similarly records the
GPS time of each DAQ trigger.  During offline data analysis,
after correcting for antenna delays, hardware offsets, and the 
\unit[2.449]{ms} time of flight from FNAL to Soudan, these two GPS measurements
identify far detector events consistent with beam neutrino arrival times.
There is a $\unit[64]{ns}$
uncertainty on the time offset between the near and far detectors
due to uncertainties in hardware delays~\cite{MINOSTOF}.  

Because the far detector does not have a hardware trigger but instead uses 
software to find events, a system was developed to use the timing 
information to promptly tag in-spill data.  The beam spill is
time-stamped at the near detector. The time-stamp is transmitted via
internet to Soudan, where this information is served to the
event-building software described in Sec.~\ref{sec:elec-daq}. 
A several second buffer 
of the far detector data makes possible this efficient online
time-stamping.

The prompt-time correlation allows for a relaxation of the usual
event-building conditions, capturing very small energy events
(\unit[$\sim$300]{MeV}).  During routine operation the system
works reliably to capture 99\% of all spills. The dominant sources of
inefficiency are temporary network outages and transient network
latencies between Fermilab and Soudan.

\subsection{Monitoring and control systems}
\label{sec:elec-dcs}

The MINOS Detector Control System (DCS) monitors and controls the
experimental hardware and environment.  There is a separate but similar
DCS system at each detector, each composed of many smaller subsystems.

LeCroy~1440 high voltage mainframes are used to supply the PMTs
(Sec.~\ref{sec:scint-pmts}) with high voltage.  Each PMT has its own
voltage channel.  These mainframes are controlled via a serial link
using EDAS brand
Ethernet to serial devices, with detector operators using a
menu-driven Linux program to manipulate the voltages.  A monitoring
program reads back both the supply and demand voltage for each channel every
\unit[10]{minutes} and sets off alarms if channels are out of tolerance
by more than \unit[10]{V} (1\%).

The currents, voltages and imbalance voltages of the far detector magnet
coils (Sec.~\ref{sec:steel-coil}) are monitored and controlled
via a National Instruments (NI) ``Fieldpoint'' unit, using a Windows XP
machine and NI's ``Measurement Studio'' Visual Basic libraries.  The near
detector magnet is controlled via Fermilab's ACNET control and
monitoring system; 
the DCS merely reads
its ACNET status and passes on the information.  At both detectors, 
thermocouples (type FF-T-20) monitor the temperature of the coils and
the steel, providing an additional software thermal cutoff.

A number of different environmental parameters are monitored by the
NI Fieldpoint system: barometric pressure
(Omega PX2760-20A5V transducers), temperature and humidity (NovaLynx
220-050Y probes), radon levels (Aware RM-80 monitors), and thermocouples
(also used to monitor the coil heat dissipation as discussed in Sec.~\ref{sec:steel-coil}) 
located in all the high-power 
near detector front-end electronics racks and in several key locations within the 
cavern (Sec.~\ref{sec:elec-fe-near}).  At the far detector, the control panel
for the cavern chiller supplies information on its components, as does
ACNET for the near detector cavern pumps.

The status of the electronics racks is  monitored and
controlled by BiRa RPS-8884 Rack Protection Systems (RPSs).  Each rack has an
RPS, which gets inputs from in-rack smoke detectors, temperature and air flow
sensors, fan pack failure status transistors, coolant drip sensors, and
DC low-voltage monitoring.  
If any of these parameters is out of
tolerance, the RPS sets a warning bit, read out via a Linux daemon over
Ethernet  to allow preemptive maintenance.
If any parameter exceeds an upper tolerance level, or the RPS detects a
serious hardware problem like smoke or a water leak, the RPS drops the
AC voltage permit signal, power to the rack is cut and alarm bits are
set on the network.  Additionally, the VME crate power supplies have a
CANbus interface, 
allowing the monitoring of the voltage and currents
supplied to the VME-based electronics.  This interface also allows power
supplies to be shut down by operators or the monitoring software.

All DCS data sources send their information to a MySQL database at each
detector.  Status web pages are generated from these databases and alert
shift-workers when something is out of tolerance.  The contents of these
databases are archived as ROOT files as well as copied to the master
MINOS database for offline use in later data analysis and calibration.


\section{Calibration}
\label{sec:calib}

The MINOS detectors measure hadronic and electromagnetic shower energy
by calorimetry.  
The relative detector energy calibration is critical for neutrino oscillation studies
that rely on comparisons of energy spectra and event characteristics in
the near and far detectors.
A calibration system is used to determine 
the calorimetric energy scale and to ensure that it is the same
in the near and far detectors. Because MINOS is intended to
measure $\Delta m^{2}_{3 2}$ 
to an accuracy of better than 10\%, shower energy scale calibration goals were set at 2\% relative systematic uncertainty
and 5\% absolute uncertainty. 

This section describes the calibration of the responses of the near,
far and calibration detectors.  This calibration corrects for
scintillator light output variations as well as nonuniformities of
light transmission and collection in the fibers, PMTs, and
readout electronics. Section~\ref{sec:calib-li} details how
the optical light-injection system is used to measure the
linearity and time variation of the readout response in all three
detectors. Section~\ref{sec:calib-muons} outlines the use of
cosmic ray muon tracks to measure scintillator strip light output variations
with time and position, specifically to record interstrip and intrastrip
nonuniformities.  In addition,
Sec.~\ref{sec:calib-muons-timing} describes the use of cosmic ray muon
data to measure the fast time response of far detector scintillator
strips, used to distinguish up-going from down-going
cosmic ray muon tracks.  Section~\ref{sec:calib-energy} provides a detailed
description of how the relative and absolute energy scales of the near
and far detectors are determined using cosmic ray muons and
calibration-detector data from charged particle test beams.

The calibration uses both the
optical light-injection system, which measures the behavior of the readout
instrumentation, and cosmic ray muon tracks, which measure the
response of the scintillator.  This calibration is a multi-stage procedure
that converts the raw pulse height $Q_{\rm raw}(s,x,t,d)$ in strip $s$, 
position $x$, time $t$ and detector $d$ into a fully-corrected signal
$Q_{\rm corr}$.  Each stage applies a multiplicative calibration
constant:

\begin{equation} \label{eq:cal}
Q_{\rm corr} = Q_{\rm raw} 
\times D(d,t)
\times L(d,s,Q_{\rm raw})
\times S(d,s,t)
\times A(d,s,x)
\times M(d)
\end{equation}
where
\begin{description}
\item[$D$] is the drift correction to account for PMT,
  electronics, and scintillator response changing with temperature and age.
\item[$L$] is the function that linearizes the response of each
  channel with pulse-height.
\item[$S$] is the strip-to-strip correction that removes differences
  in response, strip-to-strip and channel-to-channel.
\item[$A$] is the attenuation correction, which describes the
  attenuation of light depending on event position along each strip.
\item[$M$] is an overall scale factor that converts corrected pulse
  height into the same absolute energy unit for all detectors.
\end{description}

The data collected with the calibration detector at
CERN~\cite{Adamson:2005cd} are used to determine the absolute
calibration of the detector energy response by comparing the calibrated
shower energy to the known energies of the particle beams.

Two additional calibrations are required. First, a strip-by-strip
timing calibration is performed on the far detector data, primarily to
aid in reconstructing the directionality of cosmic ray muons and
atmospheric neutrino events.  Second, the PMT single-photoelectron
response is determined in order to adjust for PMT thresholds and
cross-talk.

\subsection{Light injection system}
\label{sec:calib-li}

The light-injection system~\cite{Adamson:2002ze,Adamson:2004mh} is
used to map the linearity of the instrumentation, to monitor the
stability of the PMTs and electronics over time, to evaluate the
single-photoelectron gain, and to monitor the integrity of the optical
path and readout system. Nearly identical systems are used for the
near, far and calibration detectors.

The light-injection systems use pulsed UV 
LEDs to illuminate the WLS fibers of
scintillator modules.  The LEDs are housed in rack-mounted ``pulser
boxes,'' each containing a set of 20 or 40 LEDs.  Optical-fiber fan-outs
allow each LED to illuminate multiple fibers through a set of optical
connectors on the back panel of each pulser box.

\begin{figure}[htpb]
  \begin{center}
    \includegraphics[width=0.9\columnwidth]{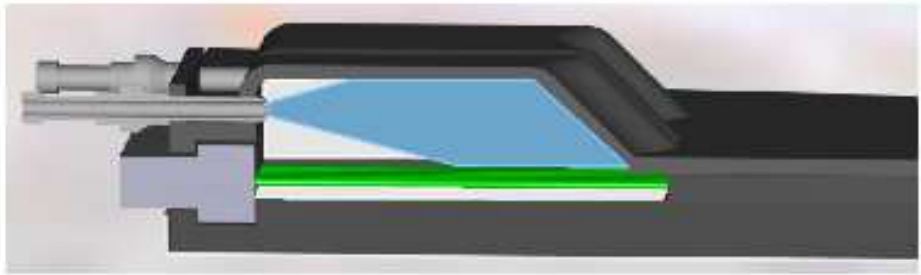}
    \caption{A cutaway view of the
      light-injection module. A curtain of up to ten green WLS fibers runs
      across the bottom of the LIM cavity, and the
      injected light illuminates them from above. The T-shaped component
      at the lower left is the bulk optical connector into which the WLS
      fibers are glued.  In the upper left foreground is a cutaway of
      one of the connectors terminating the light-injection fibers; another
      connector is visible behind it.}
    \label{fig:limconcept}
  \end{center}
\end{figure}

From the pulser boxes, optical fibers carry the light to scintillator module
manifolds on the outer surface
of the detector.  Highly-reflective cavities in the manifolds, called
light-injection modules (LIMs), allow the LED light to
illuminate the WLS fibers, as shown in Fig.~\ref{fig:limconcept}. The light
injection pulses mimic scintillation-light signals from scintillator strips.
The intensity of this injected light is monitored by PIN
photodiodes that are read out simultaneously with the detector PMTs.

\subsubsection{Gains and instrumentation drift}
\label{sec:calib-li-drift}

During data taking, the light injection system periodically pulses
the fiber at every strip end  to monitor the
stability and gain of every channel.  Each far-detector strip end is
pulsed about 300~times per hour and every near-detector strip end
is pulsed 1000~times per hour.  The pulse intensity is tuned such
that a PMT pixel receives approximately 50~photoelectrons per
pulse. Variance in injection fibers, readout fibers, and PMT
efficiencies are such that some pixels receive up to a factor of two
more or less illumination than the average.  LED pulses are monitored
by PIN diodes to correct for drift in LED intensity over time.

These light injection data provide corrections for changes in gains of 
PMTs and electronics as well as for other transient instabilities.
Experience has shown that good environmental control eliminates most
short-term variations in the instrumentation response, but
long-term studies reveal slow drifts in the response, perhaps caused by
seasonal environmental changes and aging effects, equivalent to
changes in detector gain of $\lesssim 4$\% per year.

\begin{figure}[htpb]
  \begin{center}
    \includegraphics[width=0.48\columnwidth]{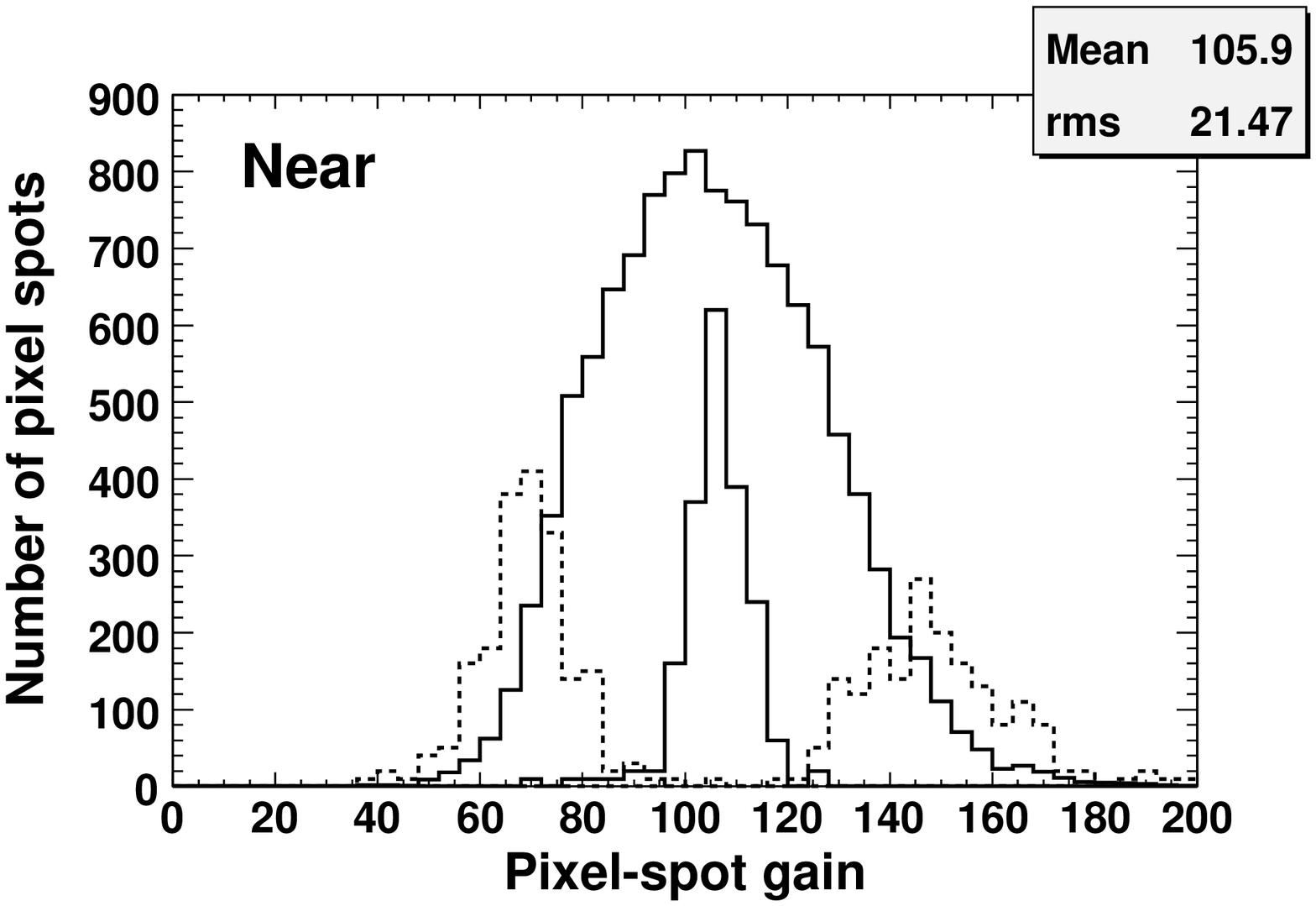}
    \includegraphics[width=0.48\columnwidth]{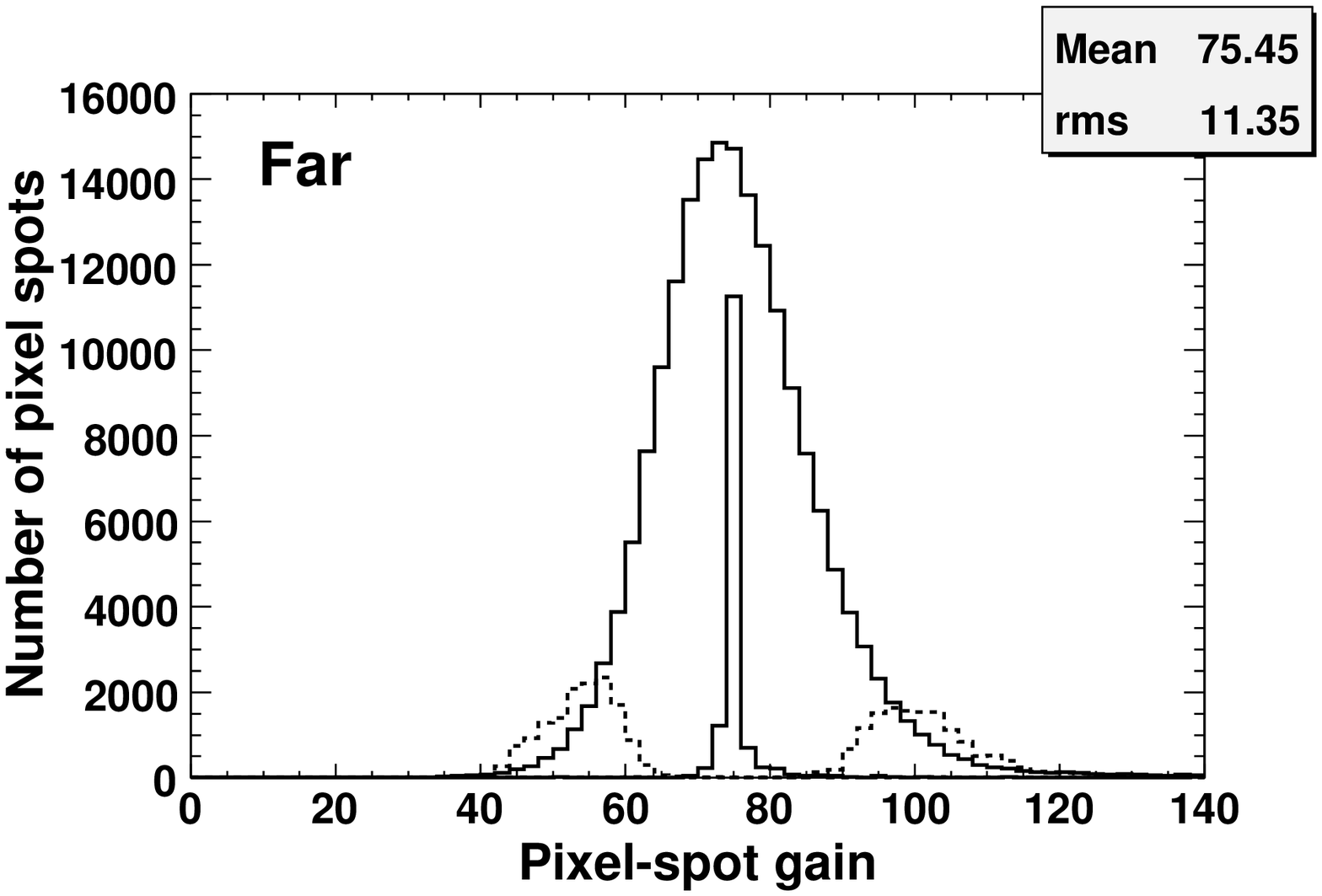}
    \caption{Gains measured by the light injection system
        for the near (left plot) and far (right plot) detectors. 
        The solid line histograms show the gains for all channels (broad
        distribution) and the 
        mean gain per PMT (narrow distribution).  The dashed-line 
        histograms are distributions of the minimum and maximum 
        pixel gains for all PMTs. 
      }
    \label{fig:cal_gains}
  \end{center}
\end{figure}

To track these variations, the data from each month are collated
and used to compute the average response per photoelectron per
channel.  This is done using photon statistics 
~\cite{Adamson:2005cd, Tagg:2004bu}. By
comparing the rms widths of many pulses to the mean, the number of ADC
counts per photoelectron is found for each channel.  Figure~\ref{fig:cal_gains}
shows the results from one such calibration.  The offline software uses
these ``gains'' for Monte Carlo
simulations, cross-talk identification, strip counting efficiency, and other
reconstruction tasks.

\subsubsection{Linearity calibration}
\label{sec:calib-li-linearity}
The light injection system is also used to map the nonlinearity of PMT
response.  The PMTs become $\sim$5--10\% nonlinear at light levels of
approximately 100~photoelectrons~\cite{Tagg:2004bu,Lang:2005xu}. In
addition, the far detector electronics have a nonlinear response of 
similar scale, so it is convenient to linearize both components with a
single correction. To this end, light is injected into the
scintillator modules in steps that cover the range from a few to
hundreds of photoelectrons.

During experiment operation, gain data are collected once a month at both
near and far detectors, interspersed with normal physics data taking.
Each scintillator strip-end fiber is pulsed 1000 times at many different
light levels. The pulse height settings for each LED are tuned so that
the average response of the strips connected to it covers the full
dynamic range of interest. Two PIN diodes, one amplified with high gain and one with low
gain, monitor the stability of the light output of each LED. Low-gain
PIN diodes are better suited for large light levels where saturation of
the PIN readout electronics might occur. At lower light levels, the
high-gain PIN diodes provide a better signal. Extensive test stand
studies have shown the response of the PIN diodes used in MINOS to be
linear to 1\%~\cite{Adamson:2002ze} over a dynamic range roughly
corresponding to 5-100 PMT photoelectrons.

These data are used to parametrize PMT response as a function of
true illumination.  The linearity correction, applied offline,  is
determined by extrapolating PMT response in the linear region to
the nonlinear region.  This is accomplished by mapping  PMT
response as a function of either of two PIN diodes or a combination of
both. For the far detector, both ends of each strip are read out, but
light is injected into only one end at a time.  Light on
the distant end is substantially attenuated and therefore in the
linear region of the response curve, giving an independent
measurement of the true illumination.

Although the PIN diodes themselves are measured to be quite
linear, it is difficult to verify the linearity of the PIN diode
readout electronics. Nonlinearities on the order of 1-2\% are
apparent in the data from both detectors. In the near detector,
noise from the PIN amplifier was exacerbated by the characteristics of
the sampling readout.  In the far detector, problems were suspected
in the analog electronics where cross-talk from PMT signals may have
contaminated the PIN signals. 

\begin{figure}[htpb]
  \begin{center}
    \includegraphics[width=0.48\columnwidth]{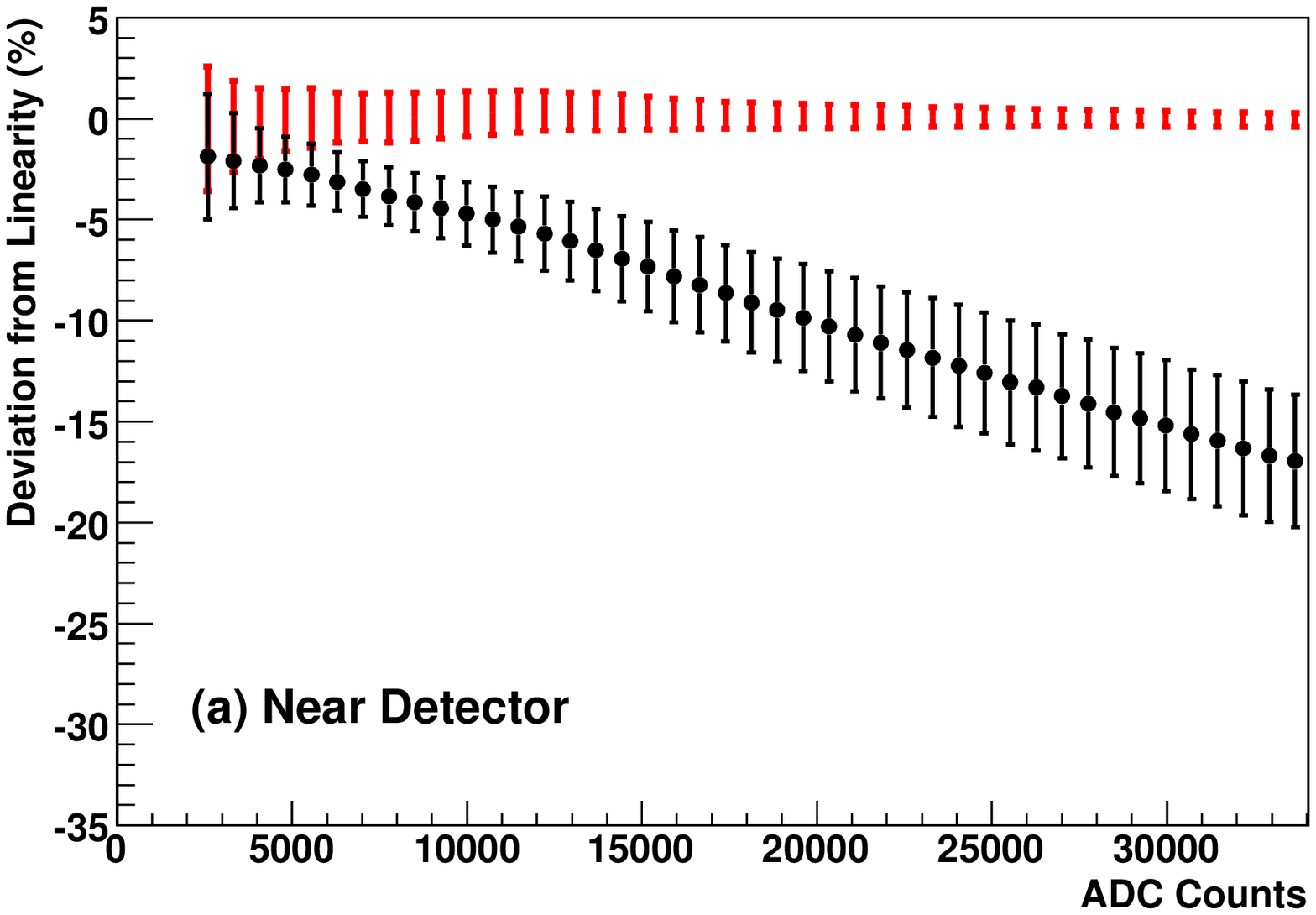}
    \includegraphics[width=0.48\columnwidth]{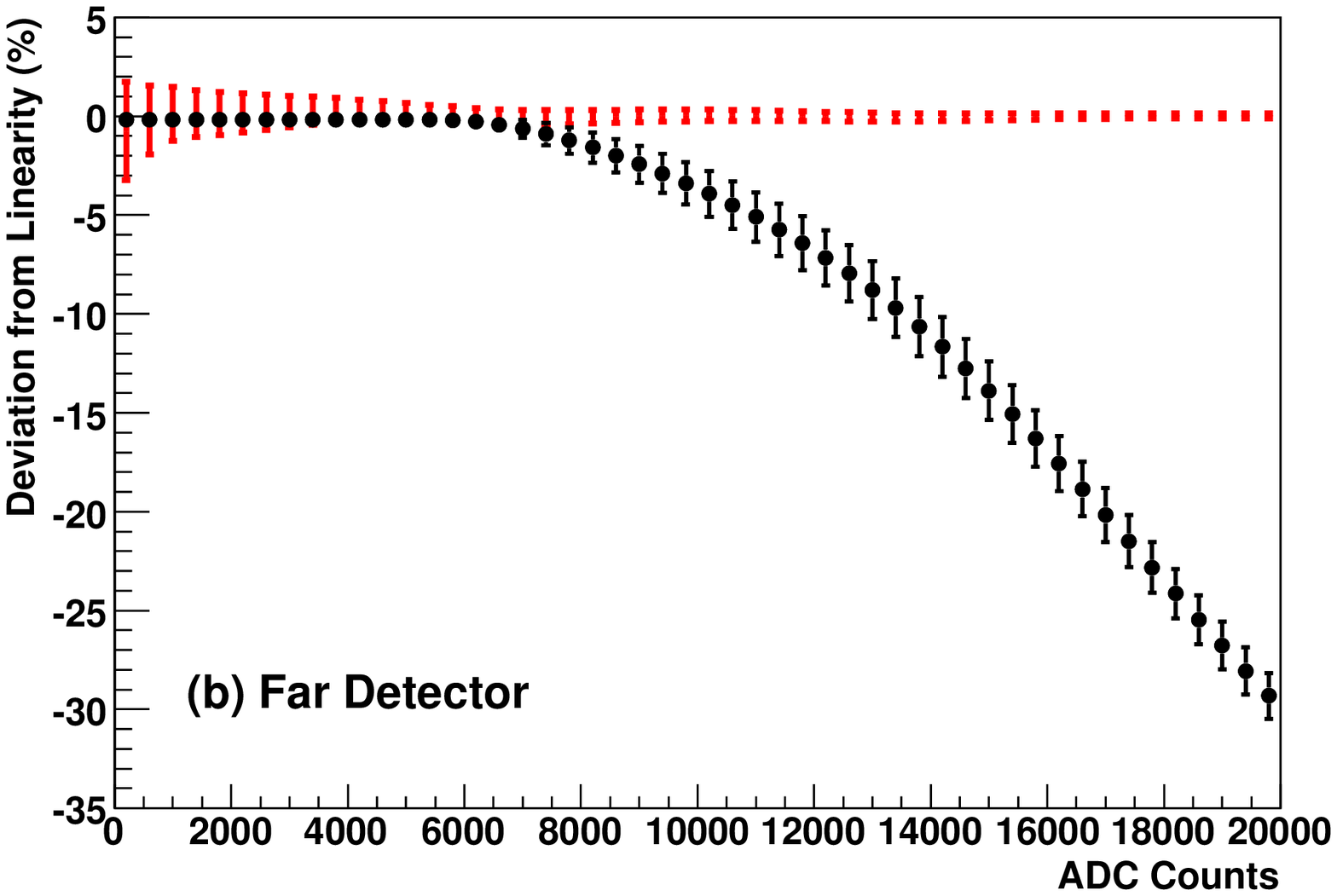}
    \caption{Nonlinearity calibration of near and far detector
      instrumentation. The intrinsic nonlinearity of PMT response 
      (circles) and its residual after calibration (error bars only) 
      are shown for the near (a) and far (b)
      detectors. The error bars depict the rms spread of
      channels in each detector. For scale, a minimum ionizing particle normal to the plane will generate roughly 500-600 ADCs of charge each detector. A single photoelectron is roughly 75 ADCs (Far) or 100 ADCs (Near). }
    \label{fig:cal_nonlin}
  \end{center}
\end{figure}

Figure~\ref{fig:cal_nonlin} shows the results of the nonlinearity calibration.
The far detector nonlinearity is measured with
the distant-end technique, while the near detector nonlinearity is
measured by the PIN diodes.  The near detector PMTs become
nonlinear at low pulse height, while the far detector
readout electronics saturate at high pulse height.  Low pulse-height
data from the near detector (not shown) are unreliable due to PIN readout
difficulties.

The flexibility of the system provides additional information for
detector debugging. High- and low-gain PIN diodes have independent light
paths to the LED and are used to diagnose problems related to the LED.
The near and far strip-end readouts at the far detector provide an
independent way to identify problems with the PMT and front-end
electronics.

\subsection{Cosmic ray muon calibrations}
\label{sec:calib-muons}

Cosmic ray muons are used at each detector to measure
scintillator related quantities.  Throughgoing cosmic ray muons have
an average energy of $\sim$\unit[200]{GeV} at the far detector and a
rate of $\sim$\unit[0.5]{Hz}. At the near detector, the mean energy is
$\sim$\unit[55]{GeV} and the rate is $\sim$\unit[10]{Hz}.

Stopping cosmic ray muons, clean examples of which which constitute 0.3\% of
the total cosmic ray events recorded at the far detector~\cite{Hartnell:2005uq}, 
are used for absolute energy calibration.

\subsubsection{Measurement of time variations}
\label{sec:calib-muons-drift}

Although the light-injection system measures the
time variations of PMT and electronics responses, the scintillator and WLS
fiber cannot be monitored in this way. The changes in 
scintillator and WLS fibers caused by temperature variations
and aging are described in Sec.~\ref{sec:scint-performance-aging}.
These are monitored with cosmic ray muons to correct for small changes
in the detector response over time.

Cosmic ray muons are
used to track the response of each detector on a daily basis.  This ``drift''
calibration is performed by measuring the total
pulse height per plane of through-going cosmic ray muons. Although the
energy deposition of these muons is not the same at each detector and
also depends on zenith angle, the average energy deposited at each
detector site is expected to be constant with time. The daily median
of the pulse height per plane is computed, and the relative change in
this quantity is used to compute the drift, $D(d,t)$:
\begin{equation}
  D(d,t) = \frac{\textrm{Median response} (d,t_0)}{\textrm{Median response} (d,t)}.
\end{equation}
This measured drift encompasses changes due to the scintillator, WLS
fiber, PMTs and electronics.

\begin{figure}[htpb]
  \begin{center}
    \includegraphics[width=\columnwidth]{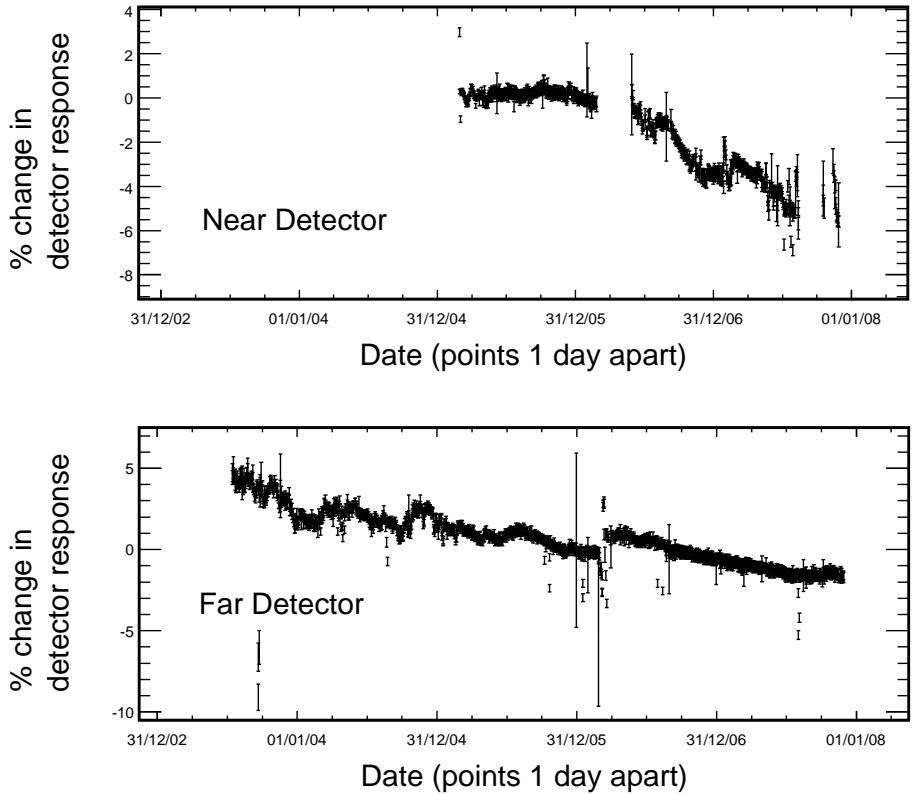}
    \caption{Drifts in the responses of the near (top) and far (bottom)
      detectors.  Each point
      is the percentage drift of one day's data from an arbitrary zero point
      of the daily median pulse height per plane of through-going
      cosmic ray muons. The errors bars show the rms spreads of the daily pulse
      heights per plane.
      The gap in the data corresponds to the summer 2006 shutdown.}
    \label{fig:cal_muon_drift}
  \end{center}
\end{figure}

Figure~\ref{fig:cal_muon_drift} shows the results of detector drift measurements
over several years of operation.
The data show that the far detector drifts at a level of $\sim 2\%$ per year.
The near detector appears more stable initially, later showing a downward 
drift.
While the long-term decrease in detector response can be attributed to aging 
of the detector components, the short-term response changes are well 
correlated with temperature variations.  During the summer 2006 shutdown the 
high voltages supplied to the far detector PMTs were retuned, causing a rise in 
the detector response.

\subsubsection{Strip-to-strip nonuniformity calibration}
\label{sec:calib-muons-strip}

Throughgoing cosmic ray muon data are used to measure the strip-to-strip
(channel-by-channel) time dependent response of the detector,
$S(s,d,t)$. This calibration relates the mean response of each strip end
to the detector average:
\begin{equation}
  S(s,d,t)=\frac{\textrm{Mean Response of Detector}(d,t)}{\textrm{Mean 
  Response of the Strip End}(s,d,t)}
\end{equation}
The normalization incorporates several detector effects that vary
channel-by-channel, including scintillator light yield, WLS collection
efficiency, readout fiber attenuation, PMT quantum efficiency and PMT gain.
Although some of these effects can be measured and corrected for
independently, it is convenient to include them in a single correction.

Cosmic ray muon tracks are used to measure the post-linearization 
mean light level at each strip end. To remove known spatial
and angular dependencies, attenuation and path-length  corrections are
applied to each hit such that the calibration constant is calculated to
be the mean response of a muon of normal incidence traveling through the
center of the strip. A statistical
approach is used to estimate the frequency that a track may clip the
corner of a scintillating strip~\cite{Smith:2002tk}.
Due to the low light level at the PMT face, $\sim$\unit[2--10]{photoelectrons},
the mean-light level calculation must account for the Poisson
probability of producing zero photoelectrons. The spatial resolution of
the detectors is not
sufficient to predict missed strips from event topology, so an
iterative technique is used to estimate the most probable light yield of
a strip. This is used to calculate the zero contribution 
probability~\cite{Symes:2005ej}.
    
\begin{figure}[htpb]
  \begin{center}
    \includegraphics[width=0.9\columnwidth]{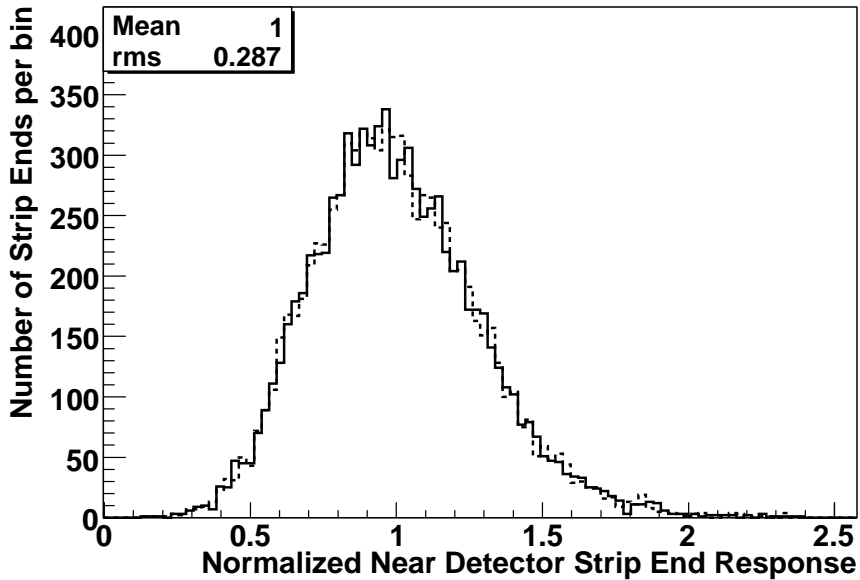}
    \caption{Mean value of the strip-end responses normalized to
      the detector average. The mean response of the strip ends varies by
      approximately 29\%. The solid and dashed lines are the calculated
      responses of two separate data sets from June~2005. The
      statistical variation in the individual calibration constants from 
      these two data sets is on the order of 2.1\%. }
    \label{fig:cal_muon_strip_top_strip}
  \end{center}
\end{figure}

The stability of this technique has been verified through the comparison
of two separate data sets spanning the same time period. Specifically a 
one month period in the near detector (two weeks equivalent per data set 
or $\sim$2400 hits per strip end) was compared to a four-month period in 
the far detector (two months equivalent per data set or $\sim$760 hits 
per strip-end).
Figure~\ref{fig:cal_muon_strip_top_strip} shows the relative response
of each strip end ($1/S(s,d,t)$) in the near detector. The mean responses
of the strip ends vary by approximately 29\% from the detector average.
The statistical variation in the calibration values determined with each
data set is on the order of
2.1\%.
The mean responses of individual strip ends in the far detector also vary by
30\%, with the calibration values stable to within 4.8\%.
     
\subsubsection{Wavelength shifting fiber attenuation correction}
\label{sec:calib-muons-attenuation}
Cosmic ray muon data could be used to correct the variation in light
caused by attenuation along the WLS fiber in a scintillator
strip. However, it is more accurate to calculate calibration constants
from the module mapper measurements described in Sec.
~\ref{sec:scint-performance-mappers}. These data
are fit to a double exponential:

\begin{equation} \label{8}
A(x) = A_{1}e^{-x/L_{1}}+A_{2}e^{-x/L_{2}}
\end{equation}

\noindent 
where $x$ is the length along the strip and $L_{1}$, $L_{2}$ stand for
two attenuation lengths. A fit is performed for each strip and the
resulting parameters are used to correct the data.

\begin{figure}[htpb]
  \begin{center}
    \includegraphics[width=0.9\columnwidth]{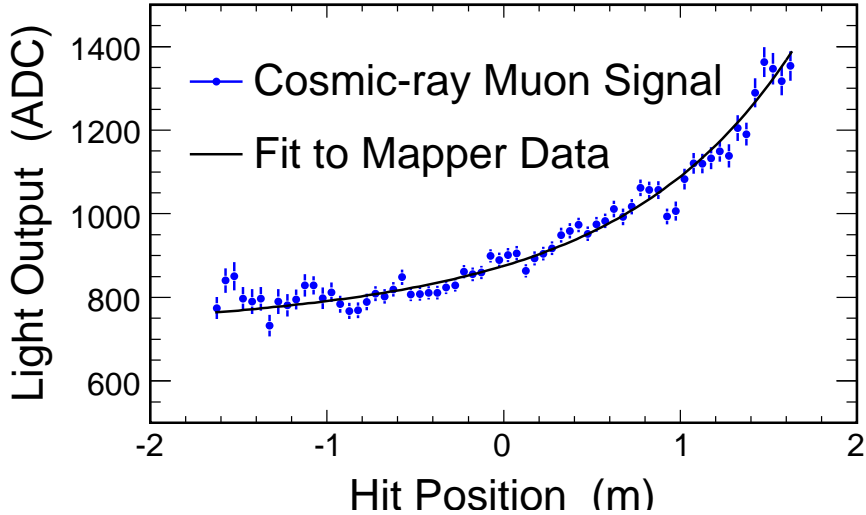}
  \end{center}
  \caption{Comparison of cosmic ray muon data (points) with module
   mapper fitting results (solid curve) for a typical strip in the
   near detector.  
}
  \label{fig:atten1}
\end{figure}

The attenuation constants were subsequently checked using
through-going cosmic ray muons. The pulse height from a strip hit by
a track is plotted as a function of the longitudinal track position
(as measured by the orthogonal plane view).  Figure~\ref{fig:atten1}
shows an example of a double exponential fit from module 
mapper data (Sec.~\ref{sec:scint-performance-mappers}), 
compared with cosmic ray muon data curve for one of the strips.

These studies show that in the near detector, the
difference between the cosmic ray muon data and the fit curve of the
mapper data is about 4\%. Thus the muon measurements are
consistent with the test-stand data. However, since fits to the mapper measurements
were sensitive to fine granularity variations, they are the primary
source of data for the far detector attenuation correction, while the
high cosmic ray statistics available at the shallow near detector allow
the use of muon data for the 
attenuation fits.

\subsubsection{Timing calibration}
\label{sec:calib-muons-timing}

A timing calibration has been performed for the far
detector~\cite{Blake:2005nr}.  Although the readout electronics are
found to be stable to within \unit[1]{ns}, the detector readout as a whole is
synchronized only to within \unit[30]{ns} due to differing cable lengths and
channel-to-channel electronics variations.  The timing system is
calibrated by measuring the size of the time offsets between readout
channels with through-going cosmic ray muons. The calibration also
tracks shifts in these offsets over time, which are largely due to
hardware replacements in the detector readout.
 
The times and
positions of the reconstructed hits on each track are compared and a
linear timing fit is applied, assuming that the muons travel at the
speed of light.  The mean offset between the measured and fitted times
is calculated for each strip end. These offsets are then tuned using
an iterative procedure to obtain the final calibration constants. The
measured times must be corrected to account for shifts in the timing
system caused by changes in readout components. Since the far detector
readout is double-ended, the size of these shifts can be calculated
from the corresponding shifts in the relative times of muon hits
recorded at opposite strip ends.
 
\begin{figure}[htpb]
  \begin{center}
    \includegraphics[width=0.9\columnwidth]{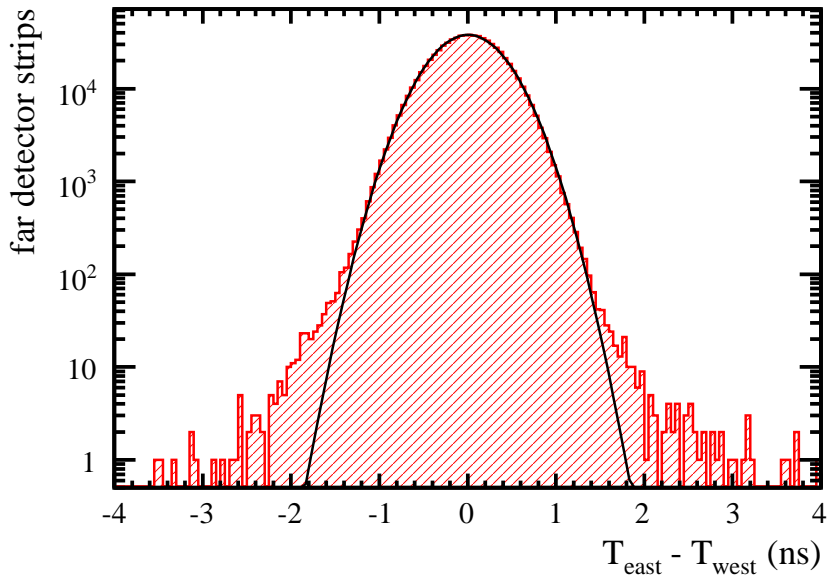}
    \caption{Distribution of mean time differences between muon hits 
      recorded at opposite ends of each of the 92,928 strips in the
      far detector. 
      The data are represented by the hatched histogram. A Gaussian 
      fit with a width of $\sigma=$\unit[0.4]{ns} is also shown.
      }
    \label{fig:TimeCal}
  \end{center}
\end{figure}

This calibration is performed independently for each side of the far detector.
The accuracy of the calibration is then calculated from the relative times
of muon hits on opposite sides of the detector. Figure~\ref{fig:TimeCal}
shows the distribution of the mean time difference between opposite
strip ends for each strip in the detector. The data have been divided
into eight sub-samples to account for any long-term drift in the timing
system. A Gaussian fit to this distribution gives an rms of \unit[0.40]{ns}.
The mean timing calibration error for a single strip end is therefore
estimated to be: $\sigma = \unit[0.40]{ns}/\sqrt{2}=\unit[0.28]{ns}$.

\subsection{Energy calibration}
\label{sec:calib-energy}

\subsubsection{Absolute calibration}
\label{sec:calib-muons-absolute}

\begin{figure}[htpb]
  \begin{center}
    \includegraphics[width=0.9\columnwidth]{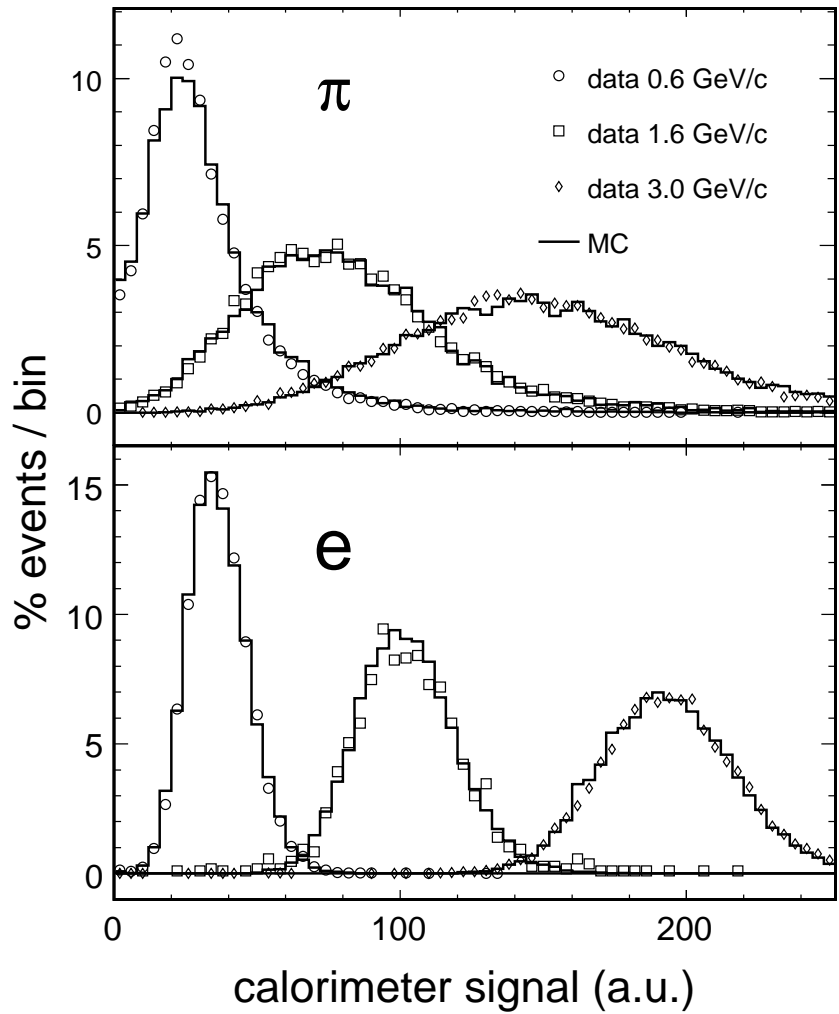}
    \caption{MINOS calorimetric response to pions and electrons at three
     momenta. The calorimeter-signal scale is in arbitrary units. The data
     (open symbols), obtained from the calibration detector exposure to
     CERN test beams, are compared to distributions from Monte Carlo
     simulations.
     }
    \label{fig:caldet_ls}
  \end{center}
\end{figure}

MINOS physics analyses require a good understanding of the detectors' response
to muons, electrons and hadrons with energies below \unit[10]{GeV}. 
The detector response includes event shape characteristics in addition to the
overall energy scale and resolution. 
The calibration detector~\cite{Adamson:2005cd} was exposed to
test beams at CERN to establish the response
to hadrons, electrons and muons with momenta in the
range \unit[0.2--10]{GeV/$c$}. The measurements were used to normalize
Monte Carlo simulations and to establish the uncertainty on the 
hadronic and electromagnetic energy scales. 

The calibration detector data were corrected to remove systematic 
effects at a level of 2\% or better.  
Data were collected at fixed beam momentum settings of both positive and
negative polarities, starting at \unit[200]{MeV/$c$} and proceeding in
\unit[200]{MeV/$c$} steps up to \unit[3.6]{GeV/$c$} and \unit[1]{GeV/$c$}
steps from \unit[4]{GeV/$c$} to \unit[10]{GeV/$c$}. Time-of-flight and
threshold \v{C}erenkov detectors were used to identify electrons, pions
or muons, and protons. Muon/pion separation was accomplished by
considering the event topology. Data were collected in two distinct
beamlines, and several momentum settings were repeated multiple times to
establish a run-to-run stability of better than 1\%.  
Details are given in Ref.~\cite{Adamson:2005cd}.

The data from the calibration detector were compared with events simulated
using the same GEANT3~\cite{Brun93aa} based Monte Carlo used for the near
and far detectors.
Beam optics were modeled with the TURTLE~\cite{Carey:1982hm,psiturtle} code. 
Upstream energy loss and
particle decays were modeled with a standalone GEANT3 program. The range of
stopping muons ($p<$\unit[2.2]{GeV/$c$}) was modeled to better than 3\%,
thereby benchmarking the combined accuracy of the muon energy loss treatment,
beam simulation and absolute scale of the beam momentum. 
Figure~\ref{fig:caldet_ls} shows the measured detector response to pions and
electrons compared with the simulation result. The simulated detector response
to electrons agreed with the data to better than 2\%~\cite{Vahle:2004mp}. Pion
and proton induced showers were compared with events simulated using the
GHEISHA, GEANT-FLUKA~\cite{Ferrari:2005zk} and GCALOR~\cite{Zeitnitz:1994bs}
shower codes. The GCALOR-based simulation was
in best agreement with the data and was adopted as the default shower code.
The Monte Carlo reproduces the response to pion and proton induced showers to
better than 6\% at all momentum settings~\cite{Kordosky:2004mn}. The energy
resolution was adequately reproduced by the simulation and may be
parametrized as $56\%/\sqrt{E}\oplus2\%$ for hadron showers and
$21.4\%/\sqrt{E}\oplus 4\%/E$ for electrons, where $E$ is expressed in GeV.

\subsubsection{Inter-detector calibration}
\label{sec:calib-muons-stopping}

The 6\% accuracy of the calibration detector hadronic energy measurement 
approaches the experiment's design goal of establishing the absolute energy 
scale to within 5\%.
The purpose of this section is to describe
the transfer of this absolute energy scale
calibration to the near and far detectors.

The calibration procedures outlined in Sections~\ref{sec:calib-li} and
\ref{sec:calib-muons} are designed to produce temporally and spatially
uniform detector responses. A relative calibration is then necessary to
normalize the energy scales at the near, far and calibration detectors
to be the same to within the 2\% goal. Stopping muons are used
for this task because they are abundant enough at all detectors and
their energy depositions in each plane can be accurately determined from
range measurements.

\begin{figure}[htpb]
  \begin{center}
    \includegraphics[width=0.9\columnwidth]{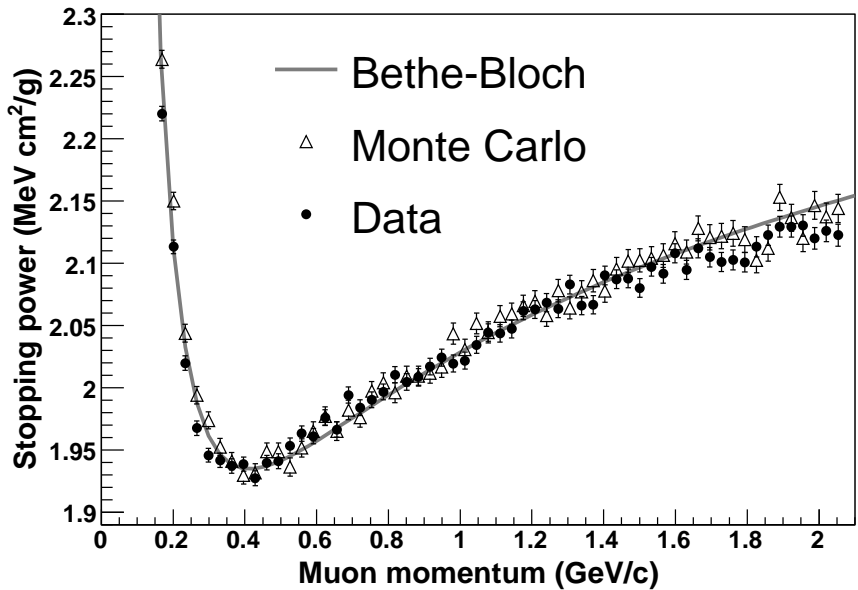}
    \caption{Stopping power for muons. The gray line shows the 
      Bethe-Bloch calculation of the stopping power for muons in
      polystyrene scintillator. The solid circles and open triangles
      show the response of stopping muons in the far detector data and
      GEANT3 Monte Carlo simulations respectively. Both data and Monte
      Carlo points have been normalized to the Bethe-Bloch
      calculation to give the expected stopping power at the minimum
      ionizing point.}
    \label{fig:betheBloch}
  \end{center}
\end{figure}

The energy loss of a muon by ionization is described by the Bethe-Bloch
equation, which determines changes with muon momentum as shown in
Fig.~\ref{fig:betheBloch}. To measure the responses of the detectors it
is necessary to know the momentum of the muons used in the calibration.
The momentum-range relations provide the simplest and most accurate
technique to determine the momentum of stopping muons, with momentum
from muon curvature in the detectors' magnetic fields serving as a
secondary measurement (see Sec.~\ref{sec:steel-coil-perform-calib}).

To improve the accuracy of the inter-detector calibration,
a ``track window technique''~\cite{Hartnell:2005uq} was developed.
The technique hinges upon the particular way in which energy loss
varies with momentum for muons.  The dE/dx of a
\unit[1.5]{GeV/c} muon increases by approximately 100\% in the last 10\%
of its range, whereas in the other 90\% of its track the dE/dx changes
by about 8\%. The track-window technique measures the response of muons
only when their momenta are between 0.5 and \unit[1.1]{GeV}. This avoids
using data from the end of the track where the rapid increase in
ionization occurs, as shown in Fig.~\ref{fig:betheBloch}. Since the
dE/dx varies so slowly in the \unit[0.5--1.1]{GeV} region, a 2\% error
on knowing where the muon stopped gives an error of approximately 0.2\%
in the energy deposition.

The calorimetric responses of the three detectors, 
quantified as $1/M(d)$ in Eq.~\ref{eq:cal} in Sec.~\ref{sec:calib} above, 
are determined by the track-window technique and used to normalize the
detectors' energy scales to within the 2\% target, thus performing the
inter-detector calibration.

The transfer of the absolute hadronic energy-scale to the near and far
detectors is dominated by the 6\% accuracy of the calibration 
detector measurement.  There are smaller, detector-specific
uncertainties such as the scintillator response uncertainties of 
0.9\% for the far detector, 1.7\% for the near
detector and 1.4\% for the calibration
detector~\cite{Hartnell:2005uq} (already included in the 6\% energy
scale accuracy) and contributions from steel thickness variations
at the $\sim$1\% level.  After adding all these
uncertainty components in quadrature, the uncertainty in the absolute 
energy scale of each detector is determined to be approximately 6\%.

\subsubsection{Comparison of near and far detector electronics response}
\label{sec:calib-near-far-elec}

The above procedure normalizes the responses of the detectors for the
energy deposition of a minimum ionizing particle. The test beam study
of calibration detector response was used to verify
that there are no energy-dependent differences introduced by the use of
different types of multi-anode PMTs and electronics in the near and far
detectors~\cite{Adamson:2008CaldetNF}.

\begin{figure}[htpb]
  \begin{center}
   \includegraphics[width=0.9\columnwidth]{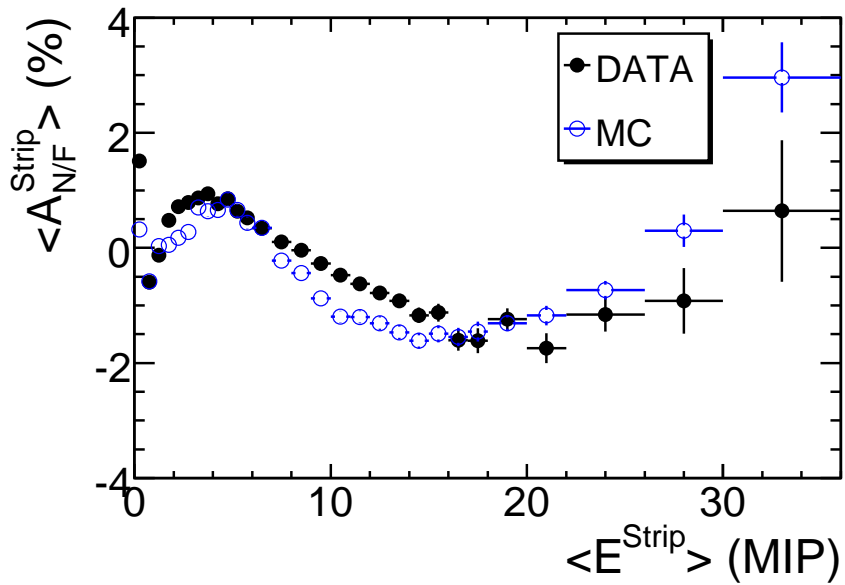}
   \caption{Near-to-far asymmetry, 
     Eq.~\ref{eq:fdnd-asym}, in the relative energy response as a
     function of the deposited energy, before linearity correction.}
    \label{fig:nfc}
  \end{center}
\end{figure}

A series of special runs was taken with the calibration detector
in which one end of each strip was read out with  near detector PMTs
and electronics while the other end was read out with far
detector PMTs and electronics~\cite{Cabrera:2005up}.  Figure~\ref{fig:nfc} shows the
relative response asymmetry in a strip
\begin{equation}\label{eq:fdnd-asym}
 A_{N/F} = {{N-F}\over{0.5 (N + F)}},
\end{equation}
measured with positrons between 0.5 and \unit[6]{GeV}. This study shows
that $A_{N/F}$ is close to 0 at an average energy deposition of around
\unit[1]{MIP}.  Energy-dependent differences of up to $2\%$ in the
relative calibration of the readout systems are introduced by the
different nonlinearities and thresholds of the two systems. However,
these are well modeled by the MC simulation that provide energy
corrections for physics analyses.  Subsequent data were taken with the
far detector electronics removed from that side of the fibers and
replaced with reflector connectors, to evaluate the effect of the
single-ended readout of mirrored fibers in the near detector
(Fig.\ref{fig:ndstrip-mirror}).  These data were in agreement with the
FD electronics data after the resulting calibration factors were
applied.

\section{Detector Laboratory Facilities and Installation}
\label{sec:install}

This section describes the underground facilities constructed to house
the near and far detectors (Sec.~\ref{sec:ops-lab-nd}
and~\ref{sec:ops-lab-fd}), and the installation procedures used to
assemble the detectors (Sec.~\ref{sec:install-nd}
and~\ref{sec:install-fd}).  The far detector is located underground 
in order to reduce counting rates due to low energy cosmic rays.
The underground location of the near detector
resulted from the 3.3$^\circ$ dip angle of the NuMI beamline. Detector
installation was somewhat challenging because components had to be moved
into the underground halls through restricted access shafts and
assembled in relatively tight quarters.
Sections~\ref{sec:install-fd-coordinates} and \ref{sec:install-align}
describe the survey procedures
used to align the detector planes and to determine the precise direction
in which the neutrino beam should be aimed.


\subsection{Near detector laboratory}
\label{sec:ops-lab-nd}

 
The near detector hall is located at the Fermi National Accelerator Laboratory,
\unit[105]{m} below the surface.  
The near detector is \unit[1040]{m} from the primary
target and \unit[330]{m} from the hadron absorber at the end of the
decay pipe.   
The
neutrino beam exits the rock \unit[40]{m} upstream of the first detector
plane.  Underground construction of the near detector hall, including
the detector support structure, was completed in March 2004.

The MINOS Service Building (MSB) 
is the primary entry point to the
underground complex for material, personnel, and services.  A
\unit[6.7]{m} diameter vertical shaft connecting the MSB to the
underground complex is divided into two sections, one containing two
elevators for personnel and the other housing conduits for electrical
power, computer networks, ventilation ducts, pipes for the removal of
ground water, and a hole large enough to lower complete MINOS near
detector planes into the experimental hall.  A \unit[13.6]{ton} capacity
overhead crane with a \unit[5.6]{m} hook height is used to move
equipment up and down this section of the shaft.  Parts of the MSB roof
can be removed to allow over-sized objects, such as the near detector
magnet coil components, to be maneuvered into and out of the shaft.

Two transformers provide electrical service to the downstream portions
of the complex, the detector and absorber halls.
In addition to \unit[1500]{kVA} of general-purpose
electrical power, \unit[750]{kVA} of ``quiet'' power is provided for
noise-sensitive uses, such as the near detector front-end electronics.

The underground complex is ventilated with air forced into it at the
base of the shaft and expelled at the hadron absorber hall 
and the downstream end of the detector hall.  An air temperature of
\unit[24$\pm1^\circ$]{C} 
and dew point of \unit[$\leq$13$^\circ$]{C} are maintained in the near detector hall at all times.

Ground water is collected from the entire underground tunnel system and
pumped to the surface through the MSB shaft. The average collection
rate, which has decreased slowly since construction was completed, was
\unit[650]{liters} per minute in 2006.  Several backup pumps with independent
electrical power are used to achieve a high level of reliability.
Before being discharged to surface ponds, a portion of the cool ground
water is directed into primary circulation loops for air conditioning,
beam absorber cooling, and heat exchanging with a secondary low
conductivity water loop (LCW) for near detector electronics and magnet power
supply systems.

The LCW loop removes most of the heat generated in the magnet power 
supply (\unit[4]{kW}), the magnet coil (\unit[47]{kW}), and
the front-end electronics racks (\unit[60]{kW}).  A Proportional
Integral Derivative (PID) controller sets the fraction of primary cooling
water bypassing a heat exchanger in order to maintain the LCW output
temperature at \unit[21.1$^\circ$]{C}.  Temperatures measured in the
electronics racks are stable to better than \unit[0.1$^\circ$]{C} in
steady-state operation.


\subsection{Near detector installation}
\label{sec:install-nd}



The access shaft to the underground detector hall at Fermilab is large enough
to accommodate the near detector planes and their transport fixture,
eliminating the need to construct planes during the underground detector
assembly process.  The near detector plane assembly process consisted of
attaching scintillation modules and alignment fixtures to the
steel planes.   This was performed in
the New Muon Lab at Fermilab while the underground detector hall
was still under construction.  The steel planes were inspected for
flatness and sorted into stacks. A plane thickness profile was measured
and recorded using an ultrasound system for the planes in the target and
calorimeter sections, where detailed knowledge of the mass of the
detector was desired (Sec.~\ref{sec:steel-planes-nd}).
Scintillation module maps 
(Sec.~\ref{sec:scint-performance-mappers}) were used to
select the best modules for use on the target and calorimeter section
planes. 

Planes were assembled flat on the floor of the New Muon Lab, supported 
on the transport fixture.   After assembly they were lifted and stored 
vertically on ``hanging file'' support structures until needed for final
construction of the near detector in the underground hall.  The scintillator
modules were positioned on each plane using a template with reference to
the magnet-coil hole and were attached to the
plane by spot welded steel clips. 
The modules were checked for light tightness immediately
prior to attachment. Small gaps between scintillation modules caused by
uneven module edges were measured and recorded.  Alignment fixtures for
use during installation were tack-welded to each plane. The assembled
planes were hung for storage in the order they would be retrieved for
underground installation.

The near detector electronics racks are positioned to the west
of the detector on two levels (Fig.~\ref{fig:MINOSnear}) to accommodate U and V
readout (Fig.~\ref{fig:NearInstFig1}).  
Fiber-optic readout cables from the scintillation modules
enter the rear (detector side) of the ``front-end'' electronics racks,
connecting to the phototube boxes.  Inside the racks, cables connect the
phototube bases to the digitizing electronics described in
Sec.~\ref{sec:elec-fe-near}. On the front of the racks, cables carry
digitized signals to the data acquisition system, housed in ``master''
racks (described in Sec.~\ref{sec:elec-fe-near}).

Installation of the detector planes began in April~2004 with the
furthest downstream plane (number~281) and was completed five months
later with the installation of plane~0 (a blank steel plane to cover
plane~1 and necessary to match the far detector's beam-view ordering of
steel-air-scintillator).  Single planes were transported horizontally
from the New Muon Lab to the MINOS Service Building on a flatbed
truck, already attached to the transport fixture to minimize flexing of
the plane. The plane was lowered vertically down the access shaft (still
attached to the transport fixture), then transferred to a similar
fixture located on a large steel frame cart.  The cart was pulled about
\unit[80]{m} from the base of the shaft into the near detector hall
where an overhead crane removed the plane from the cart and placed it
into position on the detector support structure using the transfer
fixture. The plane was then bolted into place at the corners and at the
coil hole collar fixture. A spacing alignment fixture was used to ensure
uniform plane-to-plane distance (Sec.~\ref{sec:steel-planes}). A crew of
physicists immediately inspected and tested each plane for light tightness
and installed optical readout and light-injection cables.  Alignment
fiducials, located on the scintillation modules and on the steel plane
and coil collar, were measured and recorded using the
Vulcan system~\cite{Bocean:2004dj} (Sec.~\ref{sec:install-align}).

\subsection{Far detector laboratory}
\label{sec:ops-lab-fd}

The far detector laboratory is located \unit[710]{m}
(\unit[2070]{meters-water-equivalent}) below the surface, \unit[735]{km}
from Fermilab, in Soudan, Minnesota. The facility provides the
electrical power, communications, fresh air and temperature control
required to operate the detector.  Power is supplied by a
\unit[450]{kVA} substation drawing from a \unit[2400]{kVA} station on
the surface.  A fiber optics system provides high-speed DS3
(\unit[45]{Mb/s}) internet service to the underground laboratory.



The Soudan Mine is naturally ventilated, with a flow averaging approximately
\unit[130]{l/s} from the mine tunnels to the main shaft.  About half
of this flow is intercepted, before being exhausted up the shaft, and
directed through the MINOS detector hall and the neighboring Soudan
laboratory.  The Cryogenic Dark Matter Search (CDMS II) experiment
in the Soudan laboratory uses significant amounts of helium.  As such, 
the hall air feeds are in parallel to prevent exposing any MINOS PMTs 
to this gas, which can penetrate the glass envelope and ruin the 
ultrahigh vacuum within.

An active cooling system removes heat from the laboratory generated
by lighting, electronics and the detector magnet.  A 3-stage chilled
water loop provides cooling for both the magnet coil system and the
laboratory as a whole.  It includes both a conventional
compressor-driven air-conditioning system (for the summer) and a simple
fan-coil system (for winter) on the surface and provides a flow of
\unit[13$^\circ$]{C} water to both systems.


\subsection{Far detector installation}
\label{sec:install-fd}

Installation of the MINOS far detector at Soudan began in July 2001 and
continued through September 2003.  A total of 486 planes (484 instrumented) were
installed, tested, and integrated into the detector during this two year
period.

\subsubsection{Logistics}
\label{sec:install-fd-logis}

The logistics of bringing equipment, personnel and materials to build the MINOS
detector to the underground detector hall at Soudan presented a significant
challenge.  It was especially important to mesh the delivery schedules of the
steel and the scintillator modules, as these materials comprise the bulk of the
detector and presented most stringent demands for storage and handling. 
Generally, about 55 planes worth of these materials could be stored in a
facility on the surface at Soudan.  Everything going into the Soudan
Underground Laboratory must be transported down a relatively small shaft dating
to the days of iron mining.  Equipment and personnel are transported
in cages that run on rails down the mine shaft, which is inclined by 12$^\circ$
with respect to the vertical.  The length of objects that can be
moved underground
is limited by the size of the cages and by the dimensions of the shaft stations
on the surface and underground.  The shaft size limits transverse dimensions to
slightly over \unit[2]{m}, while the lift capacity of the hoist system
limits cage loads to no more than \unit[5.5]{metric tons}.

Steel was lowered in bundles of four plates.  Each bundle was
\unit[8]{m} long  
and contained all the pieces for either a top or bottom layer of a double
layer single plane.  Upon reaching the shaft bottom, the steel bundle was 
transferred to a monorail for
transport to storage racks adjacent to the plane assembly area.  The
scintillator modules, still housed in their \unit[9]{m} long shipping crates
(the length limit of the hoist system),
were transported down the shaft in a manner similar to the steel.
Once underground the crates were
transferred to specially-designed carts and moved to a storage area near the
module mapping table.  In total, the detector comprises
\unit[5,400]{metric tons} of steel and 4,040 scintillator modules.

\subsubsection{Scintillator module tests}
\label{sec:install-fd-modtest}

Although the modules were tested and mapped with a radioactive source at the
factory, additional tests were done at Soudan to ensure that they had
incurred no damage in shipping, e.g. light leaks or broken fibers.  The modules
were found to be very robust.  In the initial stages of detector construction,
all modules were tested for light leaks and re-mapped (Sec.~\ref{sec:scint-performance-mappers}) at Soudan.  
As very few
modules showed any significant change in response from their factory mapping,
eventually shipments were checked by re-mapping only a few randomly-selected
modules.  An exception to this shortened testing occurred when there was 
reason to believe that damage may have been incurred. 
Before being installed on a plane, each module was checked for light 
tightness and repaired in the rare event of light leaks being found.

\subsubsection{Steel plane assembly}
\label{sec:install-fd-steel-assy}

Each plane was constructed on a strongback, in two layers of four plates
each as described in Sec.~\ref{sec:steel-planes-construction}.  The
assembly crew cleaned and de-burred each plate to minimize gaps between
steel layers.
During assembly of a plane it was important to ensure proper
alignment of the top and bottom layers, especially on the support ears
and in the central hole through which the coil passes.  Any protrusions
were ground away.  As the steel plane was being assembled, the cable
support brackets were affixed with a stud gun and the spacers and survey
targets were welded into place.

\subsubsection{Magnetic induction cables}
\label{sec:install-fd-induc-cab}

Magnetic induction cables for the calibration
(Sec.~\ref{sec:steel-coil-perform-calib}) were
installed on each steel plane.  Each plane had at least one 50-turn cable
wrapped around the steel through the central hole to be connected to the
calibration system.  During plane construction, only the cables were attached;
the circuit boards were connected after the plane was erected.  The upper
portion of the cable was laid out  prior to the attachment of the scintillator
modules.  The lower side of the plane was inaccessible while on the strongback,
so this side was uncoiled by a worker in a man-lift after the plane was
erected.   Six cables were installed on every twenty-fifth plane, as well as the first
ten and last ten planes of each supermodule.

\subsubsection{Scintillator mounting and cabling}
\label{sec:install-fd-scint-mount}

The alternating U-V alignment of the scintillator planes stipulated that each
of the two workstations build only planes of one orientation, completing planes
in alternation.  Because of
minor differences in the configuration of the inner scintillator modules,
a pre-sorted storage box of modules for the appropriate type of plane 
was brought adjacent to the strongback.  There the modules
were unloaded using a gantry crane with a vacuum lifting fixture.  The
scintillator layer for each plane was built from the center outward, i. e.,
beginning with the bypass modules straddling the coil hole and working toward
the edges.  As each module was positioned its retaining clips were welded into
place.

After all of the modules of a plane were installed, 
the crew checked for gaps between them
and recorded any that were found.  They then connected the optical cables for
scintillator readout (described in Sec.~\ref{sec:scint-connect}) and
the optical fibers that supplied the light injection calibration pulses
(Sec.~\ref{sec:calib-li}). All cables were
checked for correct length, orientation, light tightness and light transmission
before being coiled for the transport of the plane to the detector.

\subsubsection{Plane installation and survey}
\label{sec:install-fd-plane-surv}

After a plane was assembled and tested it was ready to be rigged for lifting
and transport to the detector support rails.  The strongback clamps were
engaged and a spreader bar attached to the strongback.  For safety, the 
assembly
area and the area through which the plane would travel were cleared of all
personnel.  The plane and strongback were lifted with an overhead bridge crane and
transported to the face of the detector.  There the crew positioned it and
balanced the strongback with lead bricks to ensure that both ears sat down on 
the rails at the same time.  Although a set of steel spacers was welded to the 
ears of the plane, it was important to avoid jolting the scintillator face 
into the detector as the plane was eased into position.  Once the plane was
resting on the rails, 6 of the 8 axial rod bolts were installed 
and the plane was shimmed if
necessary.  Then the strongback was released and returned to the assembly
area.  With the strongback out of the way, the crew used a man-lift to
install the remaining axial bolts, unroll and affix the ``bottom-side'' magnetic
induction cable, install the collar, and survey the installation (Sec.~\ref{sec:install-align}).

\subsubsection{Optical connections and testing}
\label{sec:install-fd-optcon}

Cabling for the newly-installed plane included connecting the readout cables to
the MUX boxes (Sec.~\ref{sec:scint-pmts-fd}), routing the light-injection
fibers to their
distribution box, and connecting the circuit board(s) to the magnetic induction
cable(s).  The electronics installation proceeded in phase with the plane
installation. This timely commissioning led to cosmic-ray data being routinely
used for plane diagnostic purposes within a few days of plane installation.  
In no case was a scintillator module
damaged during installation, although occasional light leaks were found and
repaired.  All 92,928 WLS fibers in the completed detector are in
operation, although a very small number have only single-ended readout.


\subsection{Far detector location}
\label{sec:install-fd-coordinates}


The success of the MINOS experiment depends on the accurate aiming of
the neutrino beam at the far detector.  The origin of the detector 
coordinate system is the upstream edge of the far detector.  This lies
\unit[15.631]{m} upstream from the midpoint between the detector's 
two supermodules.  That midpoint was surveyed and determined
to be located at latitude \unit[47.820267]{degrees} north, longitude
\unit[92.241412]{degrees} west, and ellipsoid height \unit[-248.4]{m} in 
North American Datum 1983 (NAD~83).
This survey point is \unit[705]{m} below the surface and
\unit[735.3380]{km} from the upstream end of the first horn of the
neutrino beamline, which is very close to
the (variable) target longitudinal position. The physics of the MINOS
experiment requires that the transverse position of the neutrino
beam center at the far detector be known to within \unit[100]{m}.  
Over a dozen possible
misalignments of neutrino beam elements (e.g., target, focusing horns,
decay pipe) must be contained within this tolerance, including the
transverse position of the far detector relative to the line of sight
from Fermilab.  The alignment goal of \unit[12]{m} was established
for this portion of the total tolerance; most of this tolerance window 
comes from the uncertainty in specifying the angles of the neutrino beam 
leaving Fermilab.

The relative positions of Fermilab and Soudan on the surface are
determined by making simultaneous measurements using the 
GPS satellites. Simultaneous GPS data at both
Fermilab and Soudan were recorded for a total of 26 hours in April~1999.
These data plus data from several Continuously Observed Reference Station 
(CORS) positions were analyzed by Fermilab personnel and independently by the
National Geodetic Survey (NGS)~\cite{Bocean:1999bp,Bocean:2000}.  The
agreement between the NGS result and two methods of analysis at Fermilab
was excellent.  The Fermilab to Soudan vector, averaged over the period
of the observations used, is known to better than \unit[1]{cm}
horizontally and \unit[6]{cm} vertically at 95\% confidence level, well within
requirements.  The differential earth tide effect between Fermilab and
Soudan is approximately the same as this uncertainty.  

The position of the far detector on the 27th level at the bottom of the Soudan Mine,
relative to the surface, was determined using inertial survey by the
University of Calgary in April~1999.  The inertial survey unit used
(Honeywell Laseref III IMU) contains a triad of accelerometers and
optical gyroscopes to measure force and angular velocity.  The
accelerometers are double-integrated to yield position change along each
of the three axes~\cite{Skaloud:2000}.  Independent calibration
measurements with the inertial survey unit and the internal consistency
of several inertial survey runs indicated a precision of \unit[0.7]{m}
per coordinate for the surface to bottom of the mine measurement, only a
small part of the \unit[12]{m} goal.  Interestingly, these measurements agreed to
better than \unit[4]{m} per coordinate with the miners' 1962 (and thought to be
less precise) values for the 27th level relative to the surface.

A partial check of the critical beam angles at Fermilab was made during
commissioning before the target was installed~\cite{Zwaska:2006px}.  The
proton beam was sent \unit[723]{m} past the target position and measured
to be within \unit[0.02]{m} transverse of the specified beam axis at the
downstream end of the decay pipe.  This check depends on optical survey
measurements.  At \unit[735.3379]{km}, this beam line would then be
within \unit[20]{m} transverse of the far detector, confirming that the
beam pointing was within a factor of two of the desired precision of 
\unit[12]{m} and considerably less than the \unit[100]{m} requirement.

The far detector and cavern axes are level and at an azimuth of
333.4452$^\circ$ 
(26.5548$^\circ$ west of true north).
These angles were measured with optical levels (better than
\unit[5]{microradian} accuracy) and a gyro-theodolite
(\unit[15]{microradians} accuracy). The latter is a precision theodolite
combined with an accurate gyrocompass.  The neutrino beam axis at the
far detector is at an azimuth of 333.2870$^\circ$ and is sloped
up at an angle of 3.2765$^\circ$ above the horizontal as the
beam travels approximately NNW.

\subsection{Detector alignment}
\label{sec:install-align}


The large size of the far detector planes and the potential for
mechanical instability made its survey particularly important.  The
alignment of the collars on every plane 
was also essential to allow the insertion of
the magnet coil sleeve, a copper tube which passes through 249 (237)
planes in the first (second) supermodule.  The measurements on each
plane had to be made quickly so as not to delay installation of the next
plane.  Furthermore, the retention of a professional survey crew in the
Soudan underground mine would have been cost-prohibitive.  These
criteria motivated the selection of the Vulcan Spatial Measurement
System~\cite{Bocean:2004dj}, which was capable of millimeter-level precision.
The system allowed an experienced technician to survey an entire plane
within \unit[30]{minutes} and could be operated effectively with only a
few hours training. Under these field conditions, the Vulcan system achieved
a spatial measurement accuracy of about \unit[3]{mm}.

During construction, after every plane was installed but before the
installation of the next plane, Vulcan
measurements were made on its eight axial bolts and eight collar bolts.
In addition, on every fifth plane, similar survey measurements were made
on the eight steel lugs welded to points on the perimeter
of the plane.  These lugs are visible after the erection of subsequent
planes.  These measurements helped to prevent ``creeping'' of the
detector plane positions and also established the average plane-to-plane
pitch of \unit[5.95]{cm} with a standard deviation of \unit[0.35]{cm}.

The scintillator modules were also surveyed, using built in alignment pins
as reference points.  The Vulcan measurements for
scintillator modules over the full supermodule showed that, to within
the resolution of the Vulcan, the modules were aligned with each other
and with the true $U$ and $V$ axes within approximately \unit[1]{mrad}.  This
translates to a deviation of the end of the module from the true axis of
\unit[2--3]{mm}, eliminating the need for software rotation corrections 
during event analysis.

The far detector coil comprises loops of cable
running through a copper sleeve inserted through a series of circular
collars, so any rotation of the collar from plane to plane does not matter.  
The near detector, however,
employs an aluminum coil comprising rectangular ``planks,'' which are
inserted through a series of square collars.  The square shape of the
near detector collars places limits not only on the allowable
plane-to-plane translation of the collar, but on the rotation as well.
While the Vulcan was employed for measurements of scintillator modules
and axial bolts, it did not have sufficient precision for the collar
alignment.  For this task, a team of Fermilab surveyors set up a laser
tracker in the near detector hall.  Final laser tracker measurements
showed the transverse alignment of collars along the entire length of the
detector to be within \unit[3]{mm}, which was sufficient for coil insertion.

The near detector plane spacing was set to the same \unit[5.95]{cm}
pitch as in the far detector, using an alignment fixture that was
attached to each plane during the installation process.  Survey measurements
during plane installation found the average plane-to-plane
pitch to be \unit[5.97]{cm} with a standard deviation of \unit[0.37]{cm}.


Cosmic-ray muons were used to determine the precise alignment of the far
detector scintillator modules along the $U$ and $V$
axes~\cite{align-doc1047:2004}.  This was a one-dimensional software
alignment procedure in which the positions of the $U$- and $V$-modules were
adjusted along the $V$ and $U$ axes, respectively.  At the underground
location of the far detector, the cosmic-ray flux is insufficient to do
a strip-by-strip alignment in a reasonable time.  Instead, the relative
strip positions are taken from the module mapper data
(Sec.~\ref{sec:scint-performance-mappers}) and combined with the aligned
module position to put the strip positions into the software geometry.
A straight line is fitted to a muon track using all the hits except
those in the module to be aligned.  The difference between that module's
hit position and the projected position from the fit is the residual.
The alignment program then adjusts the position and proceeds to the next
module in the track.  After multiple iterations, the residuals are
minimized and those positions are the accepted ones for the modules.
The final far detector alignment fits had residuals of
\unit[0.75--0.85]{mm}.  The near detector modules were aligned using a
similar procedure, which resulted in residuals of approximately
\unit[0.3]{mm} for the modules and \unit[1.0]{mm} for the
strips~\cite{align-doc1101:2005}.

\section{Detector Operation and Performance}
\label{sec:ops}

As the result of several years of routine data-taking, extensive operational
experience has been obtained with the MINOS detector systems.  Representative
observations are presented here.  Section~\ref{sec:ops-rel} summarizes
detector performance and reliability information.  
Section~\ref{sec:ops-quality} describes the systems and procedures used
to ensure data quality and to monitor detector system performance in real time.
Finally, Sec.~\ref{sec:ops-off-line} gives an overview of the offline software
used for detector performance measurements and data analysis.

Since the NuMI beam and near detector are located at Fermilab while
the far detector is \unit[735]{km} north in Soudan, MN,
the coordination of experimental operations is non-trivial.
This challenge has been addressed by making as much of the experiment as
possible controllable remotely over computer networks.  
Physicist shift workers are present \unit[24]{hours/day}, \unit[7]{days/week} 
in the main MINOS control room on the 12th floor of Wilson Hall at Fermilab,
where both near and far detectors are monitored and controlled.
In addition, the NuMI beam is monitored by the MINOS shift workers but
is controlled from the Fermilab accelerator control room.  Weekdays
between the hours of 7:30 and 17:30 US Central Time, four to five full-time 
technicians are present in the MINOS cavern of the Soudan Underground
Laboratory, monitoring and controlling the far detector, and, as needed, 
repairing the detector subsystems.  Both sites have
technical support on call for after-hours intervention when necessary.
Close coordination among the three control rooms has provided
high detector live times for periods when the beam is in operation (see
Sec.~\ref{sec:ops-rel-fd}).

\subsection{Detector reliability and live-time fractions}
\label{sec:ops-rel}

\subsubsection{Near detector}
\label{sec:ops-rel-nd}

The near detector was commissioned in January~2005.  Since
then the detector has been kept in an operational state
except during extended periods when the beam was not in operation.  
The near detector data taking is usually organized into an approximately
\unit[24]{hour} run sequence, consisting of \unit[210]{seconds} of
calibration runs followed by a \unit[24]{hour} physics run.
Excluding periods when the beam has been off, the fraction of
time during which the near detector has been in physics data taking mode
has averaged above 98.5\%. Typically more than 99.95\% of the
detector channels are operational.  The small fraction of the lost beam time
is due to the daily calibration runs and infrequent detector maintenance.

Most downtime with beam on is due to maintenance, usually for the
replacement of failed front-end electronics cards.  A mean number of
\unit[3.5]{cards} (out of 9,360~total) failed per week before a mass
replacement of unreliable on-board fuses was done in the summer of 2007.
After that, the failure rate dropped to less than a board per week.  The
typical intervention to replace a few front-end cards requires
approximately one hour of downtime, including the calibration of the
replacement channels.

\subsubsection{Far detector}
\label{sec:ops-rel-fd}

The far detector installation was completed in July~2003 and the
detector has been recording cosmic ray and atmospheric-neutrino data
since then.  By the time the beam arrived in the spring of 2005,
reliable detector operation had become routine. 
Similar to near detector data taking, a 
\unit[24]{hour} run sequence for physics data and calibrations is also
the operational mode at the far detector.  The overall live-time
fraction in 2005 was 96.7\% and has risen to the high 90\%'s since.  Typically all channels in the detector are
functional, with isolated failures being fixed during beam downtime,
usually within hours to days of the appearance of the problem.

After the neutrino beam turned on in March~2005 the important metric for
evaluating experimental performance was the fraction of 
protons-on-target (``POT'') while the far detector was taking good data.  The MINOS experiment's
sensitivity is driven by the statistics of neutrinos observed at the far
detector, making it crucial to keep the far detector operating as much as
possible.  Since the end of beam commissioning in March 2005,
the detector has run very smoothly, taking physics data for $>$98.7\% of all
delivered NuMI protons for the first year's beam
operations~\cite{Adamson:2007gu} and better than 99\% thereafter.

\subsection{Data quality and real-time monitoring}
\label{sec:ops-quality}

A combination of real-time
monitoring and offline or post-processing monitoring is
performed on a daily basis by physicists on shift to ensure data quality.
The systems developed
for these tasks keep track of similar parameters for both near and far
detectors.

The MINOS Online Monitoring (OM) system is designed to provide real-time
monitoring of data quality in the near and far detectors. It is based on
the system used for the CDF experiment's Run~II operations~\cite{wagner:2001is}
and consists of three main processes.
i)~The {\it Producer} process receives raw data records from the DAQ via the
MINOS Data Dispatcher, which are then processed to fill monitoring
histograms. ii)~The {\it Server} process receives monitoring
histograms from the Producer, handles connections from external GUI
processes, and serves histogram data to these processes on request.
iii)~The {\it GUI} process allows browsing and plotting of any of these
monitoring histograms.

The monitor histograms are grouped into sections, e.g., those relating to
digitized hits from the detector (channel occupancies, ADC
distributions, etc.), singles rates, and distributions relating to
electronics calibration and light injection data. A representative
subset of these histograms is checked once every six hours by
the shift crews at the detector sites and any problems are entered into 
the MINOS electronic logbook via a checklist template. All
monitoring histograms are archived to tape for future reference.

The raw event data are moved to storage at Fermilab and copied over to the
Farm Batch System for offline processing.  From there the reconstruction is
completed with a stable software release.  Offline reconstruction is
performed on data taken the previous day and used for the offline
monitoring and subsequent data quality checks. The data are divided into
separate streams in-time and out-of-time with the beam spills to
facilitate monitoring as well as analysis.

The Offline Monitoring system serves two main purposes. It allows
monitoring of the detector systems using reconstructed event data
quantities such as event rates per POT, demultiplexing and scintillator
strip efficiencies. Additionally the system provides the ability to
verify that the offline
production is proceeding normally so that unexpected changes can be
tracked down quickly.  The Offline Monitoring system has a histogram
making process which runs once per day, reading in all the reconstructed
data from the near and far detectors processed in the previous day and
producing a set of histograms for monitoring.  It also runs the
{\it OMhistory} package, a process for viewing how these histograms
change over time.   

Other tasks performed during shifts include completing a checklist of the 
DCS systems described in Sec.~\ref{sec:elec-dcs}, monitoring 
quasi-real-time event displays for both near and far detectors, and 
monitoring the NuMI beam performance ~\cite{Kopp:2006nq}.  
These checks are performed during each
shift to ensure that problems are noticed promptly and flagged for
repair. 

\subsubsection{Near detector}
\label{sec:ops-quality-nd}

\begin{figure*}[htpb]
  \centering
  \includegraphics[width=\textwidth,keepaspectratio=true,bb=0 0 740 235]{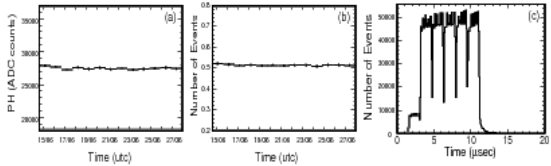} 
  \caption{Distributions examined
    in near detector data quality monitoring include: (a) the average
    spill pulse height, (b) the average number of
    reconstructed events per \unit[$1\times10^{12}$]{POT} in a
    \unit[13]{day} period, and (c) reconstructed event times in the
    $\unit[20]{\mu s}$ gate.)}
  \label{fig:ops-quality-nd}
\end{figure*}

In order to detect anomalies and trends in both the performance and data
quality of the MINOS near detector, several quantities are verified
weekly, including the total uncalibrated digitized response of the detector
activity per POT during the $\unit[\sim10]{\mu s}$ spill as a function
of time (Fig.~\ref{fig:ops-quality-nd}a), the number of reconstructed
events per POT as a function of time (Fig.~\ref{fig:ops-quality-nd}b),
and the reconstructed event time 
(Fig.~\ref{fig:ops-quality-nd}c).  Instabilities in these quantities
may indicate a detector and/or reconstruction problem. The data used for
these quantities come from the in-time spill stream, taking advantage of
the large flux of beam neutrino events at the near site.

\subsubsection{Far detector}
\label{sec:ops-quality-fd}

\begin{figure*}[htpb]
  \centering
  \includegraphics[width=\textwidth,keepaspectratio=true]{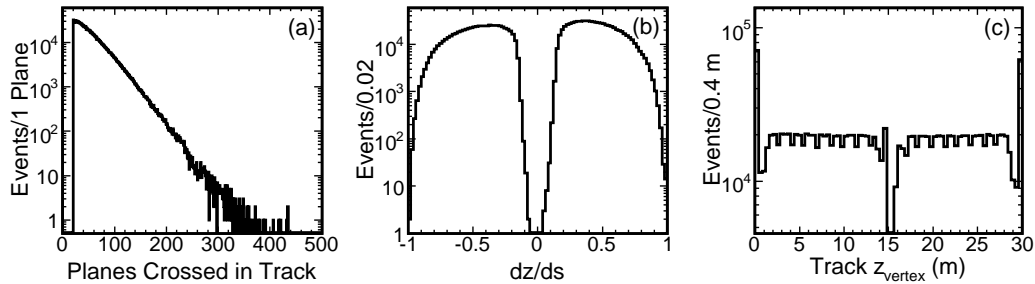} 
  \caption{Distributions examined during
    far detector data quality monitoring include:  (a) the number
    of planes crossed by cosmic ray muon tracks, (b) the incoming
    direction of the cosmic ray muons with respect to the beam direction
    and (c) the entry location for cosmic ray muons along the
    length of the detector.}
  \label{fig:ops-quality-fd}
\end{figure*}

The far detector data are checked weekly for anomalies in
the reconstruction and data quality by comparing distributions of several
reconstructed quantities to a baseline data set.
Examples of distributions monitored are the number of planes crossed 
by muons in the detector (Fig.~\ref{fig:ops-quality-fd}a), the
incoming directions of the tracks and showers 
(Fig.~\ref{fig:ops-quality-fd}b), and track entry
locations (Fig.~\ref{fig:ops-quality-fd}c).
Other quantities that help ensure the detector calibration remains stable 
are the reconstructed velocity for cosmic ray muons 
for timing calibration and pulse heights of tracks and showers for
energy calibration.  Cosmic ray muons are most useful for these
checks as they are the most abundant data source in the far detector.

\subsection{Offline software overview}
\label{sec:ops-off-line}

Although MINOS comprises three detectors (the near, far and calibration
detectors) at different depths and
latitudes and with different sizes, physical configurations, beam
characteristics and electronic readout schemes, the simplicity of the
active detector technology has allowed a single framework
of offline analysis software to be constructed for all detectors.  The object
oriented characteristics of the C++ language~\cite{Stroustrup97} have
enabled the modularity required for this task.  MINOS software is
made available to collaborators using the Concurrent
Versioning System (CVS)\footnote{\url{http://www.nongnu.org/cvs/}}
embedded in the SLAC-Fermilab Software Release Tools (SRT) code
management
system\footnote{\url{http://www.fnal.gov/docs/products/srt/}}.
The system uses software
libraries from the CERN ROOT project~\cite{Brun:1997pa}, including
ROOT tools for I/O, graphical display, analysis, geometric detector
representation, database access and networking.

Raw data from different data acquisition processes at the MINOS
detectors are written to disk as separate ROOT TTree ``streams.''  These
include physics event data, pulser calibration data, beam monitoring
data and detector control data.  This information immediately becomes available
for monitoring, calibration and event display processes 
through an online data ``dispatcher'' service.  This utility can access the
online ROOT files while they are still open for writing by the MINOS DAQ
systems.  Subsequent offline processing produces additional TTree
streams for event reconstruction results and analysis ntuples.

\begin{figure*}[htpb]
  \centering
  \begin{minipage}[b]{0.45\textwidth}
    \includegraphics[width=\textwidth,bb=155 530 375 760]{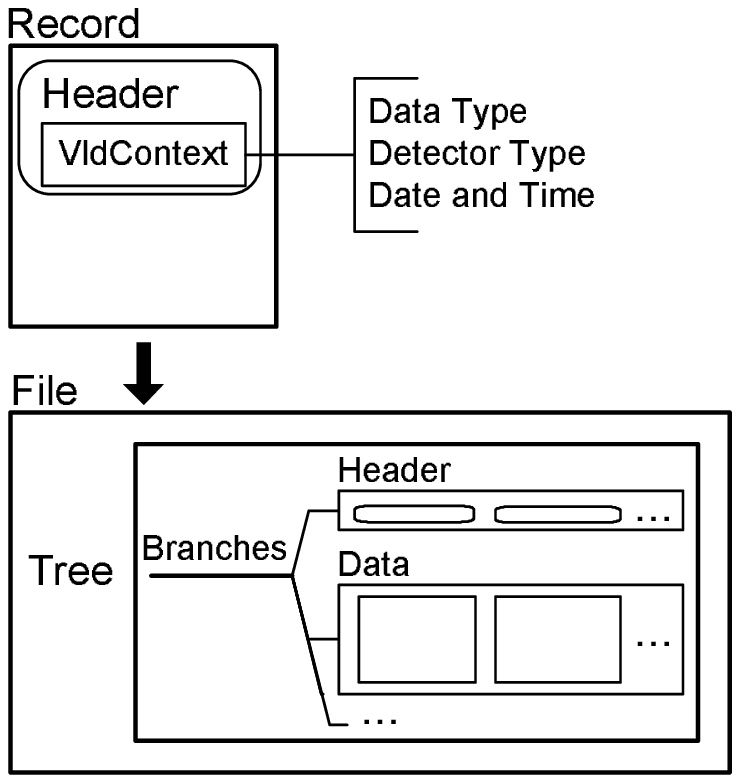}
  \end{minipage}
  \begin{minipage}[b]{0.4\textwidth}
    \includegraphics[width=\textwidth,bb=180 230 385 495]{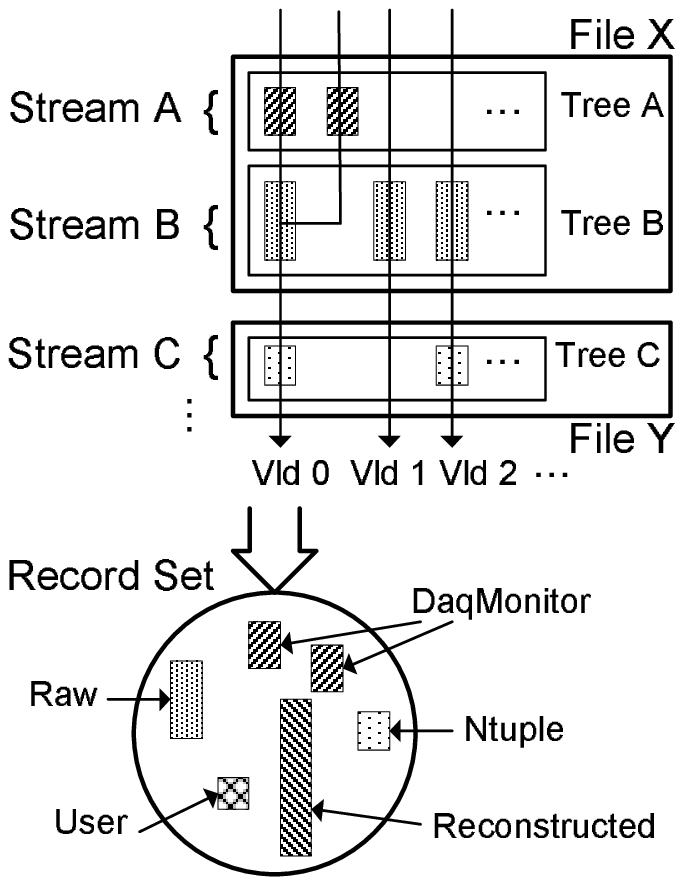}
  \end{minipage}
  \caption{
    MINOS data structure.  The schematic on the left shows a MINOS record type 
    derivation and header structure.  The schematic on the right shows 
    a MINOS stream as an ordered
    sequence of records stored in a ROOT TTree.  On input, records
    stored in different streams are associated with each other by
    VldContext and not by Tree index.}
  \label{fig:validity}
\end{figure*}

The need to correlate these streams of MINOS data with each other has
motivated a key element in the MINOS software strategy called {\it VldContext}
(``Validity Context'').  VldContext is a C++ class that encapsulates
information needed to locate a data record in time and space.  Separate
streams of data from different sources can be synchronized by comparing
their VldContext objects.  When MINOS offline software opens files
containing these streams, it indexes each stream according to the
VldContext of each record.  The indexing information can then be used
to put VldContext-matched
records into computer memory simultaneously.  The GPS timestamps
attached to raw data records enable this matching for far and near
detector data in the same offline job.  These features are illustrated
in Fig.~\ref{fig:validity}.  

All MINOS record types derive from a common
record base class with a header that derives from a common header base
class.  The minimum data content of the record header is the VldContext,
used to associate records on input.  The small record header is stored
on a separate ROOT TTree Branch from the much larger data blocks.  A
MINOS stream is an ordered sequence of records stored in a ROOT TTree
containing objects of a single record type extending over one or more
sequential files.  On input, records stored in different streams are
associated with each other by VldContext and not by Tree index.  The
default mode is that records of a common VldContext form an input record
set.  Alternative input sequencing modes by VldContext are also
supported.

The MINOS offline database contains calibration and survey data,
including component locations and connection maps from the construction
phase of the detector.  These relational tables are keyed with a notion
of ``Validity Range'' or scope of VldContext values to which a 
database record applies.  For physics data of a particular VldContext, 
the offline database interface enables retrieval of matching database
records whose Validity Ranges encompass the VldContext of the physics
data in question.


The offline software accesses the offline database through
a low-level ROOT~\cite{Brun:1997pa} interface, which allows data to be
saved to and retrieved from compliant database products.
The central database warehouse is served by
MySQL\footnote{\url{http://www.mysql.com/}} 
and is about \unit[100]{GB} in size.  Local distributed databases
are in MySQL installations and can be substantially smaller
depending on the local needs.  Data in the distributed MySQL
servers are automatically synchronized with the MySQL
warehouse through a multiple-master replication scheme.


\section{Conclusions}
\label{sec:conclude}

The MINOS far detector has been in continuous operation since July~2003
and the near detector has been operating since January~2005. Both detectors
and their magnet coils routinely operate nearly 99\% of the time when the
neutrino beam is on, with close to 100\% of their readout channels
working properly. 

Cosmic ray muons are used to calibrate the MINOS detectors, correcting
for drifts over time, for variation of response over position within a
detector, and for differences between detectors. The light injection
system monitors the PMT nonlinearities and
gains. Together, these calibration methods have established the
relative near-far shower energy scale to 2\%. The calibration detector
test-beam measurements have established the shower energy scale to a
precision of 6\%.
With units of $E$ expressed in GeV, the measured calorimetric energy 
resolutions of $21.4\%/\sqrt{E}\oplus 4\%/E$ for electromagnetic showers
and  $56\%/\sqrt{E}\oplus2\%$ for hadronic showers are in excellent agreement
with predictions. 

The far detector time resolution of \unit[2.3]{ns} is sufficient to give
useful time-of-flight information for cosmic-ray events, and allows a
clean sample of upward-going muons (from atmospheric-neutrino
interactions in the rock beneath the detector) to be identified. The
near detector time resolution of approximately \unit[5]{ns} provides a clean
separation of neutrino events from different RF buckets within the same
\unit[8--10]{$\mu$s} Main Injector spill. 
The experiment routinely
transmits spill-time information over the internet from Fermilab to
Soudan, allowing a real-time spill gate to be applied (in the trigger
software) to far detector neutrino beam data with an uncertainty of
$\unit[< 1]{\mu s}$.

As of early 2008, the far detector has measured neutrino oscillation parameters
from atmospheric-neutrino events recorded during its first two years of data
taking~\cite{Adamson:2006an} and higher-energy upward-going
muons~\cite{Adamson:2007vt} as charge-seperated samples. The experiment
has also accumulated data from more than
$4 \times 10^{20}$ Main
Injector protons on target, less than half of its total expected data sample,
and has made its first measurements of
$\Delta m^2$ and $\sin^2 2\theta$~\cite{Michael:2006rx,Adamson:2007gu}.
Thus it has been demonstrated that
the performance and calibrations of the near and far detectors 
provide a suitable foundation for achieving
the neutrino physics goals of the MINOS experiment. 

\section*{Acknowledgments}
\label{sec:conclude-ack}

This work was supported by the U.S. Department of Energy, the U.K. Particle
Physics and Astronomy Research Council, the U.S. National Science Foundation, 
the State and University of Minnesota, the Office of Special Accounts for
Research Grants of the University of Athens, Greece, and FAPESP (Fundacao de
Amparo a Pesquisa do Estado de Sao Paulo) and CNPq (Conselho Nacional de
Desenvolvimento Cientifico e Tecnologico) in Brazil. This experiment would not
have been possible without the dedicated efforts of the members of the
Fermilab Accelerator and Particle Physics Divisions in building and operating
the NuMI neutrino beamline. We thank the members of the Beam Design Group at
the Institute for High Energy Physics, Protvino, Russia for their important
contributions to the designs of the neutrino-beam target and horn systems. We
gratefully acknowledge the Minnesota Department of Natural Resources for their
assistance and for allowing us access to the facilities of the Soudan
Underground Mine State Park. We also thank the crew of the Soudan Underground
Laboratory for their tireless work in building and operating the MINOS
far detector. Students in the University of Minnesota Mechanical
Engineering Department made substantial contributions to the design of
the scintillator module crimping machine.



\bibliographystyle{elsart-num}

\bibliography{minos-nim}

\begin{thebibliography}{10}
\bibitem{ASDLite}
ASDLite is a prototype ASIC originally developed for the ATLAS muon chamber
  readout.
\end{thebibliography}

\end{document}